\documentclass[fleqn,usenatbib]{mnras}

\usepackage{newtxtext,newtxmath}

\usepackage[T1]{fontenc}
\usepackage{ae,aecompl}
\usepackage{url}
\usepackage{placeins}

\usepackage{graphicx}	
\usepackage{amsmath}	
\usepackage{pifont}
\usepackage{float}
\usepackage{color}
\usepackage{grffile} 
\usepackage{pdflscape}

\newcommand{\xmark}{\ding{55}}%
\newcommand{\lsim}{{\lower.5ex\hbox{$\; \buildrel < \over \sim \;$}}}
\newcommand{\gsim}{{\lower.5ex\hbox{$\; \buildrel > \over \sim \;$}}}
\newcommand{\nufnu}{$\nu$f$({\nu})$}

\newcommand{\gr}{$\gamma$-ray}
\newcommand{\nup}{$\nu_{\rm peak}^S$}
\newcommand{\fermi}{{\it Fermi}}

\usepackage{lineno}
\usepackage{ulem}

\definecolor{darkspringgreen}{rgb}{0.09, 0.45, 0.27}

\defcitealias{Paiano_2021}{Paper~I}

\title[The Spectra of IceCube Neutrinos - IV]{The Spectra of IceCube Neutrino (SIN) candidate sources - IV. Spectral energy distributions and multi-wavelength variability}

\author[Karl et al.]{Martina Karl$^{1, 2}$\thanks{E-mail:
martina.karl@eso.org}, Paolo Padovani$^{2,3}$, Paolo Giommi$^{4,5,6}$ \\
$^{1}$Technische Universit{\"a}t M{\"u}nchen, TUM School of Natural Sciences, Physics Department, 
James-Frank-Str. 1, D-85748 Garching bei M{\"u}nchen, Germany\\
$^{2}$European Southern Observatory, Karl-Schwarzschild-Str. 
2, D-85748 Garching bei M\"unchen, Germany\\
$^{3}$Associated to INAF - Osservatorio di Astrofisica e Scienza dello Spazio, Via Piero 
Gobetti 93/3, I-40129 Bologna, Italy\\
$^{4}$Institute for Advanced Study, Technische Universit{\"a}t M{\"u}nchen,
Lichtenbergstrasse 2a, D-85748 Garching bei M\"unchen, Germany\\
$^{5}$Center for Astro, Particle and Planetary Physics (CAP3), New York University Abu Dhabi, PO Box 129188 Abu Dhabi, United Arab Emirates\\
$^{6}$Associated to INAF, 
Osservatorio Astronomico di Brera, 
via Brera, 28, I-20121 Milano, Italy\\
}

\date{Accepted XX. Received YY; in original form ZZ}

\pubyear{2023}

\begin{document}
\label{firstpage}
\pagerange{\pageref{firstpage}--\pageref{lastpage}}
\maketitle

\begin{abstract}

We present hybrid spectral energy distributions, combining photon and neutrino fluxes, for a sample of blazars, which are candidate IceCube neutrino sources. We furthermore check for differences in our sources' variability in the near-infrared, optical, X-ray and $\gamma$-ray bands compared to a sample of non-neutrino source candidate blazars, and investigate the state of each blazar at the arrival time of high-energy neutrinos. We find no significant differences when comparing our sample with control sources, also in terms of their spectral energy distributions, and no correlation between flaring states and neutrino arrival times. Looking for signatures of hadronic production, we check for similar strengths of the $\gamma$-ray and neutrino fluxes and find a $2.2\,\sigma$ signal for our source candidates. 
The hybrid spectral energy distributions assembled here will form the basis of the next step
of our project, namely lepto-hadronic modelling of these blazars to assess the physical likelihood 
of a neutrino connection. 

\end{abstract}

\begin{keywords}
neutrinos --- radiation mechanisms: non-thermal --- galaxies: active 
--- BL Lacertae objects: general --- gamma-rays: galaxies ---  astroparticle physics
\end{keywords}

\section{Introduction}\label{sec:Introduction}

Ten years ago the IceCube Neutrino Observatory\footnote{\url{http://icecube.wisc.edu}} 
detected the first high-energy astrophysical neutrinos of likely extragalactic origin
with energies reaching more than 1 PeV ($10^{15}$ eV: \citealt{IceCube_2013}) and since then has 
produced a growing list of events \citep[e.g.][and references therein]{abbasi2022constraints}. So far, however, only 
two extragalactic objects have been associated with a significance larger than 
$\sim 3\,\sigma$ with IceCube neutrinos. Namely, the blazar TXS\,0506+056 
at $z=0.3365$ (at the $3 - 3.5\,\sigma$ level: \citealt{icfermi,neutrino}) 
and the local ($z = 0.004$) Seyfert 2 galaxy NGC\,1068 (at the $4.2\,\sigma$ level: \citealt{doi:10.1126/science.abg3395f}) (see also \citealt{GP_2021} for a recent review on
possible astronomical associations). 

\citet[hereafter G20]{Giommidissecting} carried out a detailed 
dissection of all 
the public IceCube high-energy neutrinos that were well reconstructed 
(so-called tracks) and off the Galactic
plane at the time of publication, which provided a $3.2\,\sigma$ (post-trial) correlation
excess with $\gamma$-ray detected IBLs\footnote{Blazars are sub-divided
from a spectral energy distribution (SED) point of view on the basis of the
rest-frame frequency of their low-energy (synchrotron) hump (\nup) into
low- (LBL/LSP: \nup~$<10^{14}$~Hz [$<$ 0.41 eV]), intermediate- (IBL/ISP:
$10^{14}$~Hz$ ~<$ \nup~$< 10^{15}$~Hz [0.41 -- 4.1 eV)], and high-energy
(HBL/HSP: \nup~$> 10^{15}$~Hz [$>$ 4.1 eV]) peaked sources respectively
\citep{Padovani1995,Abdo_2010}.} and HBLs. No excess was found for LBLs.
Based on these results, out of the 47 IBLs and HBLs in Table 5 of G20 
16 $\pm$ 4 could be new neutrino sources waiting to be identified. 

To test this out, we required optical spectra, needed 
to measure the redshift, which were not available for the majority of
the G20 sample, and hence derive the luminosity of the source, vital to do
any modelling, determine the properties of the spectral lines, and also 
possibly estimate the mass of the central black hole, $M_{\rm BH}$.

Our group has therefore started ``The spectra of
IceCube Neutrino (SIN) candidate sources'' project, whose main aims are to: (1) 
conduct an observational campaign to secure optical spectra; (2) determine the nature of 
the sources; (3) assemble their SEDs using all available multi-wavelength and neutrino 
data; (4) model their SEDs and, subsequently, the expected neutrino
emission from each blazar; (5) determine the likelihood of a connection 
between the neutrino and the blazar using a physical model for the 
blazar multi-messenger emissions, as done, for example, by
some of us in \cite{Petropoulou2015,Petropoulou2020}.

We have addressed point (1) in \citet[hereafter Paper I]{Paiano_2021}  and 
\citet[hereafter Paper III]{Paiano_2023}, who presented the spectroscopy of a
large fraction of the objects selected by G20, which, together with results taken 
from the literature, covered the whole sample\footnote{To the 
G20 sample we further added M87 and 3HSP J095507.9+35510, for reasons described in Paper II, for a total of 49 sources.}. Point (2) was dealt with in 
\citet[hereafter Paper II]{Padovani_2022} and Paper III, where it was shown that the 
BL Lacs under study were, in many cases (34 -- 77 per cent), masquerading BL Lacs, 
i.e. objects in which the weakness of the emission lines and their low values of the
equivalent widths were due to a very bright Doppler-boosted continuum, 
which was washing out the lines, unlike ``real'’ BL Lacs, which are intrinsically weak-lined objects. 
As discussed in Papers II and III, this is extremely important for two reasons: (1) ``real'' BL Lacs and FSRQs turn out to belong to 
two very different physical classes, i.e. objects without and with high-excitation emission lines in their optical spectra, referred to as low-excitation (LEGs) 
and high-excitation galaxies (HEGs), respectively (e.g. \citealt{AGNReview}, and references therein); (2) masquerading BL Lacs, being HEGs, benefit from 
several radiation fields external to the jet (i.e. the accretion disc, photons reprocessed in the broad-line region (BLR) or from the dusty torus), which, 
by providing more targets for the protons, might enhance neutrino production as compared to LEGs. Indeed
TXS~0506+056, which is the blazar with the most significant association with a neutrino event \citep{icfermi,neutrino} is also a masquerading BL Lac \citep{Padovani2019}. 

The purpose of this paper is to take care of point (3), that is, collect all
available multi-wavelength and neutrino data for the G20 sources and put together their hybrid SEDs.
Given the amount of available data, we also study our sources' variability in 
the near-infrared (IR), optical, X-ray, and $\gamma$-ray 
bands to see how they  behave compared to the rest of the blazar population,
and check if there are any electromagnetic flares in these bands coincident with the
neutrino arrival times. 

The paper is structured as follows: Section \ref{sec:update-tracks} presents 
updates to the G20 sources, in Section \ref{sec:photon_data} 
and \ref{sec:neutrinos} we describe the multi-messenger (photon and neutrino) 
data that we have used in this paper, 
while Section \ref{sec:results} presents the results of the SEDs, the variability properties in the
various bands, and searches for flares coincident with neutrinos. 
Finally, Section \ref{sec:discussion} discusses our results, and
Section \ref{sec: conclusions} summarises our conclusions. 
Appendix \ref{sec:appendix} provides complementary information. 
We use a 
$\Lambda$CDM concordance cosmology with Hubble constant $H_0 = 70$ km s$^{-1}$ Mpc$^{-1}$, matter
density $\Omega_{\rm m,0} = 0.3$, and dark energy density
$\Omega_{\Lambda,0} = 0.7$.

\section{An update to recent IceCube tracks}\label{sec:update-tracks}

Since 2019 the IceCube collaboration has revised the criteria for identifying neutrino events with a probable astrophysical origin \citep{blaufuss2019generation, abbasi2023icecat1}. They also introduced different classifications of tracks, differentiating between ``gold'' and ``bronze'' alerts. Gold alerts aim for a higher astrophysical purity compared to bronze alerts. The previous selection criteria for the tracks in G20 are mostly comparable to the gold stream. Based on \citet{abbasi2023icecat1}, we identified which G20 sources 
no longer belong to the updated high-energy neutrino event selection\footnote{Tracks previously used by G20 but no longer included in the revised IceCube event list are IC160331A, IC141109A, IC150428A, IC141209A, IC141126A, IC120123A, IC140216A, IC170506A, IC190104A.} and list them in the bottom section of Table \ref{Tab:FVresults}. This leaves 34 out of 49 G20 sources for this paper's analysis. However, 
we also present the multi-messenger SEDs of the fifteen excluded sources since they were already assembled when the new selection was published. 

We note that the original selection for the G20 sample included
\nup~$>10^{14}$ Hz. As described in the G20 paper, \nup~values were 
derived by first removing all data points that could be attributed 
to components unrelated to the jet synchrotron emission, namely 
the host galaxy (usually in the IR and optical bands), the blue bump (blue + UV), 
and inverse Compton emission (X-ray). The remaining data were then 
fitted with a 2nd-order polynomial using the SSDC–SED tool\footnote{\url{https://tools.ssdc.asi.it/SED/}}. 
We have re-estimated \nup~for all our sources as done for
the control sample (where applicable) using BlaST \citep{Blast} after retrieving the SEDs for each of these blazars with VOU-Blazars V1.96 \citep{VOU-Blazars}. The average logarithmic difference between the G20 \nup~and BlaST \nup~is $0.2 \pm 0.1$.
This is within the mean uncertainty of the BlaST log \nup~= $0.54 \pm 0.02$. As a result, and
because now we also have more data available compared to G20,
some sources turned out to have \nup~$<10^{14}$ Hz. However, 
we follow \cite{GP_2021} and consider only IHBLs, i.e., sources
with rest-frame (i.e., multiplied by [1+z]) 
\nup~$\geq 10^{13.5}$~Hz (within their uncertainties),
which means no source gets excluded due to their \nup. 
Tab. \ref{Tab:FVresults} lists
the current sample, their rest-frame \nup, and the sources excluded because
of the revised IceCube selection criteria.

\section{Photon data}\label{sec:photon_data}

\subsection{Multi-wavelength SEDs}

We obtain the multi-frequency data with VOU-Blazars, which queries many different catalogues (listed in Appendix \ref{sec:appendix}) providing flux measurements across a wide range of the electromagnetic spectrum. TeV data for two sources (3HSP J023248.5+20171, a.k.a. 1ES 0229+200, and TXS 0506+056) were added by hand \citep{icfermi, 10.1093/mnras/sty857}.

For the $\gamma$-ray band, we use data from the \fermi~large area telescope (\fermi-LAT: \citealt{Ackermann_2012}) from August 2008 to August 2022 as available. We use the python package Fermipy \citep{2017ICRC...35..824W} for the analysis. We apply the recommended selections for an off-plane point source analysis\footnote{\url{https://fermi.gsfc.nasa.gov/ssc/data/analysis/documentation/Cicerone/Cicerone_Data_Exploration/Data_preparation.html}} for Pass 8 data (P8R3). This includes filtering for photons (evclass~=~128, evtype~=~3) with energies 
$0.1 - 100$~GeV within a radius of 9$^{\circ}$ around the source. Furthermore, we remove Earth limb contamination by requiring a zenith angle~$\leq 90^{\circ}$ and requiring good data quality ((DATA\_QUAL~>~0) and (LAT\_CONFIG~=~1)). The procedures for the preparation of the data are described in the \fermi~Cicerone\footnote{\url{https://fermi.gsfc.nasa.gov/ssc/data/analysis/documentation/Cicerone/}}.

After selecting the data, we define the model in the following way. The background hypothesis comprises known sources within a radius of 95 per cent of the point spread function, the diffuse galactic and the diffuse isotropic $\gamma$-ray background. The signal hypothesis is the background scenario with an additional signal by a point source at the centre. We assume a signal emission following a power law in most cases, unless a log-parabola is a better fit. The test statistic is the ratio of the maximised background likelihood $\mathcal{L}_{\text{max,}0}$ versus the maximised signal likelihood $\mathcal{L}_{\text{max,}1}$: $\text{TS}=-2\ln \left( \mathcal{L}_{\text{max,}0}/\mathcal{L}_{\text{max,}1} \right)$. A larger TS value indicates that the background hypothesis can be rejected. The best-fit parameters for the central point source are then plotted in the SED. We provide the output of the fits in a GitHub repository\footnote{\url{https://github.com/mskarl/SINIV-Fermi-SED}}.

\subsection{Variability study}\label{sec:variabilityData}

For the variability study of blazars, we collect data in the IR, optical, X-ray, and $\gamma$-ray bands. By construction, all the candidate neutrino blazars in the G20 sample are also \gr\, sources. Therefore, the natural control sample is the set of blazars in the 4LAC-DR3 catalogue 
\citep{Ajello_2020, Lott_2020}. 
We consider all objects classified as blazars (FSRQs, BL Lacs, and BCUs) without entries in the 
analysis flag (other than 0). 
Since the redshift is 
not known for all background blazars, we use the observed \nup~(and not the rest-frame one) when binning for the variability analysis for both the background and the G20 samples. We also apply the \nup$~\geq 10^{13.5}$ Hz cut to the {\it observed} values to both samples, as 
this is a good value to separate LBLs from IHBL blazars \citep[e.g.][]{GP_2021}. 
This removes three sources (5BZU J0158+0101, CRATESJ024445+132002, and CRATESJ232625+011147) from our sample and leaves us with 31 sources for the variability study when binning in \nup\, for the comparison to the background sample. We note that these three sources still meet the \nup\, criterion within their uncertainties when considering their rest-frame \nup. 
Table \ref{Tab:LCDataOverview} lists the time span covered and the sampling of the light curves used to calculate the FVs in the IR, optical, and X-ray bands.

\subsubsection{IR band}

The intermediate location of IR wavelengths within the overall SED is particularly beneficial for blazar studies since the spectral slope in this energy band is a direct measurement of whether \nup~is located at frequencies \lsim $10^{13}$ Hz, like in LBL sources, or $\gsim 10^{13.5}$ Hz, like in the case of IBL and HBL blazars. The amount of IR variability over time is also useful in quantifying the dynamics of the multi-frequency emission in blazars.

The WISE satellite \citep{WISEmission} observed the entire sky at IR frequencies during its main mission in 2010 and as part of the NEOWISE reactivation phase \citep{Mainzer2014}, which started in 2013 and is still ongoing. Although the main purpose of NEOWISE is to track near-Earth objects, it also detected a large number of blazars at two infrared wavelengths (3.4 and 4.6 $\micron$), generating a very valuable blazar variability database covering a period of over nine years.

We found WISE and NEOWISE data for 47 out of the original 49 sources using the 
VOU-Blazars tool \citep{VOU-Blazars} V1.96. After removing sources as described above, we have 30 sources with IR data left. 
Furthermore, we retrieved WISE and NEOWISE data for the control sample in the 4LAC-DR3 catalogue with \nup$~\geq 10^{13.5}$ Hz, i.e., 1479 blazars.

\subsubsection{Optical band}

Whenever possible, we retrieved flux measurements from the Zwicky Transient Facility \citep[ZTF:][]{ztf}. This sky monitoring survey offers magnitudes measurements with a three-day cadence in three optical filters for astronomical sources at $\delta \gtrsim
-30^{\circ}$. 

We generated optical light curves with data from the ZTF for the G20 sources and the blazars in the 4LAC-DR3 catalogue falling within
the sampled declination range, deriving results for 43 out of the original sample of 49 G20 sources. After applying the updated criteria as described above, we have 29 optical light curves and fractional variabilities (FVs: see equation \ref{eq:FV}). The background sample consists of 1135 blazars. 

\subsubsection{X-ray band} 
Swift-XRT\footnote{\url{https://www.swift.ac.uk/swift\_portal/}} observations are available for 37 of the 49 G20 sources. Of these, 17 sources have sufficient measurements to calculate the FV. After cleaning the sample as described above, eleven sources remain. As for the control sample, most 4LAC-DR3 blazars have not been observed multiple times in the X-ray band. Hence, only for the X-ray band, we use the 63 sources observed by
Swift-XRT at least fifty times \citep{GiommiXRTspectra} as a control sample. Selecting sources with \nup$~\geq 10^{13.5}$ Hz leaves 34 background objects. 

\subsubsection{$\gamma$-ray band}

The FVs for 28 out of 34 G20 sources are available from 
the 4FGL-DR3 catalogue \citep{2022ApJS..260...53A}. The 4FGL-DR3 catalogue covers 12 years of data in the energy range of 50~MeV-1~TeV and calculates the FVs based on yearly fluxes; in 15 cases FV $>0$. 
As a background sample, we retrieve the FVs for the 4LAC-DR3 blazars with \nup$~\geq 10^{13.5} $ Hz, for a total
of 1628 blazars.

\subsection{Flaring state of G20 sources}
In addition to their variability, we study the state of the G20 sources at the time of the neutrino arrival. For this, we use the light curves associated with the data described in the previous section. In the optical, however, the ZTF was only commissioned in 2018. Most of the G20 alerts were detected before that date so we additionally use light curves from the All-Sky Automated Survey for Supernovae (ASAS-SN) \citep{asasn2017} provided by \citet{deJaeger2023} where available. In the $\gamma$-ray band, we use the \fermi-LAT light curve repository (\fermi-LAT LCR) \citep{2023ApJS..265...31A}. The \fermi-LAT LCR includes only sources that exceed a minimum threshold of the variability index (21.67), where the variability index is related to the average FV over one year. 

\section{Neutrino data for multi-messenger SEDs}\label{sec:neutrinos}

Besides electromagnetic data, we also want to add neutrino information to our
SEDs in two complementary ways, namely by estimating fluxes and sensitivities.
We then initially use ten years of published IceCube through-going muon tracks 
\citep{IceCube_2021} to fit, for the first time, each source 
candidate's neutrino flux. We employ an unbinned maximum likelihood approach \citep{Braun:2008bg}, where we compare a background and a signal hypothesis:
\begin{itemize}
    \item background hypothesis $H_B$: the neutrino emission is due to atmospheric background and diffuse astrophysical neutrino emission;
    \item signal hypothesis $H_S$: an additional signal component comes from the source and clusters around it. We assume it follows a power law $\propto E^{-\gamma}$. The total neutrino emission observed is the sum of the signal and background components. 
\end{itemize}
This likelihood approach is detailed in Appendix \ref{appendix:neutrinos}. In most cases, however, the 68 per cent uncertainties are compatible with no neutrino emission and therefore, we provide only a 68 per cent upper limit (based on \citealt{Feldman_1998}).

Furthermore, we show the sensitivity and the discovery potential from  
\cite{Aartsen2020}. The former describes the 90 per cent confidence upper flux 
limit IceCube can set when there are no events from the source direction. The 
discovery potential describes the source flux at which IceCube has a 50 per cent chance 
to make a $5\,\sigma$ discovery (since the number of signal neutrinos produced in a 
source with a specific flux is Poissonian distributed). For both the sensitivity and the discovery potential, it is assumed the neutrino emission follows a power law with $E^{-2}$ \citep{Aartsen2020}. Since the analysed data set contains only muons, the inferred fluxes are for muon neutrinos and muon antineutrinos. The all-flavour fluxes can be estimated by multiplying the single-flavour flux by a factor of 3.

\section{Results}\label{sec:results}
\subsection{SEDs}
We present the hybrid SEDs in Appendix \ref{appendix:SEDS}. There are 28 cases of the original 49 G20 sources where the best-fit neutrino flux is greater than zero. This applies to 23/34 of the still included sources. However, only in two of these cases, the 68 per cent contours are incompatible with no neutrino flux at all: TXS~0506+056 (4FGL~J0509.4+0542) and 3HSP~J152835.7+20042 (4FGL~J1528.4+2004). In all other cases, we present the upper bound of the 68 per cent uncertainty as a 68 per cent upper flux limit compatible with the best fit. There are three cases where the best-fit spectral index of the neutrino flux is harder than $-2$. However, in most cases the number of expected signal neutrinos is quite small, i.e. of the order of one neutrino or less. Thus, the spectra derived with such small numbers of events might change
considerably with more data. 

\subsection{Variability}

Variability is one of the properties common to all blazars. Quantifying the amount of this defining property is a fundamental way to investigate  
the physical mechanisms that power these objects. An estimator that is often used to measure the degree of variability in blazars is
the FV parameter \citep[e.g.][]{Schleicher2019}.
In the following, we calculate the FV of the G20 sources and of a large control sample in the IR, optical, X-ray, and $\gamma$-ray bands.

Following \cite{Vaughan2003} we calculate FV as follows 

\begin{equation}
\text{FV} = \sqrt{\frac{S^2-\langle ~\sigma_{err}^2 \rangle}{ \langle ~f~ \rangle^2}},
\label{eq:FV}
\end{equation}
where $S^2$ is the variance of the observed fluxes ($f$) in a given source and $\sigma_{err}$ is the uncertainty of the single flux measurements.

The statistical error of FV can be calculated as follows

\begin{equation}
\sigma_{\text{FV}} = \sqrt{
\text{FV}^2+\sqrt{\frac{2~\langle ~\sigma_{err}^2\rangle^2}{N ~\langle ~f~ \rangle^4} 
+ \frac{4~\langle ~\sigma_{err}^2\rangle ~FV^2}{N ~\langle ~f~ \rangle^2}}}-FV .
\end{equation}

We now compare the G20 sources FV in different wavelengths 
with a background sample of blazars as described in Section \ref{sec:variabilityData}. Table \ref{Tab:FVresults} gives all FVs for the G20 sources. A comparison between FVs in different 
bands would require the same time window and a time
binning to properly sample the variability in all bands, which is not our case.

\begin{scriptsize}
\begin{table*}
\setlength{\tabcolsep}{1.7pt}
\caption{FVs of G20 sources in different bands.}
\begin{center}
\begin{tabular}{llcrccccl}
\hline\hline
4FGL Name & Name 
& RA & Dec.~~~~~~ &\multicolumn{4}{c}{Fractional Variability} & log (\nup)  \\
     &       &    
     &     &   infrared & optical   &  X-ray    & $\gamma$-ray  &  rest-frame [Hz]\\ 
    (1)    & (2) 
    &  (3)  & (4)~~~~~~~~~& (5)  & (6) & (7) & (8) & (9)  \\
\hline
\textbf{4FGL J0158.8+0101}  &  \textbf{5BZU J0158+0101}  &  01 58 52.776  &  $01$ 01 32.880  &   0.18 $\pm$ 0.08  &   0.35 $\pm$ 0.06  &   -  &   -  & 13.1 $\pm$  0.5 \\ 
4FGL J0224.2+1616  &  VOU J022411+161500  &  02 24 11.808  &  $16$ 15 00.000  &   0.23 $\pm$ 0.04  &   0.38 $\pm$ 0.03  &   -  &   -  & 14.7 $\pm$  0.4 \\ 
\textit{4FGL J0232.8+2018}  &  \textit{3HSP J023248.5+20171}  &  02 32 48.600  &  $20$ 17 18.600  &   0.06 $\pm$ 0.05  &   0.10 $\pm$ 0.03  &   0.28 $\pm$ 0.04  &   -  & 17.6 $\pm$  0.5 \\ 
4FGL J0239.5+1326  &  3HSP J023927.2+13273  &  02 39 27.216  &  $13$ 27 38.520  &   0.2 $\pm$ 0.1  &   0.2 $\pm$ 0.1  &   -  &   1.4 $\pm$ 0.4  & 14.2 $\pm$  1 \\ 
4FGL J0244.7+1316  &  CRATESJ024445+132002  &  02 44 45.696  &  $13$ 20 07.080  &   0.54 $\pm$ 0.02  &   0.16 $\pm$ 0.05  &   -  &   0.6 $\pm$ 0.2  & 13.3 $\pm$  0.5 \\ 
\textit{4FGL J0344.4+3432}  &  \textit{3HSP J034424.9+34301}  &  03 44 24.936  &  $34$ 30 17.280  &   0.41 $\pm$ 0.02  &   0.38 $\pm$ 0.03  &   -  &   -  & 15.5 $\pm$  0.5 \\ 
\textit{\textbf{4FGL J0509.4+0542}}  &  \textit{\textbf{TXS 0506+056}}  &  05 09 25.964  &  $05$ 41 35.334  &   0.376 $\pm$ 0.007  &   0.355 $\pm$ 0.006  &   0.64 $\pm$ 0.04  &   0.6 $\pm$ 0.1  & 14.6 $\pm$  0.4 \\ 
3FGL J0627.9-1517  &  3HSP J062753.3-15195  &  06 27 53.376  &  $-15$ 19 57.000  &   0.06 $\pm$ 0.09  &   0.16 $\pm$ 0.10  &   -  &   -  & 17.4 $\pm$  0.4 \\ 
4FGL J0649.5-3139  &  3HSP J064933.6-31392  &  06 49 33.528  &  $-31$ 39 20.160  &   0.25 $\pm$ 0.04  &   -  &   0.3 $\pm$ 0.1  &   0.6 $\pm$ 0.2  & 17.2 $\pm$  0.5 \\ 
\textit{4FGL J0854.0+2753}  &  \textit{3HSP J085410.1+27542}  &  08 54 10.176  &  $27$ 54 21.600  &   0.08 $\pm$ 0.09  &   0.1 $\pm$ 0.1  &   -  &   -  & 15.7 $\pm$  0.6 \\ 
\textit{4FGL J0946.2+0104}  &  \textit{3HSP J094620.2+01045}  &  09 46 20.208  &  $01$ 04 51.600  &   0.1 $\pm$ 0.1  &   0.13 $\pm$ 0.09  &   0.2 $\pm$ 0.2  &   0.4 $\pm$ 0.2  & 17.6 $\pm$  0.4 \\ 
\textit{4FGL J0955.1+3551}  &  \textit{3HSP J095507.9+35510}  &  09 55 07.882  &  $35$ 51 00.885  &   0.11 $\pm$ 0.09  &   0.19 $\pm$ 0.05  &   0.34 $\pm$ 0.06  &   -  & 17.6 $\pm$  0.5 \\ 
\textit{4FGL J1003.4+0205}  &  \textit{3HSP J100326.6+02045}  &  10 03 26.616  &  $02$ 04 55.560  &   0.21 $\pm$ 0.06  &   0.20 $\pm$ 0.04  &   0.7 $\pm$ 0.2  &   -  & 15.3 $\pm$  0.4 \\ 
4FGL J1055.7-1807  &  VOU J105603-180929  &  10 56 03.528  &  $-18$ 09 30.240  &   0.85 $\pm$ 0.02  &   0.2 $\pm$ 0.1  &   -  &   -  & 15.3 $\pm$  1 \\ 
\textit{\textbf{4FGL J1117.0+2013}}  &  \textit{\textbf{3HSP J111706.2+20140}}  &  11 17 06.216  &  $20$ 14 07.080  &   0.20 $\pm$ 0.03  &   0.24 $\pm$ 0.01  &   0.42 $\pm$ 0.09  &   0.5 $\pm$ 0.1  & 16.3 $\pm$  0.5 \\ 
4FGL J1124.0+2045  &  3HSP J112405.3+20455  &  11 24 05.352  &  $20$ 45 52.920  &   0.20 $\pm$ 0.05  &   0.29 $\pm$ 0.03  &   -  &   -  & 15.6 $\pm$  0.4 \\ 
4FGL J1124.9+2143  &  3HSP J112503.6+21430  &  11 25 03.552  &  $21$ 43 04.080  &   0.16 $\pm$ 0.06  &   0.19 $\pm$ 0.08  &   -  &   -  & 16.0 $\pm$  0.5 \\ 
\textit{3FGL J1258.4+2123}  &  \textit{3HSP J125821.5+21235}  &  12 58 21.456  &  $21$ 23 51.000  &   0.1 $\pm$ 0.1  &   0.0 $\pm$ 0.2  &   -  &   -  & 16.2 $\pm$  0.7 \\ 
4FGL J1258.7-0452  &  3HSP J125848.0-04474  &  12 58 48.048  &  $-04$ 47 45.240  &   0.21 $\pm$ 0.05  &   0.23 $\pm$ 0.03  &   0.38 $\pm$ 0.07  &   0.8 $\pm$ 0.3  & 17.6 $\pm$  0.4 \\ 
4FGL J1300.0+1753  &  3HSP J130008.5+17553  &  13 00 08.520  &  $17$ 55 37.560  &   0.19 $\pm$ 0.09  &   0.20 $\pm$ 0.07  &   -  &   0.1 $\pm$ 0.6  & 14.2 $\pm$  1 \\ 
\textit{\textbf{4FGL J1314.7+2348}}  &  \textit{\textbf{5BZB J1314+2348}}  &  13 14 43.800  &  $23$ 48 26.640  &   0.539 $\pm$ 0.007  &   0.28 $\pm$ 0.01  &   -  &   0.23 $\pm$ 0.07  & 14.3 $\pm$  0.6 \\ 
\textbf{4FGL J1321.9+3219}  &  \textbf{5BZB J1322+3216}  &  13 22 47.400  &  $32$ 16 08.760  &   0.18 $\pm$ 0.04  &   0.27 $\pm$ 0.02  &   -  &   0.8 $\pm$ 0.3  & 14.0 $\pm$  0.8 \\ 
\textbf{4FGL J1507.3-3710}  &  \textbf{VOU J150720-370902}  &  15 07 20.808  &  $-37$ 09 02.880  &   0.25 $\pm$ 0.02  &   -  &   -  &   -  & 14.4 $\pm$  0.3 \\ 
\textit{\textbf{4FGL J1528.4+2004}}  &  \textit{\textbf{3HSP J152835.7+20042}}  &  15 28 35.784  &  $20$ 04 20.280  &   0.2 $\pm$ 0.1  &   0.28 $\pm$ 0.06  &   0.6 $\pm$ 0.2  &   -  & 16.9 $\pm$  0.5 \\ 
\textit{4FGL J1533.2+1855}  &  \textit{3HSP J153311.2+18542}  &  15 33 11.256  &  $18$ 54 29.160  &   0.08 $\pm$ 0.06  &   0.11 $\pm$ 0.08  &   0.1 $\pm$ 0.2  &   -  & 17.4 $\pm$  0.5 \\ 
\textit{4FGL J1554.2+2008}  &  \textit{3HSP J155424.1+20112}  &  15 54 24.120  &  $20$ 11 25.080  &   0.07 $\pm$ 0.05  &   0.10 $\pm$ 0.04  &   0.40 $\pm$ 0.09  &   -  & 17.0 $\pm$  0.8 \\ 
4FGL J1808.2+3500  &  CRATESJ180812+350104  &  18 08 11.544  &  $35$ 01 18.840  &   0.29 $\pm$ 0.01  &   0.426 $\pm$ 0.004  &   -  &   0.5 $\pm$ 0.1  & 14.4 $\pm$  0.4 \\ 
\textbf{4FGL J1808.8+3522}  &  \textbf{3HSP J180849.7+35204}  &  18 08 49.704  &  $35$ 20 43.080  &   0.16 $\pm$ 0.03  &   0.29 $\pm$ 0.03  &   -  &   -  & 14.9 $\pm$  0.6 \\ 
\textit{4FGL J2030.5+2235}  &  \textit{3HSP J203031.6+22343}  &  20 30 31.096  &  $22$ 34 37.019  &   0.08 $\pm$ 0.04  &   0.25 $\pm$ 0.06  &   -  &   -  & 16.2 $\pm$  0.4 \\ 
\textit{\textbf{4FGL J2030.9+1935}}  & \textit{\textbf{3HSP J203057.1+19361}}  &  20 30 57.120  &  $19$ 36 12.960  &   0.26 $\pm$ 0.04  &   0.15 $\pm$ 0.02  &   -  &   0.3 $\pm$ 0.1  & 15.9 $\pm$  0.5 \\ 
\textit{4FGL J2133.1+2529}  &  \textit{3HSP J213314.3+25285}  &  21 33 14.352  &  $25$ 28 59.160  &   0.09 $\pm$ 0.06  &   0.16 $\pm$ 0.05  &   -  &   -  & 14.6 $\pm$  0.5 \\ 
\textit{4FGL J2223.3+0102}  &  \textit{3HSP J222329.5+01022}  &  22 23 29.520  &  $01$ 02 26.160  &   -  &   0.31 $\pm$ 0.03  &   -  &   0.4 $\pm$ 0.6  & 14.9 $\pm$  0.4 \\ 
\textbf{4FGL J2227.9+0036}  &  \textbf{5BZB J2227+0037}  &  22 27 58.152  &  $00$ 37 05.520  &   0.16 $\pm$ 0.04  &   0.19 $\pm$ 0.03  &   -  &   0.3 $\pm$ 0.1  & 14.4 $\pm$  0.4 \\ 
\textbf{4FGL J2326.2+0113}  &  \textbf{CRATESJ232625+011147}  &  23 26 25.632  &  $01$ 12 08.640  &   0.82 $\pm$ 0.01  &   1.071 $\pm$ 0.006  &   -  &   1.0 $\pm$ 0.3  & 13.8 $\pm$  0.5 \\ 
\hline
\textbf{4FGL J0103.5+1526}  &  \textbf{3HSP J010326.0+15262}  &  01 03 26.016  &  $15$ 26 24.720  &   0.05 $\pm$ 0.06  &   0.14 $\pm$ 0.05  &   -  &   -  & 15.1 $\pm$  1 \\ 
\textit{4FGL J0339.2-1736}  &  \textit{3HSP J033913.7-17360}  &  03 39 13.703  &  $-17$ 36 00.783  &   0.08 $\pm$ 0.03  &   0.19 $\pm$ 0.01  &   0.1 $\pm$ 0.3  &   0.2 $\pm$ 0.1  & 15.5 $\pm$  0.5 \\ 
\textbf{4FGL J0525.6-2008}  &  \textbf{CRATESJ052526-201054}  &  05 25 28.032  &  $-20$ 10 48.360  &   -  &   0.18 $\pm$ 0.02  &   -  &   0.3 $\pm$ 0.3  & 13.4 $\pm$  0.5 \\ 
\textbf{4FGL J1040.5+0617}  &  \textbf{GB6 J1040+0617}  &  10 40 31.632  &  $06$ 17 21.840  &   0.698 $\pm$ 0.008  &   1.011 $\pm$ 0.008  &   0.5 $\pm$ 0.3  &   0.9 $\pm$ 0.2  & 13.7 $\pm$  0.5 \\ 
4FGL J1043.6+0654  &  5BZB J1043+0653  &  10 43 23.880  &  $06$ 53 09.960  &   0.1 $\pm$ 0.1  &   0.2 $\pm$ 0.1  &   -  &   -  & 15.6 $\pm$  0.6 \\ 
\textit{4FGL J1230.8+1223}  &  \textit{M87}  &  12 30 49.440  &  $12$ 23 30.120  &   0.08 $\pm$ 0.02  &   0.340 $\pm$ 0.004  &   0.24 $\pm$ 0.03  &   0.14 $\pm$ 0.06  & 13.8 $\pm$  0.4 \\ 
\textit{4FGL J1231.5+1421}  &  \textit{3HSP J123123.1+14212}  &  12 31 23.928  &  $14$ 21 24.120  &   0.15 $\pm$ 0.03  &   0.20 $\pm$ 0.05  &   -  &   0.3 $\pm$ 0.2  & 15.7 $\pm$  0.5 \\ 
\textit{4FGL J1359.1-1152}  &  \textit{VOU J135921-115043}  &  13 59 21.312  &  $-11$ 50 43.800  &   0.06 $\pm$ 0.08  &   -  &   0.3 $\pm$ 0.5  &   -  & 16.9 $\pm$  0.5 \\ 
4FGL J1404.8+6554  &  3HSP J140449.6+65543  &  14 04 49.632  &  $65$ 54 30.960  &   0.18 $\pm$ 0.03  &   0.18 $\pm$ 0.02  &   0.5 $\pm$ 0.1  &   0.5 $\pm$ 0.1  & 16.0 $\pm$  0.6 \\ 
\textit{\textbf{4FGL J1439.5-2525}}  &  \textit{\textbf{VOU J143934-252458}}  &  14 39 34.656  &  $-25$ 24 58.320  &   0.06 $\pm$ 0.06  &   0.14 $\pm$ 0.06  &   -  &   -  & 15.1 $\pm$  0.5 \\ 
4FGL J1440.0-2343  &  3HSP J143959.4-23414  &  14 39 59.424  &  $-23$ 41 40.200  &   0.21 $\pm$ 0.04  &   0.18 $\pm$ 0.05  &   -  &   0.3 $\pm$ 0.2  & 15.1 $\pm$  0.8 \\ 
4FGL J1447.0-2657  &  3HSP J144656.8-26565  &  14 46 56.832  &  $-26$ 56 58.200  &   0.13 $\pm$ 0.07  &   0.15 $\pm$ 0.09  &   0.2 $\pm$ 0.2  &   -  & 17.6 $\pm$  0.4 \\ 
4FGL J2350.6-3005  &  3HSP J235034.3-30060  &  23 50 34.368  &  $-30$ 06 03.240  &   0.12 $\pm$ 0.04  &   -  &   -  &   0.5 $\pm$ 0.1  & 16.1 $\pm$  0.4 \\ 
4FGL J2351.4-2818  &  IC 5362  &  23 51 36.672  &  $-28$ 21 53.280  &   0.07 $\pm$ 0.03  &   -  &   -  &   -    & 14.5 $\pm$  0.6 \\ 
\textbf{4FGL J2358.1-2853}  &  \textbf{CRATESJ235815-285341}  &  23 58 16.968  &  $-28$ 53 34.080  &   0.24 $\pm$ 0.03  &   -  &   -  &   -  & 14.0 $\pm$  0.5 \\ 
\hline\hline

\end{tabular}
\end{center}
\footnotesize {\textit{Notes.} FVs of G20 sources for different wavelengths (columns 5 to 8). The $\gamma$-ray FVs are taken from the 4FGL-DR3 catalogue and only listed here if $\rm{FV} > 0$. \nup~is given 
in the last column as estimated with BlaST in the rest-frame, corrected for the redshift. The first section lists the 34 G20 sources that remain after removing excluded sources. Masquerading sources are listed in bold font, sources where the extension of the best-fit $\gamma$-ray flux meets the neutrino best-fit or the 5$\,\sigma$ discovery potential line are listed in italics (see Section \ref{sec:discussion} and SEDs in Appendix \ref{appendix:SEDS}). 
The bottom section lists sources excluded because of the revised alert criteria (see Section \ref{sec:update-tracks}). 
} 
\label{Tab:FVresults}

\end{table*}
\end{scriptsize}

\subsubsection{IR band}

\begin{figure}
    \centering
    \includegraphics[width=\columnwidth]{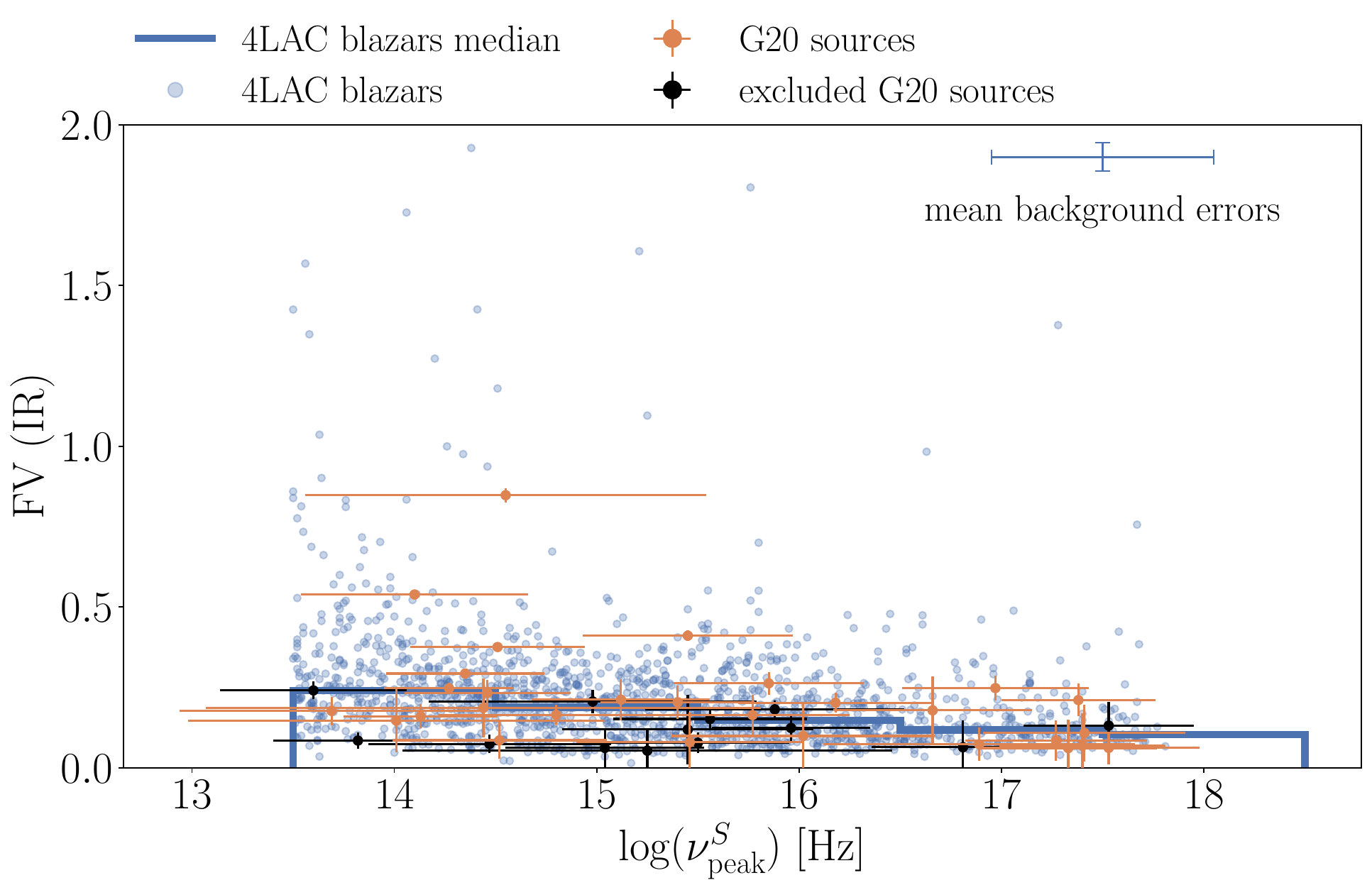}
    \caption{IR FV vs. observed \nup~fitted with BlaST. The G20 sources are shown in orange with the respective uncertainties on FV and \nup. We show the excluded G20 sources in black. The background points (in blue) are the 4LAC-DR3 blazars. 
    Average errors for the background sample are depicted in the top right corner. The blue histogram shows the median FV for each decade in \nup. For better readability, the y-axis stops at 2 but there are four background blazars with an FV between 2 and 5.5 not shown (with $13.8 < \log$\nup~$< 16$).}
    \label{fig:ir-var}
\end{figure}

We have calculated the 
FV using the \nufnu\, flux at 3.4 $\micron$. The results are reported in Table \ref{Tab:FVresults}. 
The FVs of the G20 sources show no significant deviation from the background sample (two-sample Kolmogorov--Smirnov (K-S) test p-value is 0.56). Also, when investigating each decade of \nup, the K-S test does not show significant deviations. There are 14 of 30 source candidates above the median (see Figure \ref{fig:ir-var}), which is compatible with the background expectation.

For the background sample, we find the most strongly variable sources with high FVs at lower \nup. The median FV decreases at higher \nup. Hence, in the IR, the median variability shows an anti-correlation with \nup. 

\subsubsection{Optical band}

\begin{figure}
    \centering
    \includegraphics[width=\columnwidth]{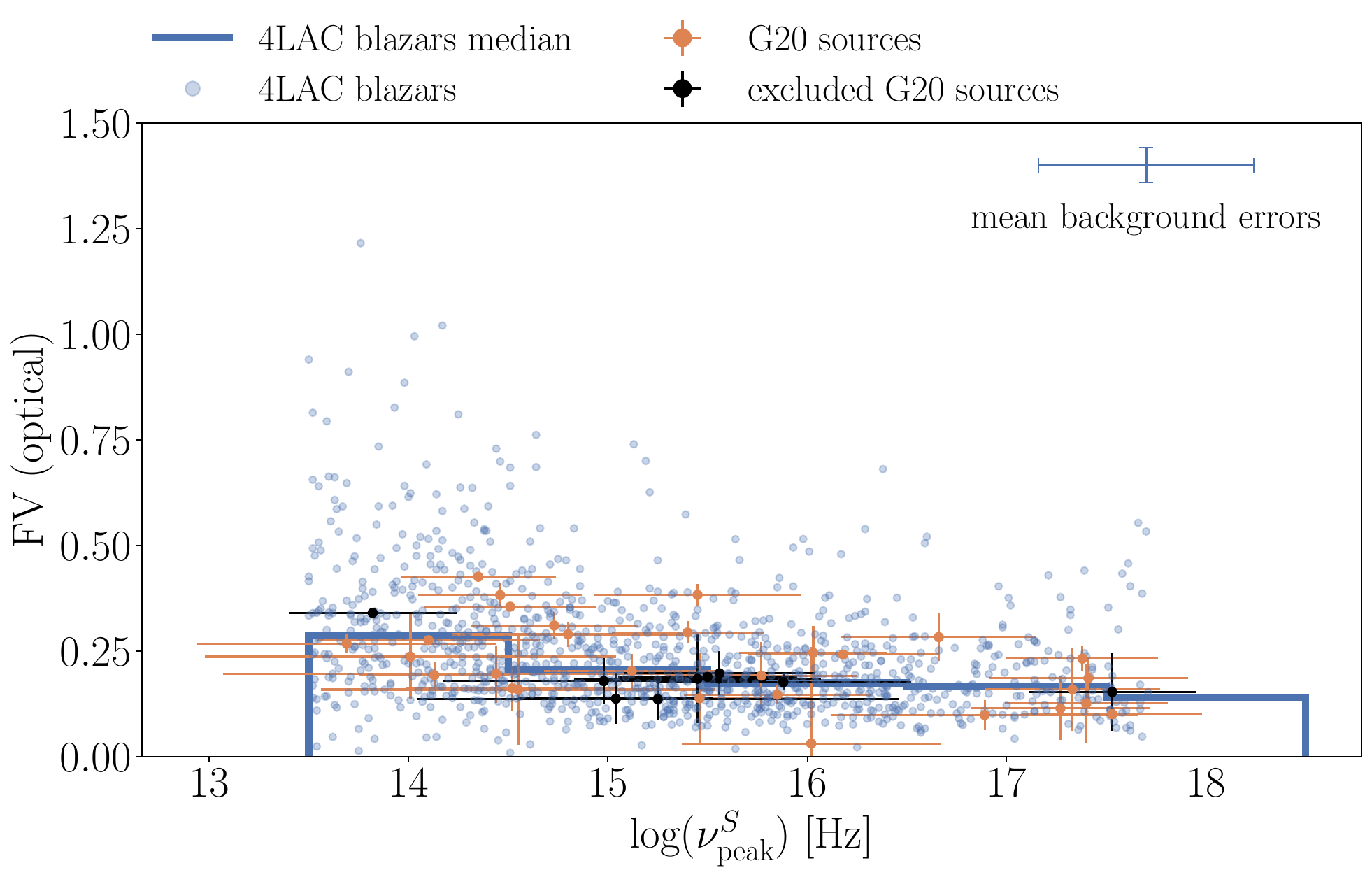}
    \caption{Optical FV vs. observed \nup~fitted with BlaST. The G20 sources are shown in orange with the respective uncertainties for FV and \nup. We show the excluded G20 sources in black. The background points (in blue) are the 4LAC-DR3 blazars with optical data from the ZTF (blue dots). 
    Average errors for the background sample are depicted in the top right corner. The blue histogram shows the median FV for each decade in \nup. For better readability, the y-axis stops at 1.5 but there is one background blazar with FV = 3.3 and $\log$\nup~$\sim 14$.}
    \label{fig:optical-var}
\end{figure}

The optical FVs are reported in Table \ref{Tab:FVresults}. The FVs of the G20 sources are compatible with those of the background sample (K-S test yields a p-value of 0.91); see Fig. \ref{fig:optical-var}. We find 13 out of 29 G20 sources above the background median, which agrees with the background expectation. We observe the same trend as in the IR, with the median FV decreasing for higher \nup.

\subsubsection{X-ray band}

\begin{figure}
    \centering
    \includegraphics[width=\columnwidth]{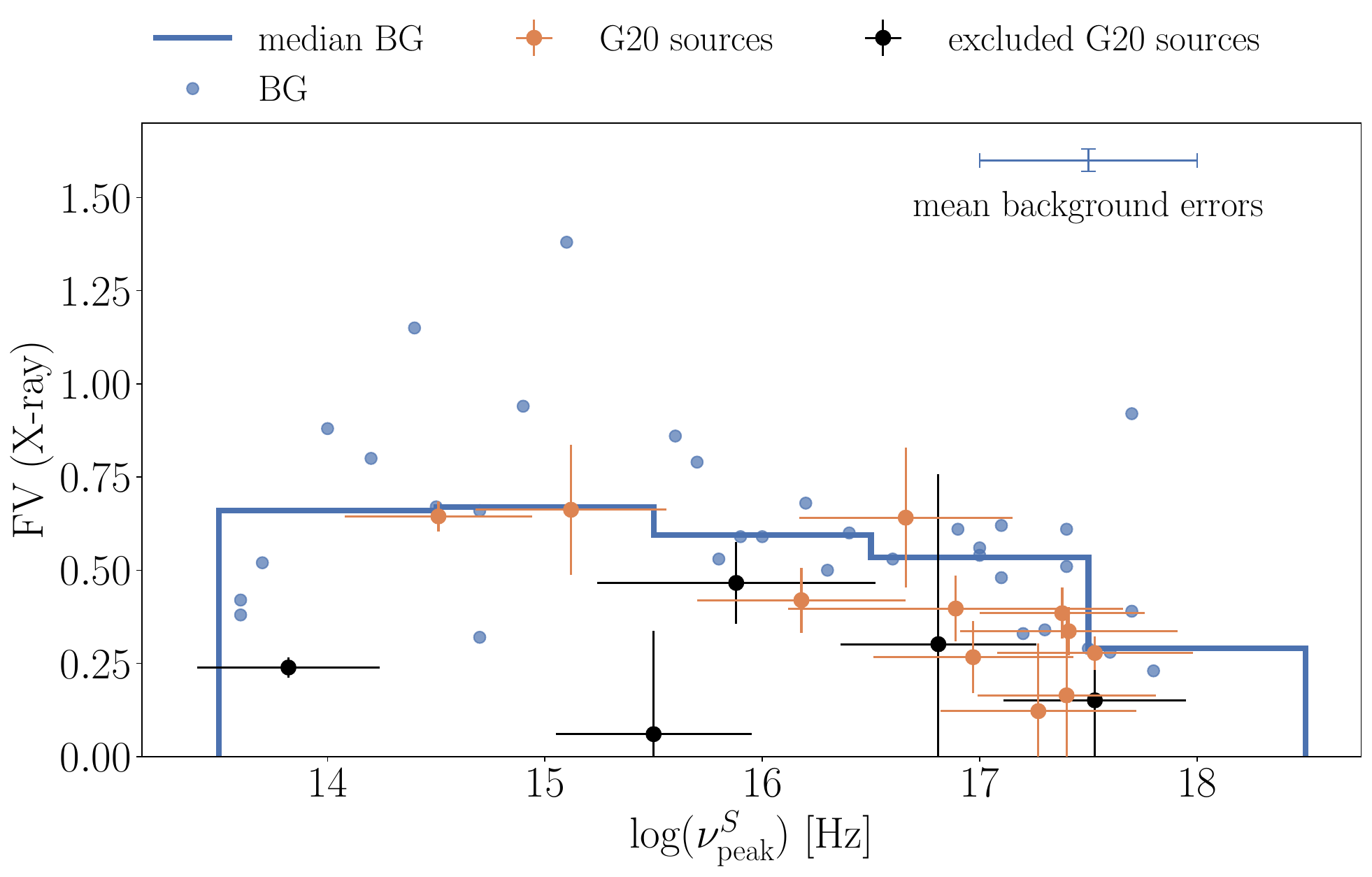}
    \caption{FV in X-rays vs. observed \nup~fitted with BlaST. The G20 sources are shown in orange with the respective uncertainties for FV and \nup. We show the excluded G20 sources in black. The background points (in blue) are from the sample of sources observed at least 50 times with SWIFT \citep{GiommiXRTspectra}. 
    Average errors for the background sample are depicted in the top right corner. The blue histogram shows the median FV for each decade in \nup.}
    \label{fig:fv-xray-var}
\end{figure}

The FV X-ray results are reported in Table \ref{Tab:FVresults}. 
Notice that, as in the $\gamma$-ray case, the errors here are larger
than in the optical/IR bands, where the individual uncertainties are
smaller and the number of data points is higher. 
For the X-ray light curves, the FV varies for different SWIFT snapshots, which indicates short-term variability for the G20 sources. Here, we take the flux of the total observation for the FV calculation (see Fig. \ref{fig:fv-xray-var}). A two-sample K-S test comparing the background sample with the G20 sources yields a p-value of 0.025. We find significantly less variability in the G20 sources compared to the background sample (10 out of 11 sources are below the median background FV). This difference is likely to be the result of a selection bias in the background sample for highly variable sources, as less variable sources might have been 
of less interest for repeated Swift-XRT observations. 

\subsubsection{$\gamma$-ray band}
\begin{figure}
    \centering
    \includegraphics[width=\columnwidth]{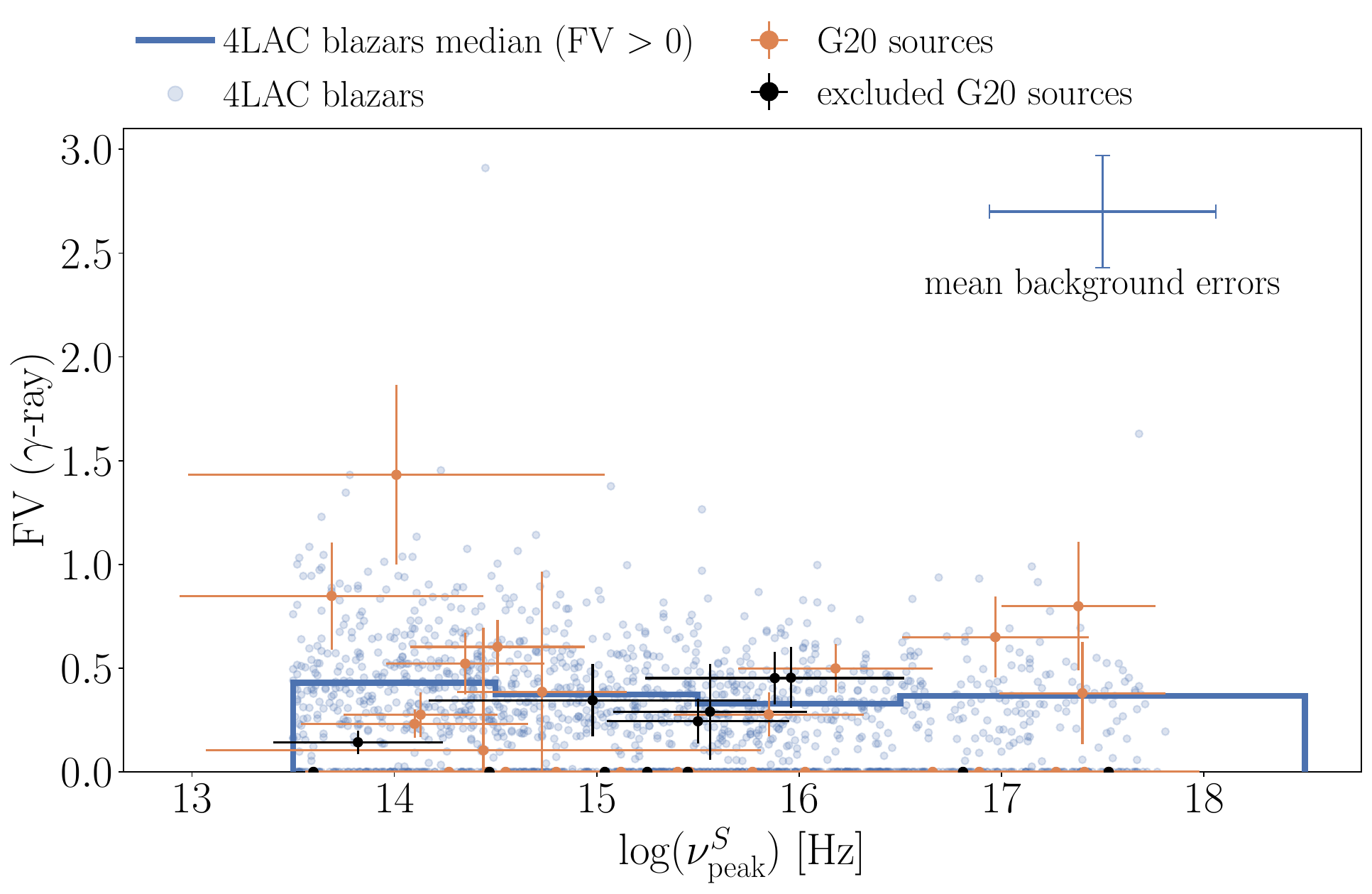}
    \caption{The FV, as given in the 4FGL catalogue, vs. observed \nup~as estimated with BlaST. The orange dots with error bars show the 4FGL FV 
    for the G20 sources. We show the excluded G20 sources in black. The background sample (blue) are all blazars in the 4LAC catalogue. 
    Average errors for the background sample are depicted in the top right corner. 
    The blue histogram shows the median FV for each decade in \nup~for the background sources with $FV > 0$.}
    \label{fig:variability-gamma-ray}
\end{figure}

 In the $\gamma$-ray band we compare the variability fraction of the G20 sources with that of the background sample (Table \ref{Tab:FVresults} and Fig. \ref{fig:variability-gamma-ray}). For this, we take all events within one decade of \nup~and
 run a two-sample K-S test comparing background and G20 sources. In total, we get a p-value of 0.19. 

We also compare the fraction of objects that have no reported FV. For the background sample, 66 per cent of objects have FV = 0. For the G20 objects, there are 46 per cent of objects without a reported FV. This is compatible with the background assumption (p-value of 0.04). 

When considering only objects with FV $> 0$, we test if there is a significant number of G20 sources above the background sample's median FV
for each \nup~band (see Fig. \ref{fig:variability-gamma-ray}). We find 9 out of 13 G20 sources above the background median FV, corresponding to a p-value of 0.13. Hence, we do not find any difference between the background sample and the G20 sources for the $\gamma$-ray FV. 

\subsection{State of G20 sources during neutrino arrival times}

The high-energy neutrino IceCube-170922A was detected while TXS~0506+056 was flaring \citep{icfermi,neutrino}. 
To search for similar patterns with all the G20 alert events and G20 sources we investigate the state of the source candidate with the light curves used in the variability study. We define flares in a similar fashion to \citet{deJaeger2023}. For each light curve, we determine flux bins with Bayesian blocks \citep{1998ApJ...504..405S, 2013ApJ...764..167S}. We use the same parameters for each source, assuming point measures (as in \citealt[Section 3.3]{2013ApJ...764..167S} with the default value of $p_0 = 0.05$ in equation 21 of \citealt{2013ApJ...764..167S}) with the flux uncertainties as $\sigma$. We define a flare where the flux in a bin, $F_{\rm bin}$, exceeds the average flux plus the standard deviation over the whole light curve.  We calculate the chance probability, $P_{\text{BG}}$, as the fraction of flaring time vs. the total data taking time. Finally, we do a binomial test for the total number of neutrinos observed coincident with a flaring source based on the chance probability for all sources combined. The following analysis excludes the objects in the bottom section of Table \ref{Tab:FVresults}. We list all results for each source in all bands in the appendix in Table \ref{Tab:COINCresults} and show example light curves for TXS~0506+056 in Figure \ref{fig:lightcurveTXS}.

\subsubsection{IR band}
In the IR band we examine the NEOWISE light curves as described in Section \ref{sec:variabilityData}. Out of 33 light curves, we find three sources where the neutrino arrival coincides with a flare per the above definition (TXS~0506+056 [4FGL~J0509.4+0542], 3HSP~J064933.6-31392 [4FGL~J0649.5-3139], and 3HSP~J125848.0-04474 [4FGL~J1258.7-0452]). The chance probability is 0.131. The binomial probability $p_{\text{binom}}$, of finding three sources out of 33 given this chance probability is compatible with the background assumption ($p_{\text{binom}} = 0.79$). 

\subsubsection{Optical band}

In total, there are nine optical light curves (with three light curves from ZTF and six light curves from ASAS-SN) during the respective neutrino arrival time. Out of these nine, we find two sources where the neutrino is coincident with a flare (TXS~0506+056 [4FGL~J0509.4+054] and 3HSP~J100326.6+02045 [4FGL~J1003.4+0205]). The chance probability is 0.112. Another binomial test of getting two sources out of nine given this chance probability yields a p-value compatible with background ($p_{\text{binom}} = 0.27$). 

\subsubsection{X-ray band}

In X-rays, nine out of 25 sources with X-ray data have sufficiently sampled light curves to allow flare 
identification. Two sources are identified where the neutrino arrival is coincident with an X-ray flare (3HSP J023248.5+201717 [4FGL J0232.8+2018] and 3HSP J095507.9+355101 [4FGL J0955.1+3551]). For all sources combined, the chance probability is 0.097. A binomial test of getting two sources based on this chance probability yields a p-value compatible with background ($p_{\text{binom}} = 0.22$).

\subsubsection{$\gamma$-ray band}

We find eight light curves for the G20 sources in the \fermi-LAT LCR database. Out of these eight, there is one case where the neutrino arrives simultaneously with a $\gamma$-ray flare: TXS~0506+056. The chance probability for finding a coincidence is 0.006. The binomial probability of finding one source out of eight based on this chance probability is 0.045 or 1.7$\,\sigma$.

Table \ref{tab:flarce_coincident_table_binomial} summarises the results for each wavelength, listing the number of available light curves, the chance probabilities, and the result of the binomial test for finding neutrinos arriving coincident with a flaring source. Only in $\gamma$-rays we find a non-background binomial p-value of 1.7$\sigma$, due to the neutrino IceCube-170922A arriving during a flaring period of TXS~0506+056. 

\begin{table}
\caption{Results of the search for neutrino alerts arriving coincident with flares in different bands.}
    \begin{center}
   \begin{tabular}{lcccc}
    \hline\hline
       Wavelength  & $N_{\text{LC}}$ & $P_{\text{BG}}$ & $N_{\text{coinc}}$ & $p_{\text{binom}}$  \\
         \hline
       IR  &  33 & 0.131 & 3 & 0.79 \\
       Optical & 9 & 0.112 & 2 & 0.27 \\
       X-ray & 9 & 0.097 & 2 & 0.22 \\
       $\gamma$-ray & 8 & 0.006 & 1 & 0.045 \\
       \hline\hline
    \end{tabular}
    \end{center}
    \footnotesize {\textit{Notes.} We list the number of available light curves,  $N_{\text{LC}}$, suitable to identify flares, the chance probability, $P_{\text{BG}}$, of a neutrino arriving during a flare, the number of coincidences, $N_{\text{coinc}}$,  found in these light curves and the p-value of the binomial test, $p_{\text{binom}}$, for finding $N_{\text{coinc}}$ given $P_{\text{BG}}$.}
    \label{tab:flarce_coincident_table_binomial}
\end{table}

\section{Discussion}\label{sec:discussion}

\subsection{Hybrid SEDs}

For the first time we have a sizeable number of blazars, 34 per cent of which
might be neutrino sources, with an estimate of their neutrino fluxes. 
Based on purely hadronic models we would expect $E_{\gamma} \approx 2 \times 
E_{\nu}$ and $F_{\gamma} \approx 2 \times F_{\nu}$ \citep[e.g.][]{Kelner_2008}.
Therefore, we can have three extreme cases: (1) $F_{\gamma} \approx F_{\nu}$, 
which implies that in both bands the emission is hadronic; (2) $F_{\gamma} 
\ll F_{\nu}$, which would suggest that $\gamma$-ray emission is absorbed 
and high-energy emission is degraded to lower energies through cascades; 
(3) $F_{\gamma} \gg F_{\nu}$, which would suggest that $\gamma$-ray emission 
is mainly leptonic. However, all the above assumes that the sampled 
$\gamma$-ray and neutrino energies are similar, which is not the case, 
as while all sources have $F_{\gamma}$ in the $0.1 - 100$ GeV range, 
the neutrino energies used to derive the best fit flux are higher and 
vary from source to 
source (see Fig. \ref{fig:includedSEDs} and Appendix \ref{appendix:neutrinos}). 

We then took the alternative but relatively robust approach of looking at the 
hybrid SEDs to see if the extension of the best-fit $\gamma$-ray flux met the 
neutrino best-fit or the 5$\,\sigma$ discovery potential line (somewhat similarly 
to what was done by \citealt{Padovani_2014}). It turned out that in 18 out of 
34 cases this was the case (those objects are highlighted in italic in Table \ref{Tab:FVresults} and in the SEDs in Appendix \ref{appendix:SEDS}). As a first approximation, we take the 15 excluded 
sources as a (very small) background sample. In that case we find 5 out of 15 
sources fulfilling the same requirement, yielding a background probability of 0.33. 
A binomial test calculating the chance of getting 18 out of 34 objects with 
such a probability is incompatible with the background assumption 
only at the 2.2$\,\sigma$ level. 

As an alternative approach, we check where the best-fit neutrino flux exceeds 0. In the background sample, we find 3 objects with a best-fit neutrino flux $> 0$. This yields a background probability of 0.2. Comparing this with the investigated sources we find 27 objects with a best-fit neutrino flux $>0$. A binomial test of getting 27 objects out of 34 with a background probability of 0.2 yields a highly significant p-value above the $5\,\sigma$ level. This result, however, depends strongly on the statistics and changes significantly when adding/removing a few sources from the signal/background sample. Furthermore, in 11 of the 27 cases with a best-fit neutrino flux > 0, the lower 68 per cent uncertainty of the neutrino flux is compatible with a flux of 0 (and in 16 cases it is incompatible with a flux of 0 at the 68 per cent confidence level). As a further test, we can compare this with the background sample where we fit a flux incompatible with 0 at the 68 per cent level for 2 sources out of 15, and the binomial test results in $4.6\,\sigma$. We emphasize that this result is based on a small background sample of 15 sources. Hence, we exercise caution concerning the reliability of these tests because of the small background sample size.

Paper III did not find any significant difference between the G20 sources and 
a control sample of IHBLs in terms of their radio and $\gamma$-ray powers, 
and \nup, although it was stated that the lack of such differences could 
have been due to the relatively small number of G20 sources, in which only 
less than half of the sources were expected to be associated with astrophysical neutrinos. A redshift difference was found but was explained in terms of a
selection effect. In this work, we investigate the flux at \nup, as an approximation of the source's bolometric flux and hence a measure of the blazar's total power, as neutrino emission might be related to the total emitted radiation. We find no differences between the fluxes at \nup~between the G20 sources and the blazars in the 4LAC-DR3 catalogue (p-value of two-sample K-S test is 0.73).

In Paper III, masquerading BL Lacs turned out to be more 
powerful than the non-masquerading ones in the radio and $\gamma$-ray 
band, with a smaller \nup. This is to be expected, as these are the properties, 
which allow them to dilute their strong, FSRQ-like emission lines effectively. When comparing the flux at \nup~of masquerading sources with non-masquerading sources in our sample, we find no difference between the two classes (with a p-value of 0.14 for a two-sample K-S test).

\subsection{Variability}

We find no significant difference in variability in the IR, optical, and $\gamma$-ray bands for the G20 sources compared to the 4LAC-DR3 background sample or the 63 sources in \cite{GiommiXRTspectra} for the X-ray band. Moreover, when comparing masquerading BL Lacs with non-masquerading BL Lacs, we also find no significant differences in FVs (the best two-sample K-S test p-value was 0.048 in the X-ray band).

However, several caveats should be
considered. For example, light curves for the majority of the G20 sources are only available in the IR and
optical bands. In the X-rays, only a limited number of observations are available with sufficient flux measurements for generating light curves. In some cases, we find rapid short-term variability within one observing period. However, the number of measurements where detecting such short-term variability is possible is limited in our sample. Hence, we take the average flux levels for one observation to generate the light curve. In the X-ray background sample, we are limited by the lack of observations for a broad population of blazars. 

In the $\gamma$-rays we used the available FVs in the 4FGL-DR3 catalogue, which are calculated based on yearly fluxes. Hence, we have differently binned light curves for different bands, which hinders the comparison of FVs between different bands. A custom generation of \fermi-LAT light curves and calculation of FVs would be needed. 

In general, the results of the different bands cannot be compared with each other since that would require similar binning and data coverage. Some light curves are only sparsely sampled (for example in the X-ray band as listed in Table \ref{Tab:LCDataOverview}). Additionally, also within one band, the light curves are not uniformly sampled and vary greatly in sampling frequency (see again Table \ref{Tab:LCDataOverview}), which can impact the individual FV comparison. However, since we look at the FV distribution of many objects and compare them with a large background population, we expect the impact to be averaged out.

\subsection{Flares}

We do not find a significant correlation of flares in the IR, optical, X-ray or $\gamma$-ray bands, and the arrival times of high-energy neutrinos. This agrees with the results of \cite{Chang2022}, who studied the temporal correlations of blazar flares in the
millimetre, IR, X-ray, and $\gamma$-ray bands with IceCube neutrino alerts, with
no significant results. Also \cite{Hovatta2021} did not find a significant correlation
between strongly radio flaring blazars and IceCube events.

In this work, TXS~0506+056 is the only source where flares in the IR, optical, and $\gamma$-rays coincide with the neutrino arrival; in the X-rays there is no flare at 1~keV, only when looking at larger energies the source shows flaring behaviour \citep{icfermi}. All other cases are restricted to one wavelength only. However, there were very few cases when relevant data were available directly at the neutrino arrival time; hence the state of most G20 sources at the time of neutrino arrival is simply not known. This emphasises the importance of immediate multi-wavelength follow-up observations of IceCube alert events. 

\section{Conclusions and Summary}\label{sec: conclusions}

We have presented the multi-wavelength and neutrino SEDs for a selection of potential neutrino sources based on \cite{Giommidissecting}. The selection was updated 
by excluding objects not matching the revised criteria of \cite{abbasi2023icecat1}. For the SEDs, we obtained the low-energy multi-frequency data with VOU-Blazars. We included $\gamma$-ray data from the \textit{Fermi}-LAT telescope and a best-fit of the $\gamma$-ray SEDs. Additionally, we analysed 10 years of publicly available neutrino data from the IceCube Neutrino Observatory. We derived for the first time the best-fit neutrino fluxes (assuming a power-law spectrum) in 27 cases where the best-fit flux is greater than 0. The 68 per cent confidence levels are compatible with no flux in 11 out of 27 cases. We included the IceCube $5\,\sigma$ discovery potential and sensitivity in the SEDs based on \citet{Aartsen2020}. 

Looking for signatures of hadronic production, we took the extension of the best-fit $\gamma$-ray flux and checked if it met the neutrino best-fit or the $5\,\sigma$ discovery potential. Comparing the updated selection with the excluded objects yielded a $2.2\,\sigma$ association of sources with matching $\gamma$-ray and neutrino fluxes in our sample. Otherwise, comparing the number of sources where we fit fluxes incompatible with 0 at the 68 per cent confidence level, we find a $4.6\,\sigma$ over fluctuation for our source selection. However, we emphasize that this result is based on a small background selection of 15 sources and requires confirmation with better statistics.

As a next step, we used the obtained data to study the sources' variability in the IR, optical, $\gamma$-ray, and X-ray bands. For the first three bands we used the blazars of the 4LAC-DR3 catalogue as a background sample. For the latter, we used a selection of 63 blazars observed with Swift-XRT at least fifty times \citep{GiommiXRTspectra}. We found no significant differences in the variability of our sources compared with the background samples. 

Furthermore, we checked for coincidences of the high-energy neutrino arrival time and flaring states of the source candidates, finding no significant correlation in any of the bands. However, in most cases there were no observations available at the time of neutrino arrival, which emphasises the importance of fast follow-up observations of the astronomical community. 

Finally, we found no significant differences in the SEDs of the G20 sources compared with selections of a general population of blazars. 
We compared the flux at \nup~as a measure of the blazar's total emission. The G20 sources show no differences in their flux distributions compared to blazars of the 4LAC-DR3 catalogue. 
Both results, however, could be related to the relatively small 
number of G20 sources. 
Furthermore, when comparing masquerading with non-masquerading sources, we also find no differences in their flux distributions at \nup.

The next step for the SIN project will be the modelling of the hybrid
SEDs put together in this paper with a lepto-hadronic code to assess 
the physical likelihood of a connection between blazars and neutrinos 
(Rodrigues et al., in preparation). 

\section*{Acknowledgments}
We thank Maria Petropoulou and Xavier Rodrigues for helpful discussions
and the referee for her/his useful comments and suggestions. 
We acknowledge the use of data and software facilities from the SSDC,
managed by the Italian Space Agency, and the United Nations ``Open
Universe'' initiative. This work is supported by the Deutsche
Forschungsgemeinschaft through grant SFB\,1258 ``Neutrinos and Dark Matter
in Astro- and Particle Physics''. 

\section*{Data Availability}
Most of the data used in this paper are publicly available from the VOU-Blazars package, from the Firmamento online tool at https://firmamento.hosting.nyu.edu, and from the Zwicky Transient Facility web pages at https://www.ztf.caltech.edu/. Specific data can be provided by the lead author upon reasonable request.

\bibliographystyle{mnras}
\bibliography{mybibliography}

\appendix

\section{}\label{sec:appendix}
 
\section*{VOU-Blazars catalogues}

The catalogues queried in VOU-Blazars are: 
NVSS, 
FIRST, 
SUMSS, 
VLASSQL, 
2SXPS, 
SDS82, 
1OUSX, 
RASS, 
XMMSL2, 
4XMM-DR11, 
BMW, 
WGACAT, 
IPC2E, 
IPCSL, 
Chandra-CSC2, 
MAXI, 
eROSITA-EDR, 
ZWCLUSTERS , 
PSZ2, 
ABELL, 
MCXC, 
5BZCat, 
SDSSWHL, 
SWXCS, 
3HSP, 
FermiGRB, 
MilliQuas, 
BROS, 
MST9Y, 
PULSAR, 
F2PSR, 
F357cat, 
XRTDEEP, 
WISH352, 
GLEAM, 
TGSS150, 
VLSSR, 
LoTSS, 
PMN, 
GB6, 
GB87, 
ATPMN, 
AT20G, 
NORTH20, 
CRATES, 
F357det, 
KUEHR, 
PCNT, 
PCCS44, 
PCCS70, 
PCCS100, 
PCCS143, 
PCCS217, 
PCCS353, 
PCCS2, 
ALMA, 
SPIRE, 
H-ATLAS-DR1, 
H-ATLAS-DR2, 
H-ATLAS-DR2NGP, 
H-ATLAS-DR2SGP, 
AKARIBSC, 
IRAS-PSC, 
WISE, 
WISEME, 
NEOWISE, 
2MASS, 
USNO, 
SDSS, 
HSTGSC, 
PanSTARRS, 
GAIA, 
SMARTS, 
UVOT, 
GALEX, 
XMMOM, 
CMA, 
EXOSAT, 
XRTSPEC, 
OUSXB, 
OUSXG, 
OULC, 
BAT105m, 
BEPPOSAX, 
NuBlazar, 
3FHL, 
2FHL, 
3FGL, 
2BIGB, 
4FGL-DR3, 
2AGILE, 
FermiMeV, 
FMonLC.

\section*{Neutrino fluxes}\label{appendix:neutrinos}
As mentioned in Section \ref{sec:neutrinos}, we use an unbinned maximum likelihood approach \citep{Braun:2008bg}, to compare a signal, $H_S$, and a background hypothesis, $H_B$.

We take the ratios of these likelihoods and optimise the signal likelihood by varying the source flux, $\Phi$, which is described by the number of observed signal neutrinos, $n_S$, and their energy distribution. Neutrinos are produced via particle interaction in a source, and the number of produced neutrinos is thus Poissonian distributed. The fitted number of signal neutrinos, $n_S$, is hence the mean number of neutrinos we expect to see in the detector for a respective flux. We fit the energy spectral index, $\gamma$, for the energy distribution. The optimised expression is \citep{Braun:2008bg}:

\begin{equation}\label{eq:TS_timeint_exp}
\text{TS} = -2 \ln \frac{\sup \mathcal{L}(H_B)}{\sup \mathcal{L}(H_S)} = 2 ~ \sum_{i} \ln \left[\frac{n_S}{N}\left(\frac{S_i}{B_i}-1\right) +1\right],
\end{equation}

with $N$ as the total number of observed events and a signal, $S_i$, and background, $B_i$, probability density function (pdf) for each event $i$. The signal pdf includes a spatial clustering around the source position $\vec{x}_S$ and the probability of detecting a neutrino with reconstructed energy, $E_i$, originating from declination, $\delta_i$, for a source emitting neutrinos with a spectrum $E^{-\gamma}$. The latter part is extracted from the effective area and the smearing matrix in the data release. We use the reconstructed event properties, i.e., origin direction, $\vec{x}_i = (\alpha_i, \delta_i)$, the uncertainty on the directional reconstruction, $\sigma_i$, and the reconstructed muon energy, $E_i$, to define the signal pdf \citep{Braun:2008bg}

\begin{equation}\label{eq:neutrino_signal_timeint}
\begin{split}
S_i &= S_{\rm{spatial}} (\vec{x}_i, \sigma_i | \vec{x}_S) \cdot S_{\rm{energy}} (E_i | \delta _i , \gamma ) \\
&= \frac{1}{2 \pi \sigma_i ^2} \exp \left( \frac{- | \vec{x}_i - \vec{x}_S | ^2}{2 \sigma _i ^2} \right) \cdot S_{\rm{energy}} (E_i | \delta _i, \gamma ) .
\end{split}
\end{equation}
The background pdfs, $B_i$, are \citep{Braun:2008bg}
\begin{equation}\label{eq:bg_timeint_exp}
\begin{split}
B_i  &= B_{\rm{spatial}} (\vec{x}_i) \cdot B_{\rm{energy}} (E_i| \delta _i) = \frac{1}{2 \pi} \cdot P(\delta _i) \cdot B_{\rm{energy}}(E_i| \delta _i).
\end{split}
\end{equation}
Here, the spatial term is uniform in right ascension, $\alpha_i$, for the background for integration times greater than a day since IceCube is located at the South Pole. The energy term describing the probability of detecting an event with energy $E_i$ at declination $\delta_i$ is extracted from the effective area and the smearing matrix published in the data release. With the detector's effective area and the smearing matrix, we calculate the neutrino flux corresponding to $n_S$.

We then use the approach in \citet{Feldman_1998} to estimate confidence intervals. We repeatedly simulate neutrino emission of various strengths for each source, following a power-law energy distribution of $\propto E^{-\gamma}$, with $\gamma \in [1, 3.7]$. We stop at 3.7 since such soft spectral indices simulate a neutrino flux similar to the atmospheric background. With equation \ref{eq:TS_timeint_exp}, we get a distribution of TS values for each simulated flux. Based on this TS distribution, we determine the 68 per cent confidence belt. In most cases, the lower limit is compatible with no neutrino emission from the source, and we show the upper level as 68 per cent upper limit. The parameter optimisation and the flux simulations are done with the open-source framework SkyLLH\footnote{https://github.com/icecube/skyllh} \citep{Bellenghi:20230u}.

The data provides the reconstructed properties, $\theta_{r}$, i.e. reconstructed origin in declination, $\delta_{i}$, the reconstructed energy, $E_{i}$, and the reconstructed spatial uncertainty, $\sigma_{i}$. However, we want to know for which true neutrino energies, $E_{\rm{true}}$, our flux estimation is valid. With the data given in the public data release, we construct the probability to observe the true energy given the reconstructed parameter: 

\begin{equation}
P(E_{\rm{true}} | \theta_{r}) = \frac{P(\theta_{r} | E_{\rm{true}}) P(E_{\rm{true}})}{P(\theta_{r})},
\end{equation}

always for a given flux $\Phi$ at a given declination $\delta_i$. We get $P(\theta_{r} | E_{\rm{true}})$ from the smearing matrix in the data release as the fractional counts $(N_{\text{reco}, E_{\rm{true}}} / N_{E_{\rm{true}}})$. The probability for the true energy, $P(E_{\rm{true}})$, can be calculated with the effective area. The effective area, $A_{\rm{eff}}$, yields the number of expected events, $N_{\rm{obs}}$, for a given flux, $\Phi $, in the energy range $E_0$ to $E_1$ and during detection time $T$:

\begin{equation}
N_{\rm{obs}} = \int _{E_0}^{E_1} \int_{t_0}^T \Phi(E) A_{\rm{eff}}(E) dE dt.
\end{equation}

The probability to get an event with true Energy, $E_{\rm{true}}$, is $P(E_{\rm{true}}) = N_{E_{\rm{true}}} / N_{\rm{all}}$. We assume $E_{\rm{true}}$ lies within the energy bin with edges $E_0$ and $E_1$ of the effective area matrix:

\begin{equation}\label{eq:trueEnergyProbability}
P(E_{\rm{true}}) = \frac{N_{E_{\rm{true}}}}{N_{\rm{all}}} = \frac{ \int _{E_0}^{E_1} \int_{t_0}^T \Phi(E) A_{\rm{eff}}(E) dE dt } 
{\int _{E_{\rm{min}}}^{E_{\rm{max}}} \int_{t_0}^T \Phi(E) A_{\rm{eff}}(E) dE dt},
\end{equation} 

where the time integral is given by the total uptime of the detector in each respective data sample. $P(\theta_{r}) = \sum_{E_{\rm{true}}} P(\theta_{r} | E_{\rm{true}}) P(E_{\rm{true}})$ and can be calculated with the above quantities. 

With this, we can get $P(E_{\rm{true}} | \theta_{r})$ for each sample. The IceCube public data is published as distinct data samples with each their own effective area distribution and smearing (describing the reconstruction in the detector). 

Suppose we want one combined probability over all samples. In that case, we weigh each contribution with the ratio of the sample effective area over the summed effective area of all samples (for a respective declination and true energy). We then scale each sample contribution with this factor:

\begin{equation}
     \text{sample contribution} = \frac{\int _{E_0}^{E_1} A_{\rm{eff,sample}}(E) dE}{\sum_{\rm{samples}} \int _{E_0}^{E_1} A_{\rm{eff,sample}}(E) dE}.
\end{equation}

Hence, each bin entry is $P(E_{\rm{true}} | \theta_{r}) \times \text{sample contribution}$, and we can take the sum over all samples for each bin to get the true energy distribution. We evaluate the true energy distribution for all $n_S$ events that contribute most to the flux (i.e., those events where the signal pdf divided by the background pdf, $S_i/B_i$, are the highest). The energy range of the fitted flux is then the central 90 per cent quantile of the true energy distribution.

\section*{State of G20 sources during neutrino arrival times}\label{appendix:SourceState}

\begin{figure*}
    \centering
    \includegraphics[width=0.8\textwidth]{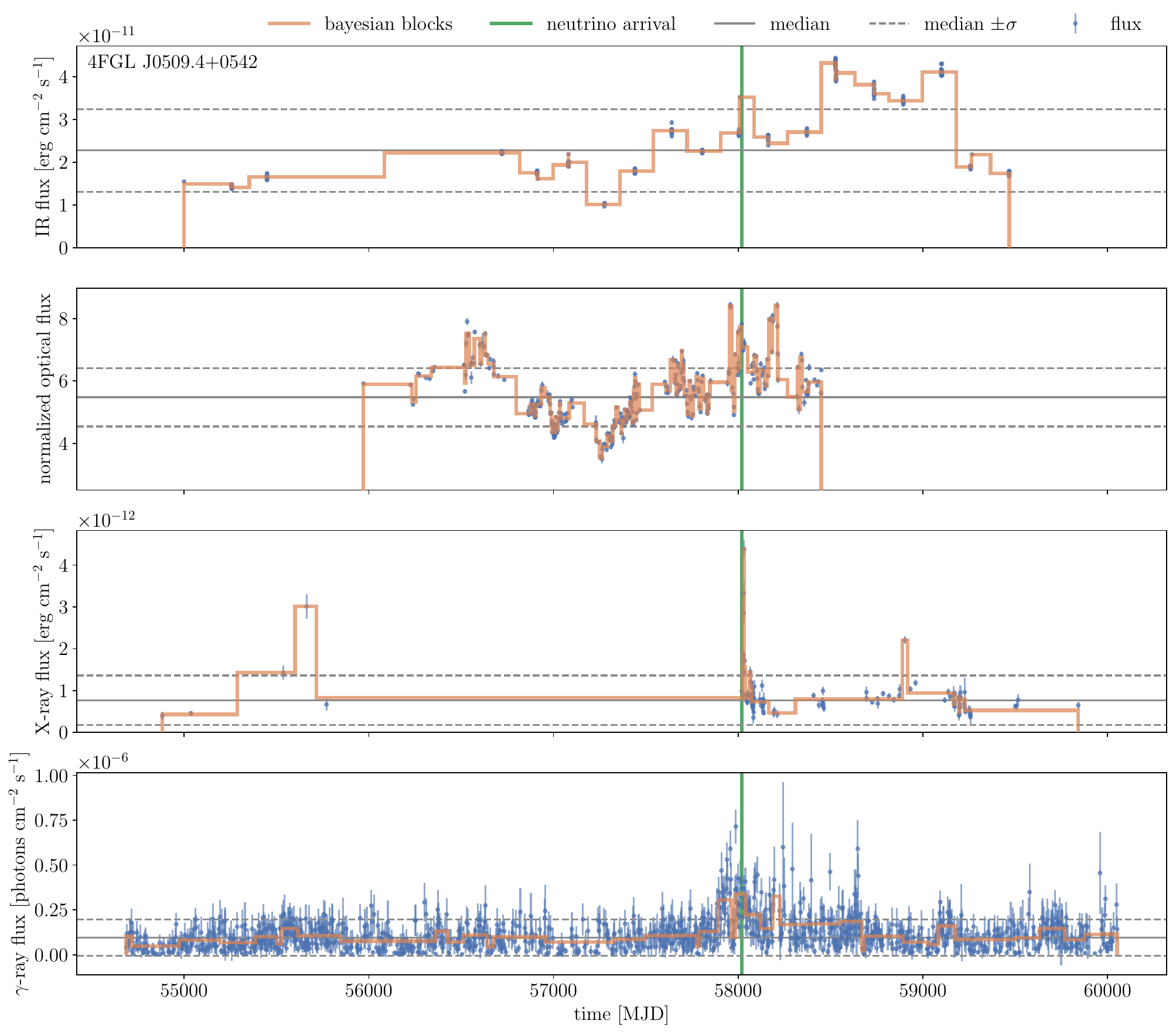}

 \caption{Light curves and their bayesian block decomposition for TXS~0506+056 in the IR, optical, X-ray, and $\gamma$-ray bands (top to bottom), where time is given in Modified Julian Date (MJD) format. We show the median flux as the horizontal solid grey line and the standard deviation as the horizontal dashed grey lines. The neutrino arrival time is indicated by the green vertical line. The orange histogram shows the bins identified by bayesian block decomposition. At the time of the neutrino arrival, the source was found to be in a flaring state in the IR, optical and $\gamma$-rays. In the X-ray band (1~keV), the flare begins right after the neutrino arrival time and is not identified as a coincident arrival.}
    \label{fig:lightcurveTXS}
\end{figure*}

We provide example light curves and their bayesian block decomposition for the case of TXS~0506+056 in Figure \ref{fig:lightcurveTXS}.

\section*{SEDs}\label{appendix:SEDS}
We show the SEDs for the 34 G20 sources in Figure \ref{fig:includedSEDs}. SEDs for excluded sources due to updated alert event criteria are displayed in Figure \ref{fig:excludedSEDs}. 
\begin{figure*}
\includegraphics[width=0.49\textwidth]{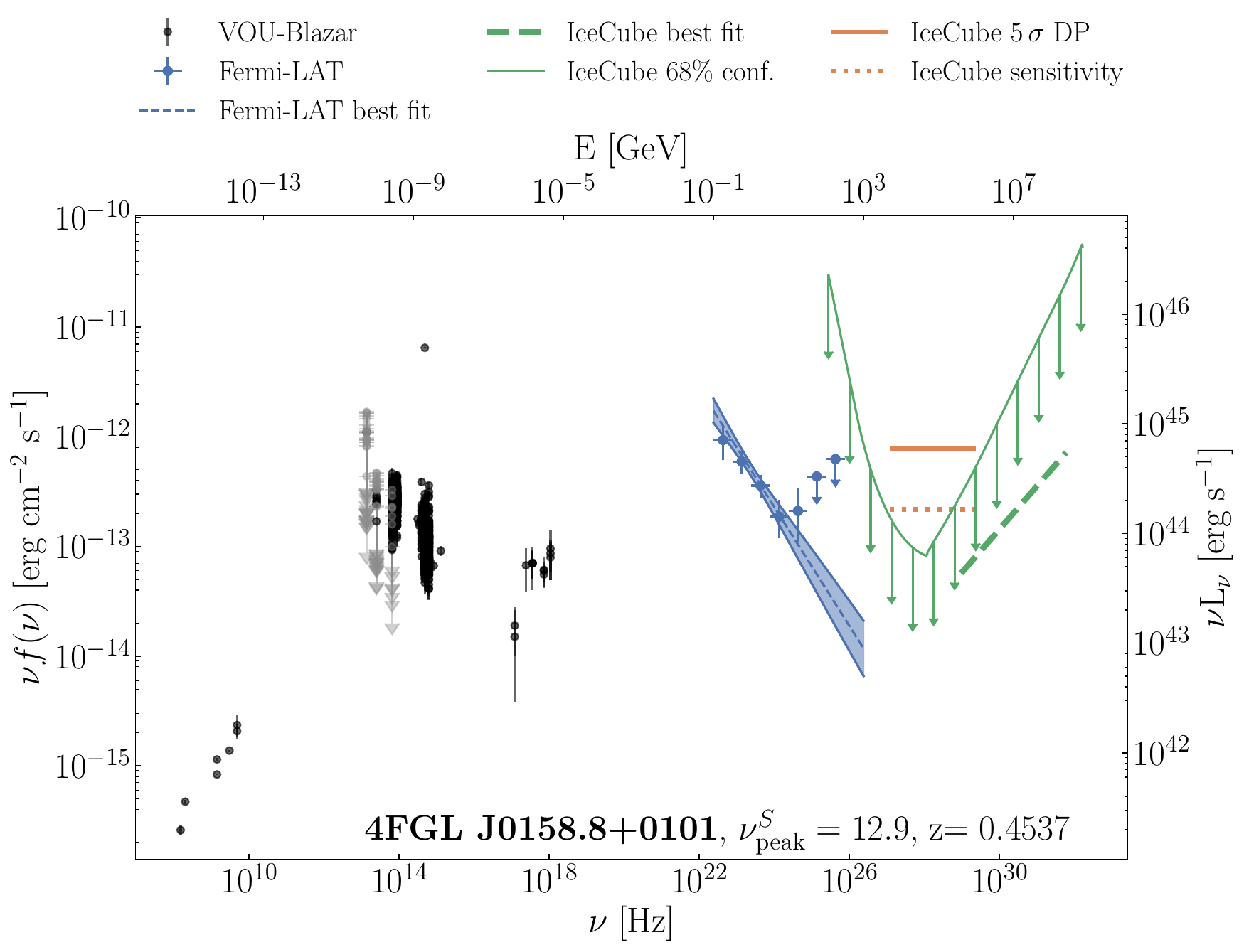}
\includegraphics[width=0.49\textwidth]{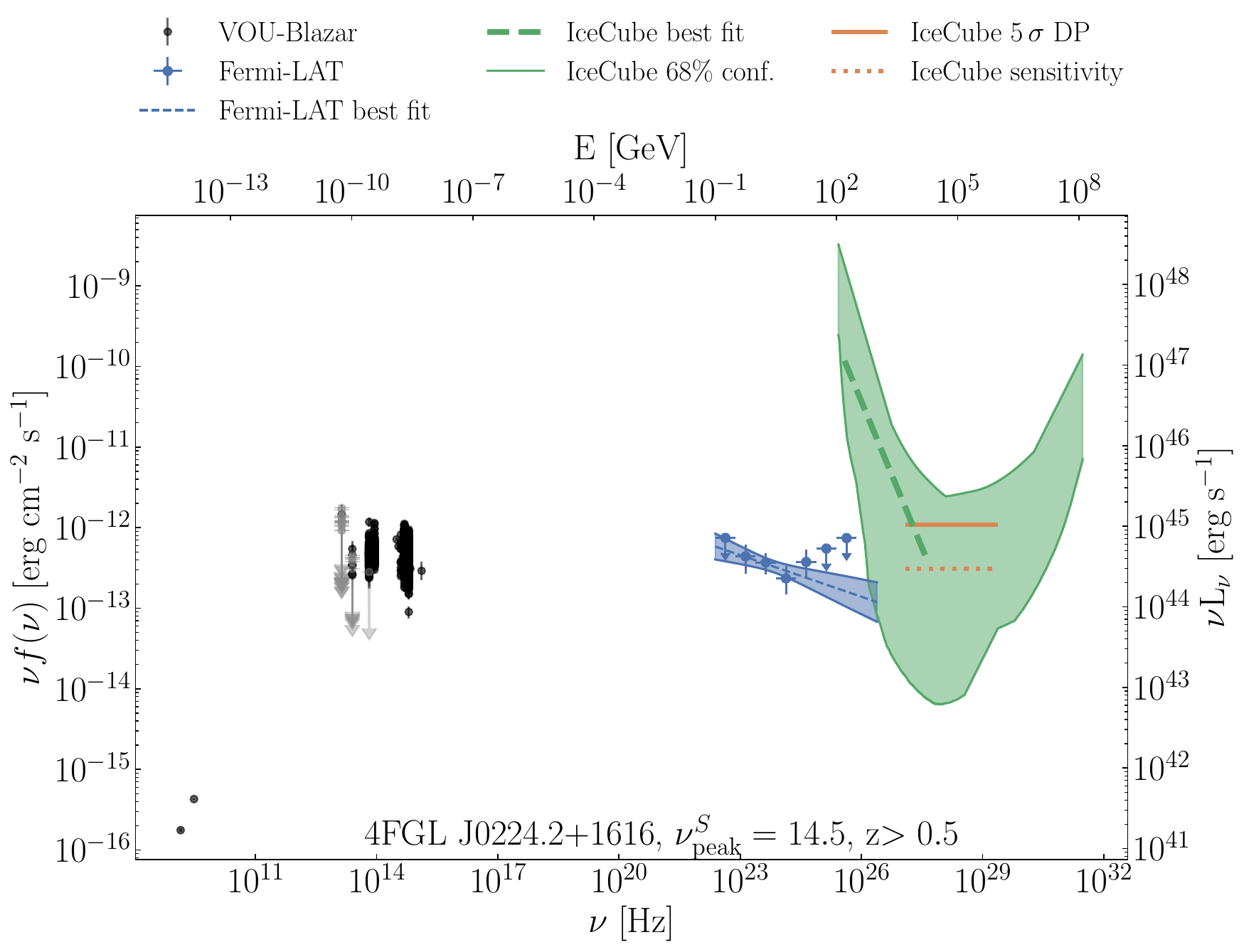}
\includegraphics[width=0.49\textwidth]{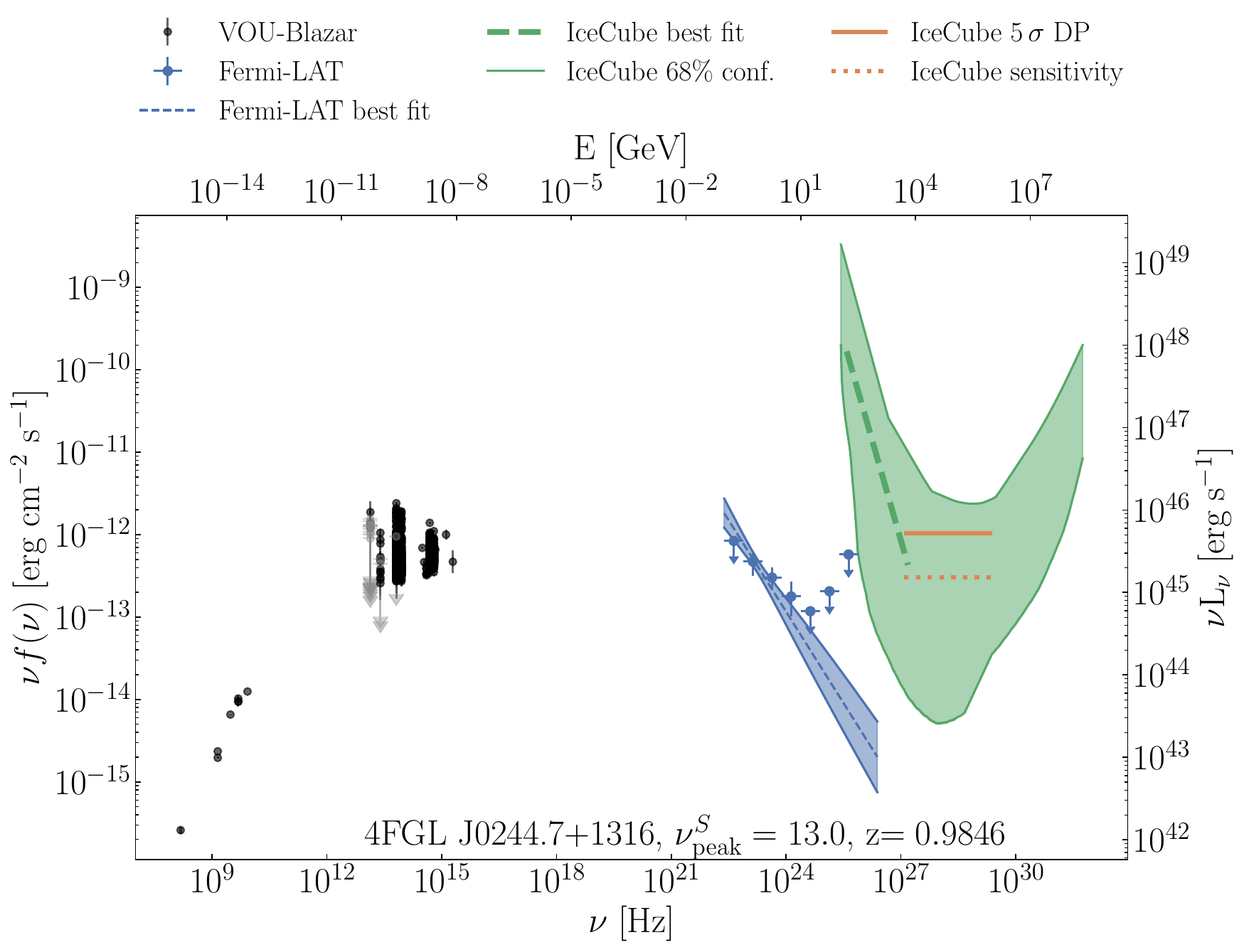}
\includegraphics[width=0.49\textwidth]{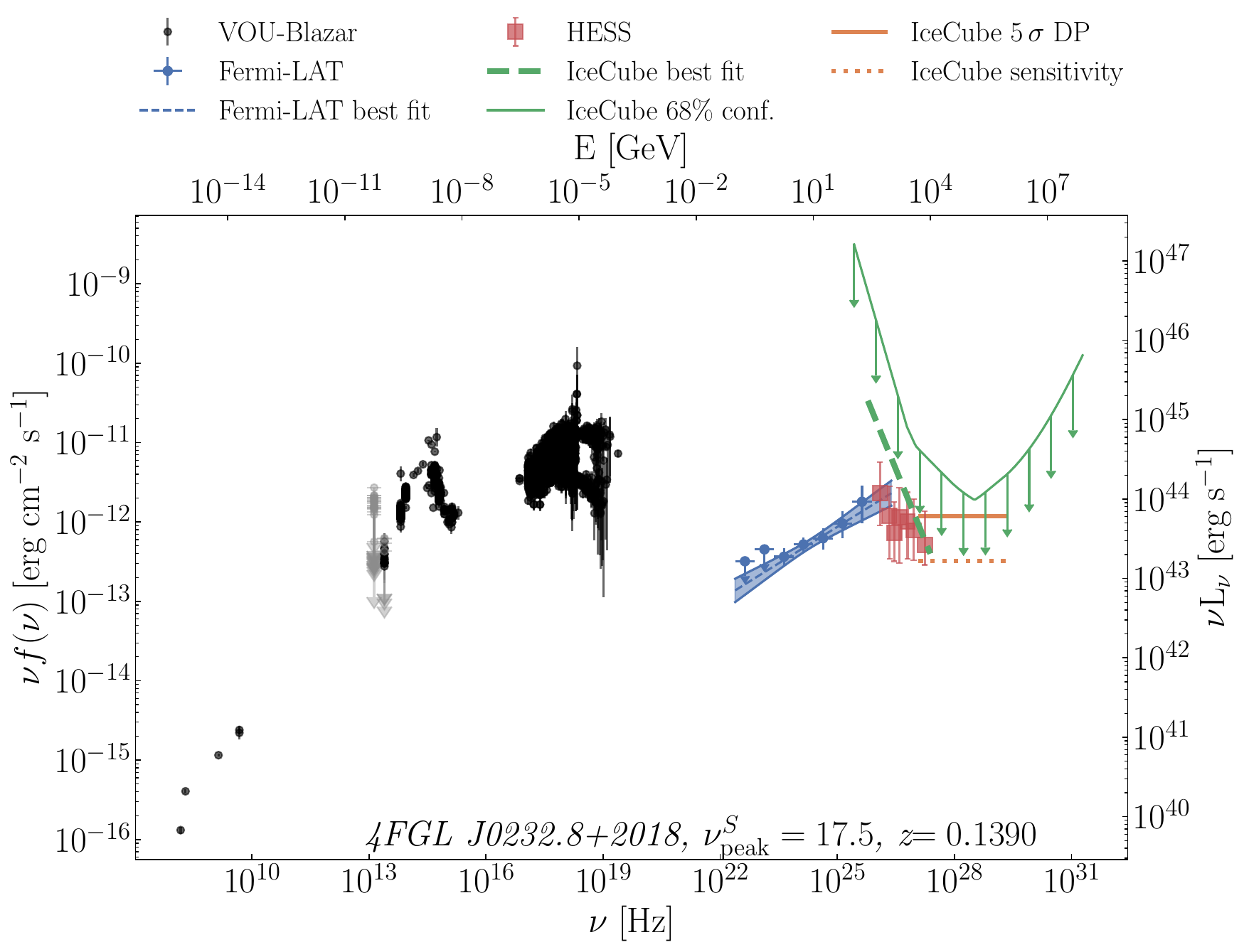}
\includegraphics[width=0.49\textwidth]{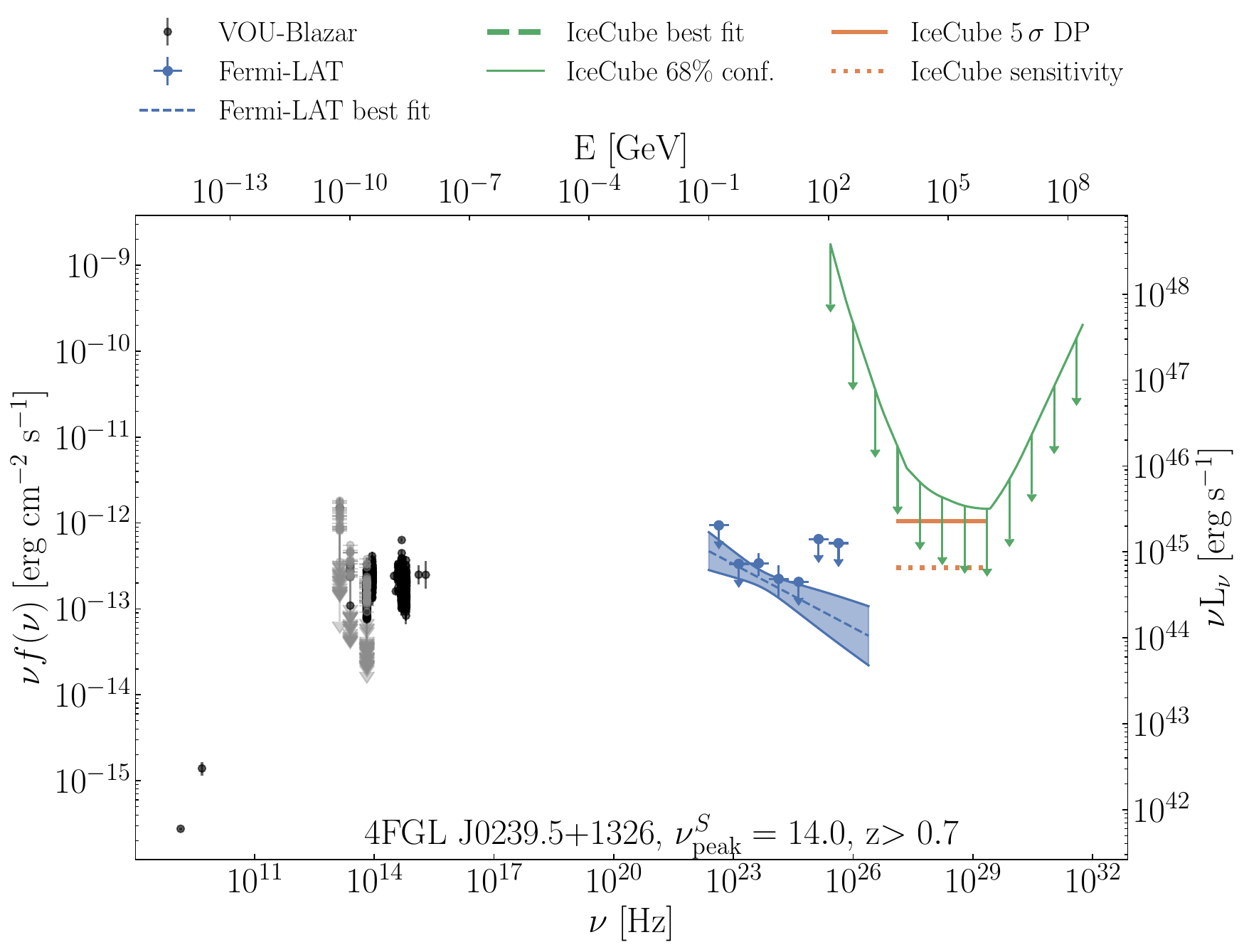}
\includegraphics[width=0.49\textwidth]{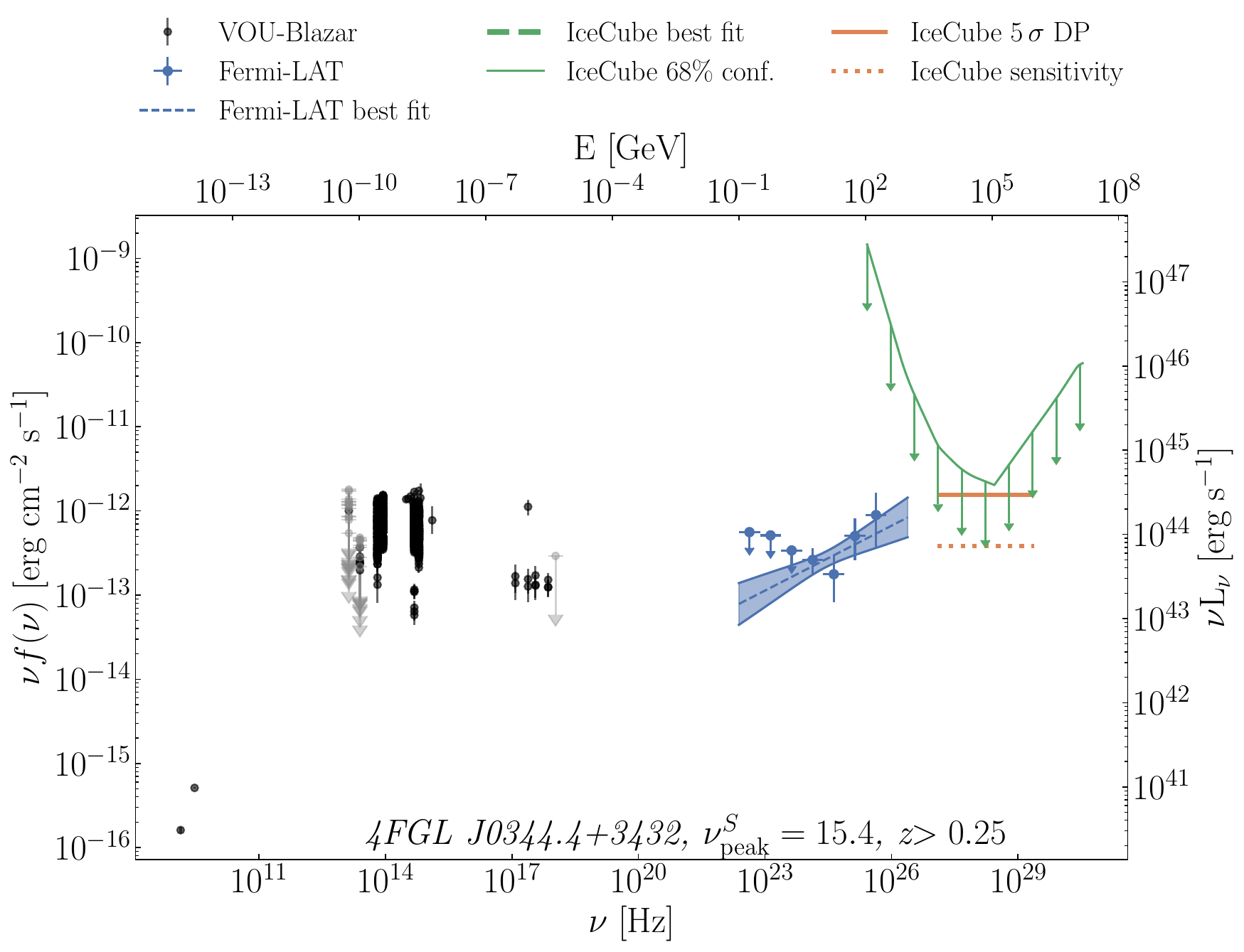}

\caption{SEDs of all investigated objects. The black dots show the multi-wavelength data, and upper limits are displayed in grey. The blue dots display the best fit of the \fermi-LAT SED, and the blue bowtie indicates the uncertainties for the fitted gamma-ray flux. For two objects (4FGL J0232.8+2018 [3HSP J023248.5+20171] and 4FGL J0509.4+0542 [TXS 0506+056]) we include MAGIC \citep{icfermi} and HESS \citep{10.1093/mnras/sty857} data in red (uncorrected for extragalactic background light absorption). The solid (dashed) orange line shows the $5\,\sigma$ discovery potential (the sensitivity) for neutrino emission from \citet{Aartsen2020}. The discovery potential (DP) is the flux (with an $E^{-2}$ spectrum) necessary to detect the source with $5\,\sigma$ significance at 50 per cent confidence level. The sensitivity shows the 90 per cent confidence flux limit in case of no neutrino flux (also assuming an $E^{-2}$ spectrum). The green line (band) shows the 68 per cent confidence limits (band) on the best-fit neutrino flux based on public IceCube data \citep{IceCube_2021}. If the best-fit neutrino flux was $>0$, we show it as a green dashed line. All neutrinos fluxes are single flavour (muon neutrino and antineutrino) fluxes. We list the observed 
\nup~and the redshift. Masquerading sources are marked in bold, and sources where the extension of the $\gamma$-ray flux meets the $5\,\sigma$ discovery potential or the best-fit neutrino flux are marked in italics. 
}
\label{fig:includedSEDs}
\end{figure*}

\setcounter{figure}{1}

\begin{figure*}
\includegraphics[width=0.49\textwidth]{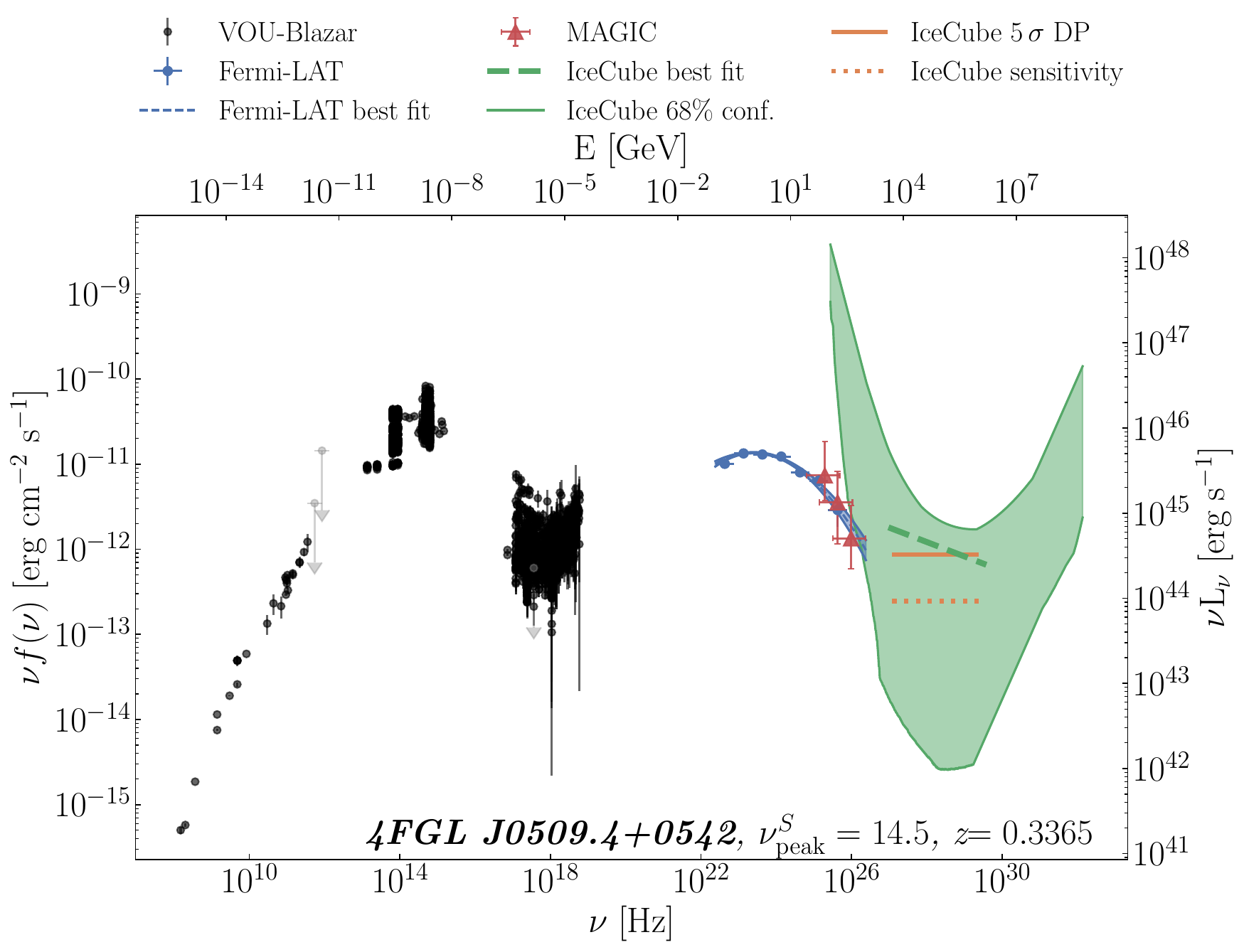}
\includegraphics[width=0.49\textwidth]{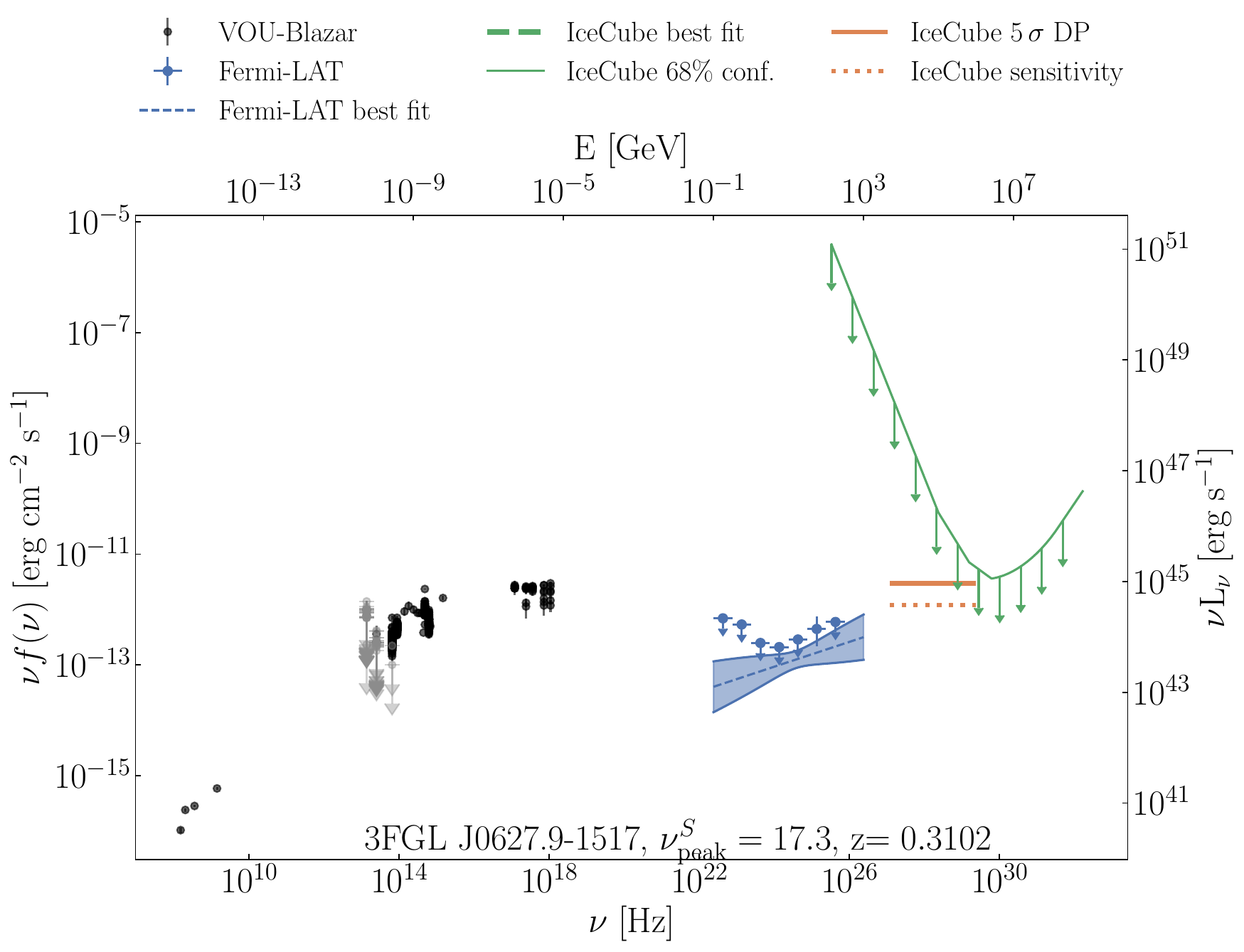}
\includegraphics[width=0.49\textwidth]{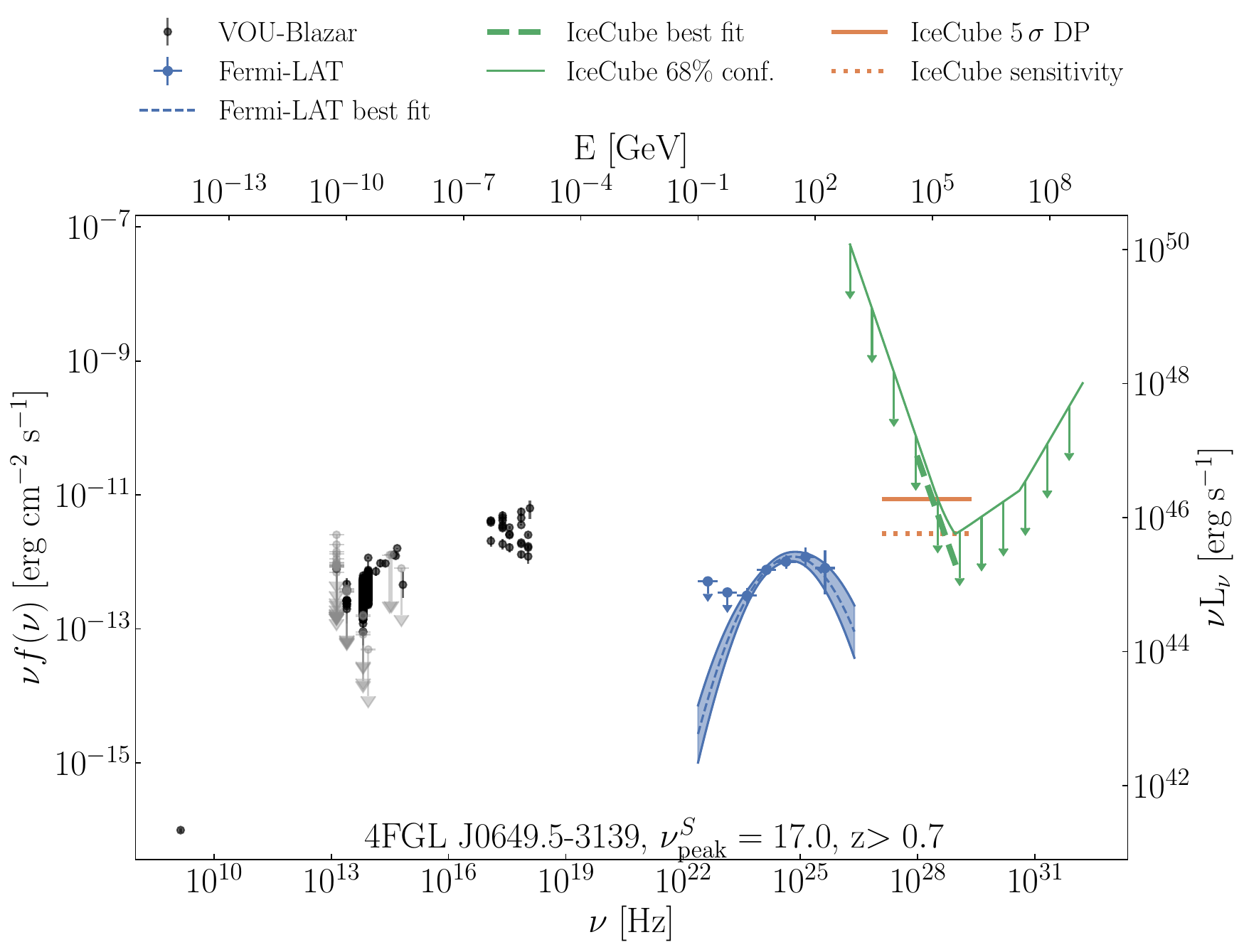}
\includegraphics[width=0.49\textwidth]{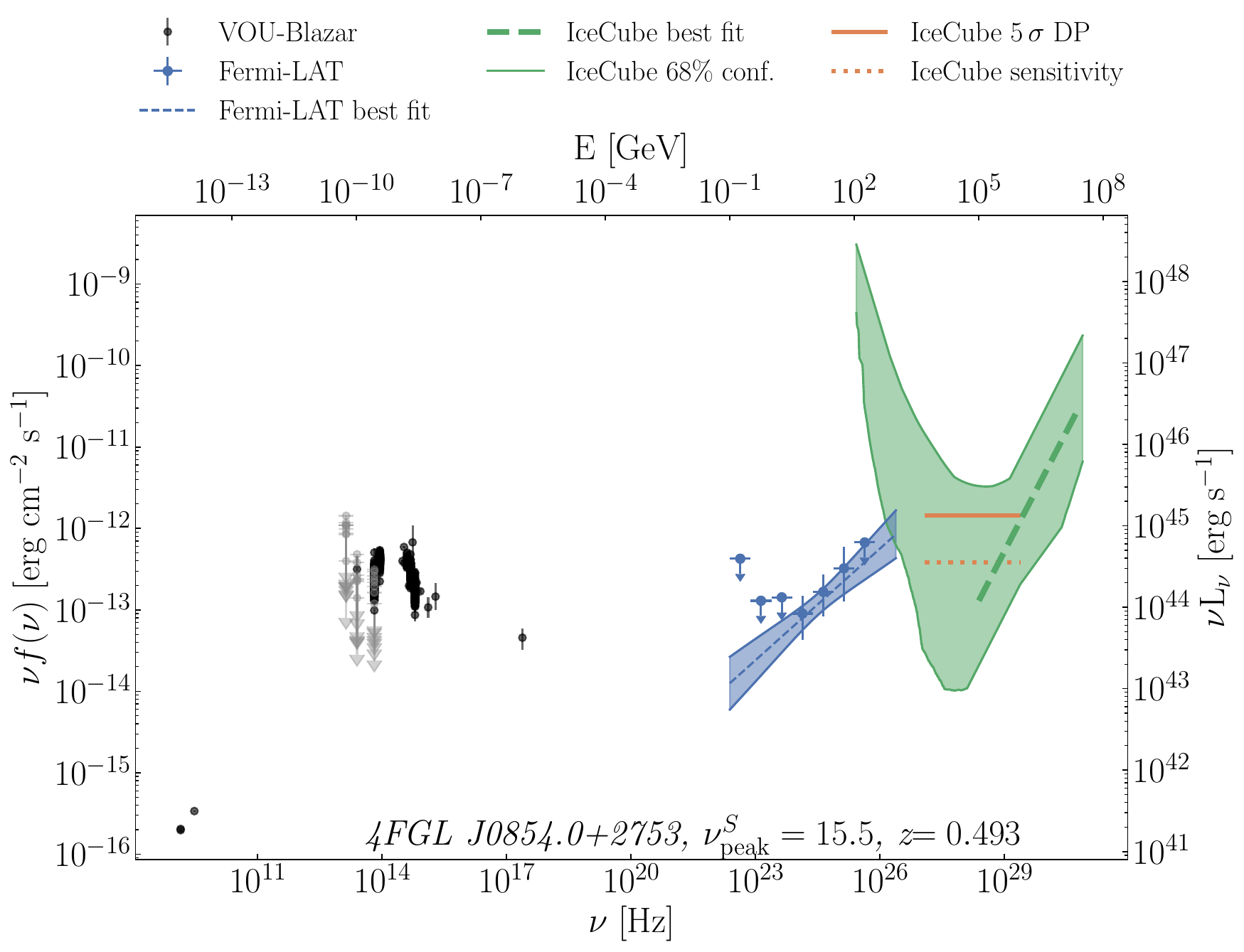}
\includegraphics[width=0.49\textwidth]{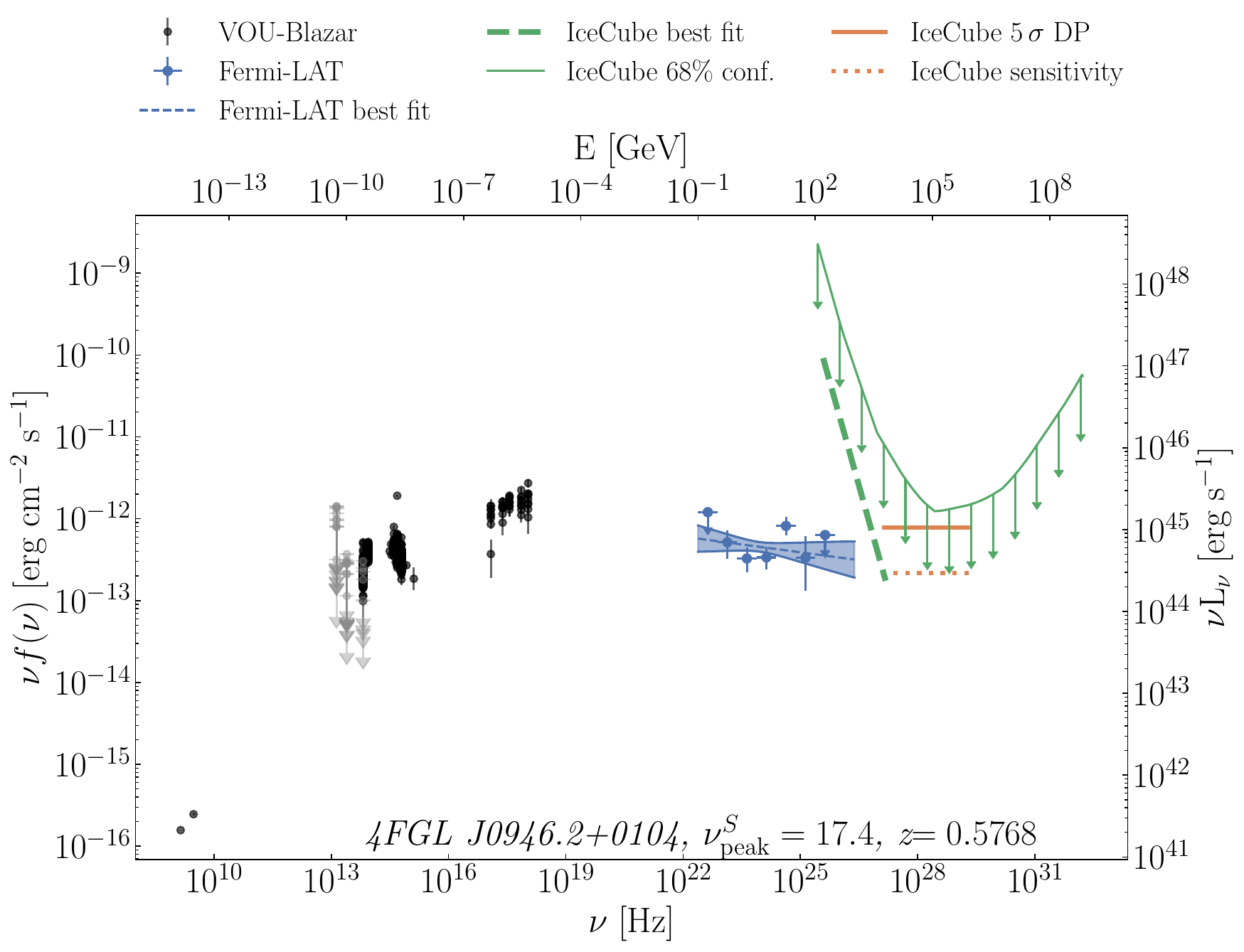}
\includegraphics[width=0.49\textwidth]{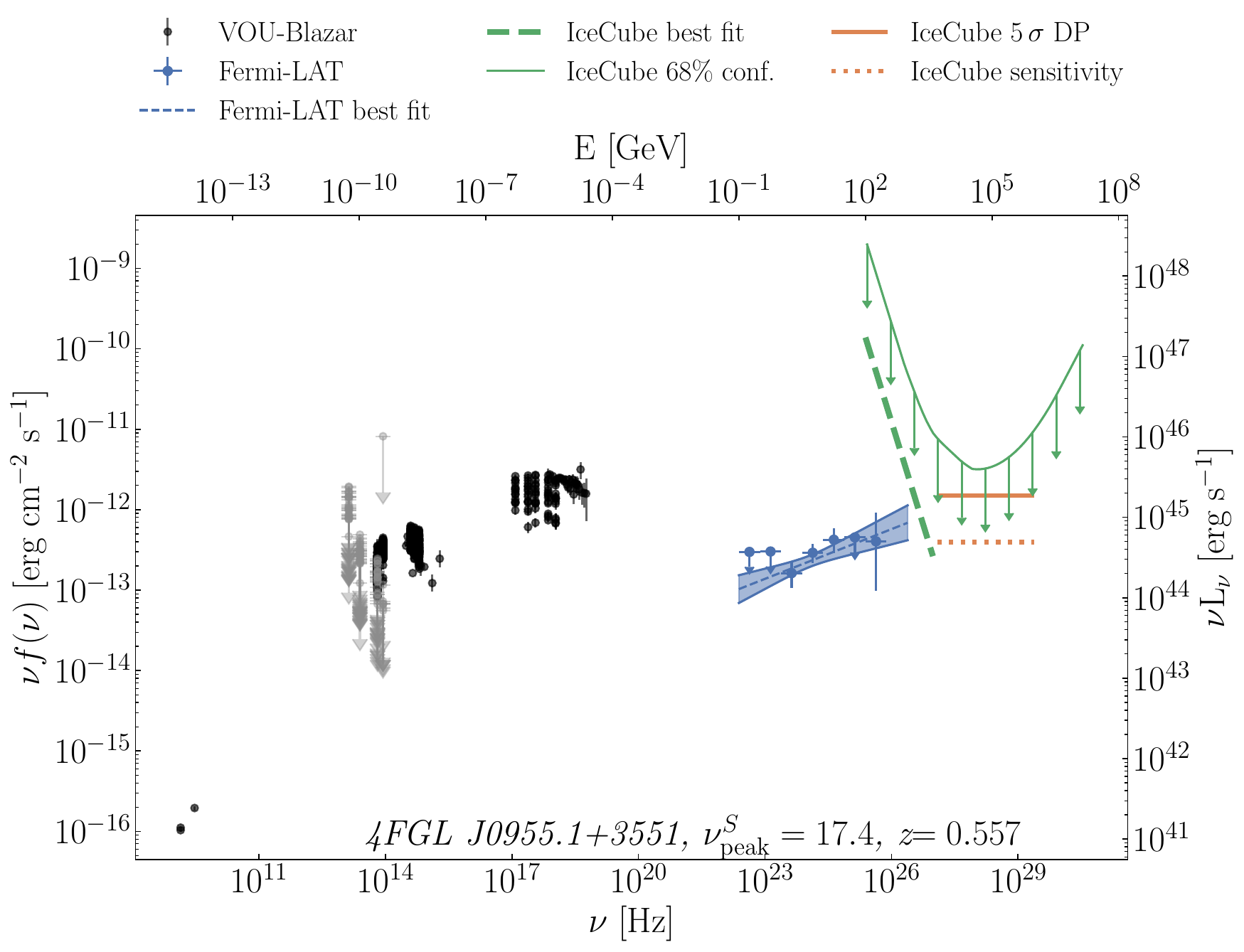}

\caption{- \textit{Continued}}
\end{figure*}

\setcounter{figure}{1}

\begin{figure*}
\includegraphics[width=0.49\textwidth]{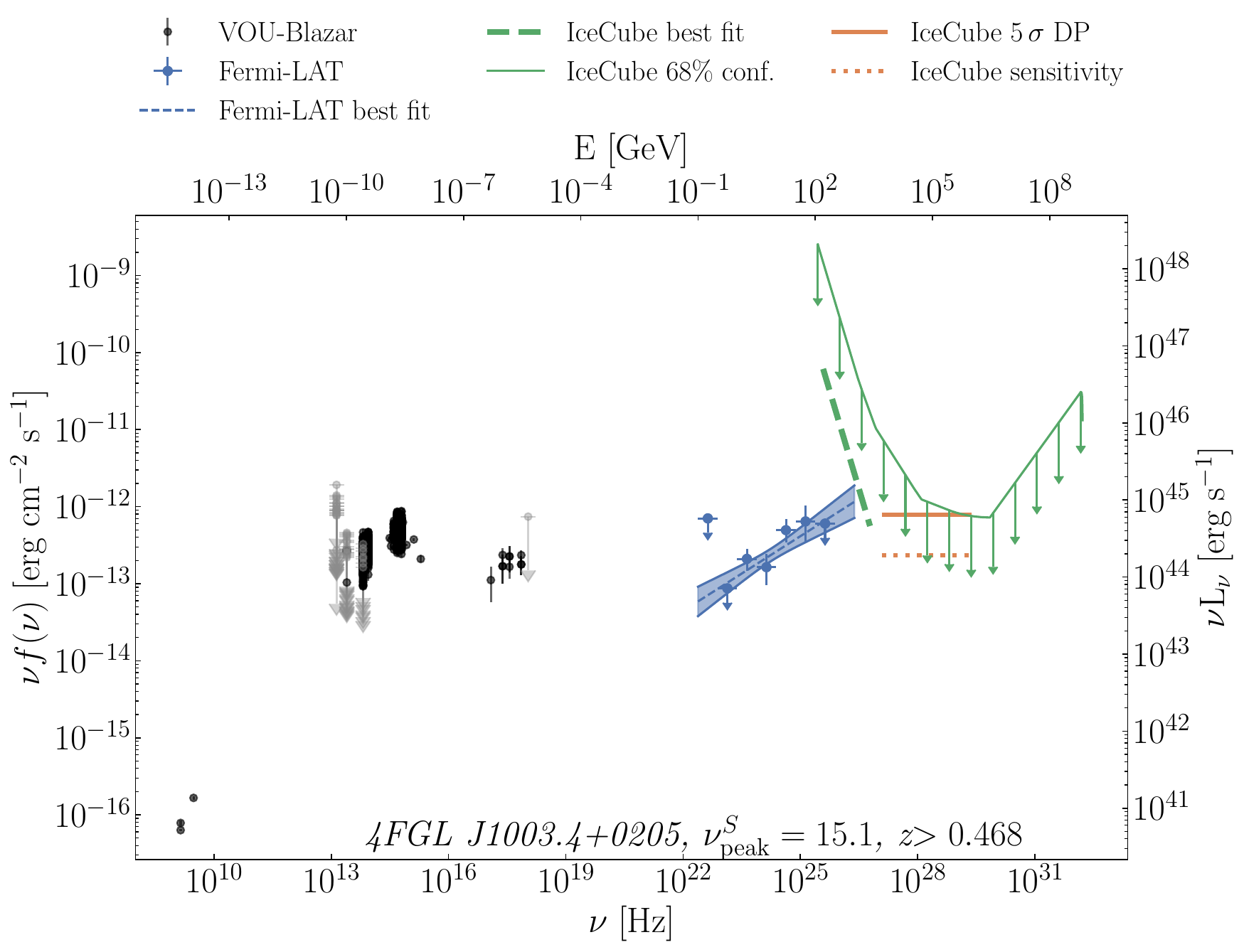}
\includegraphics[width=0.49\textwidth]{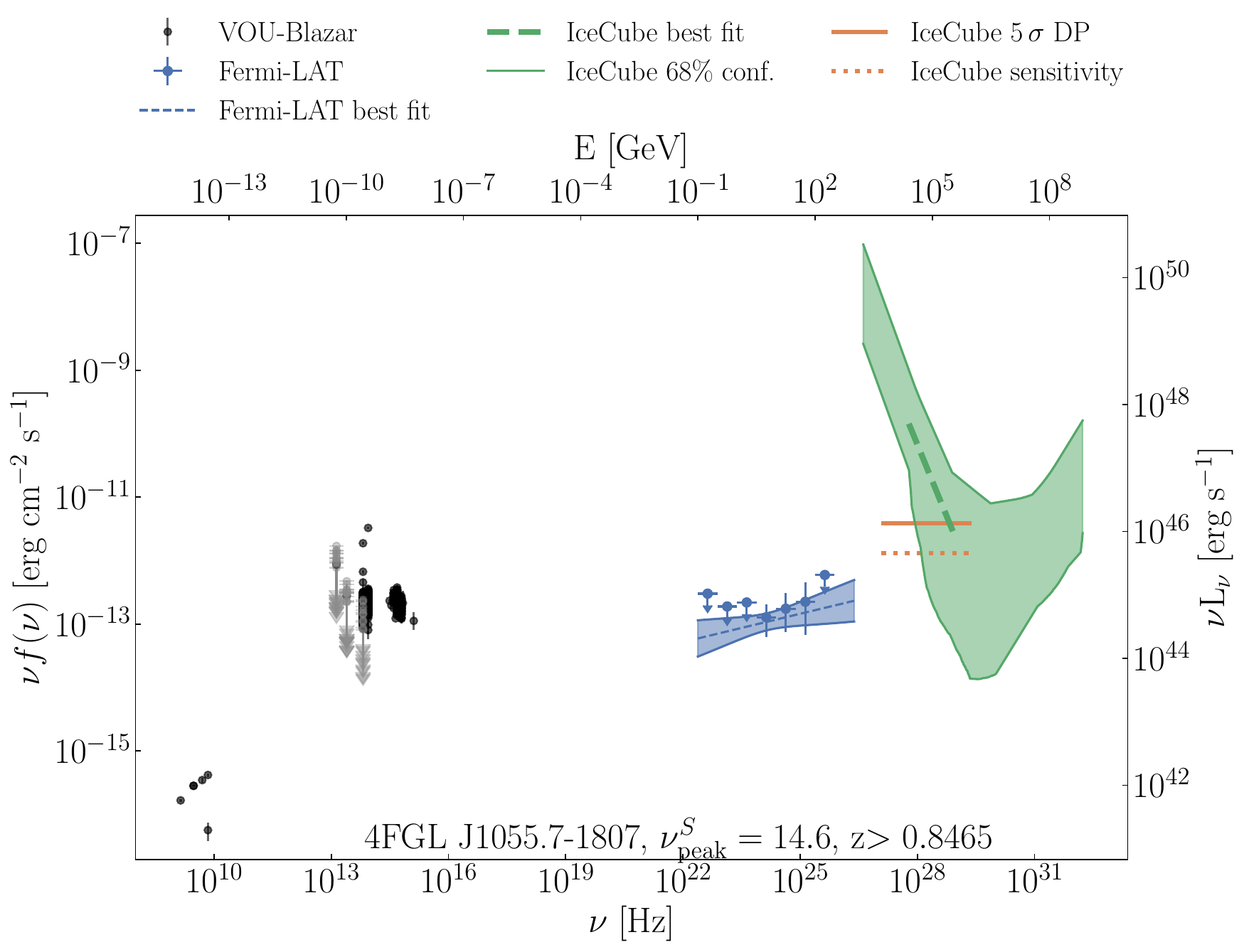}
\includegraphics[width=0.49\textwidth]{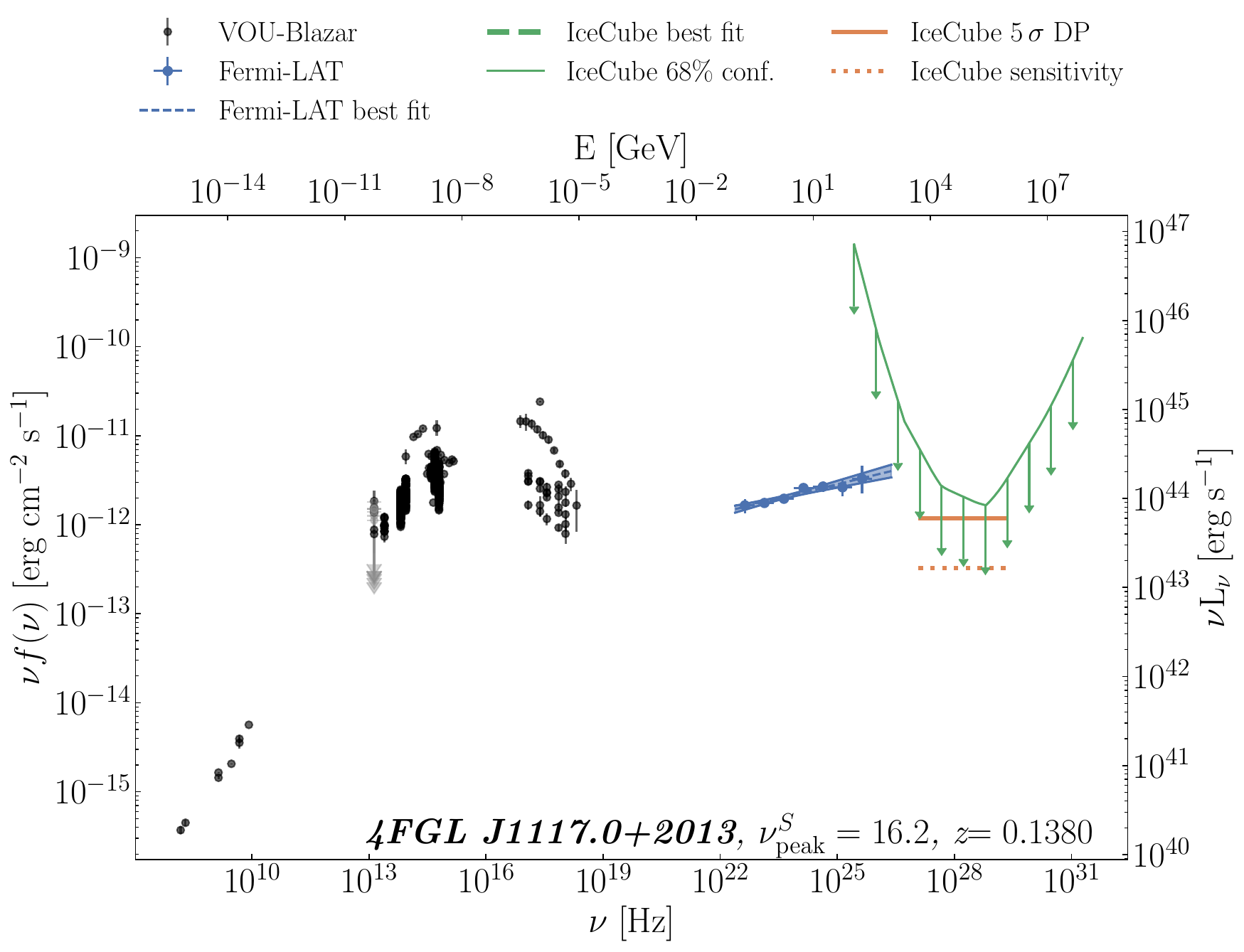}
\includegraphics[width=0.49\textwidth]{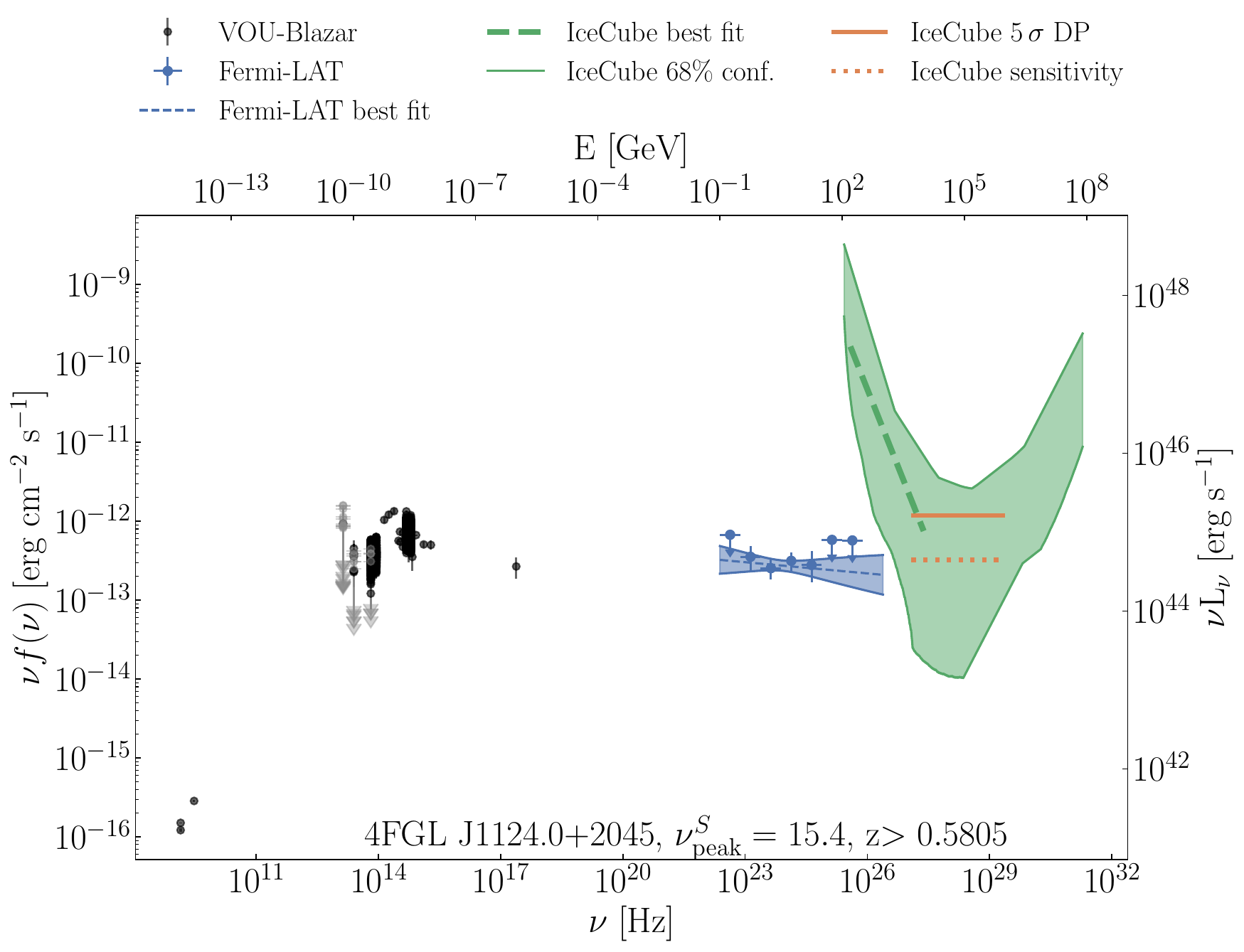}
\includegraphics[width=0.49\textwidth]{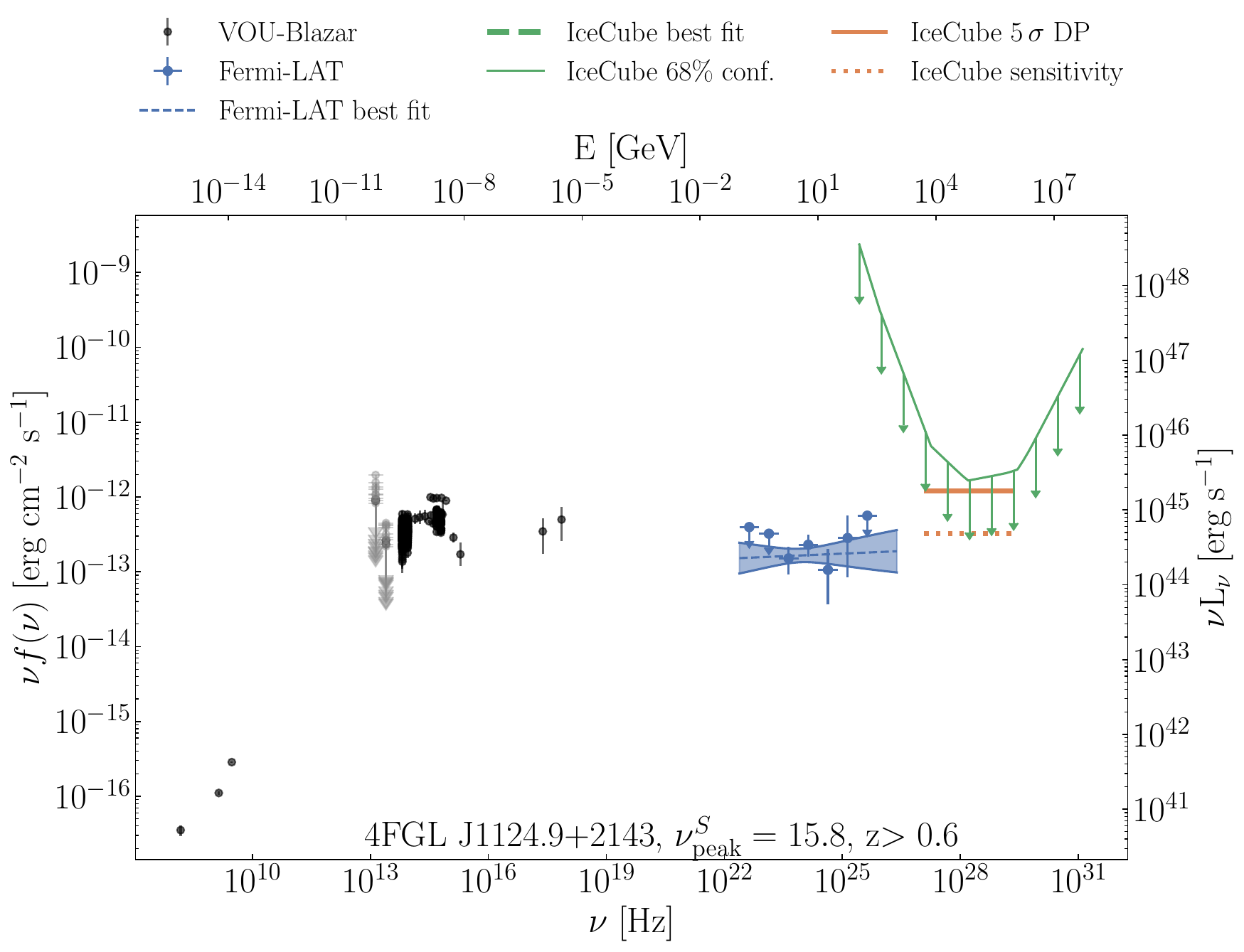}
\includegraphics[width=0.49\textwidth]{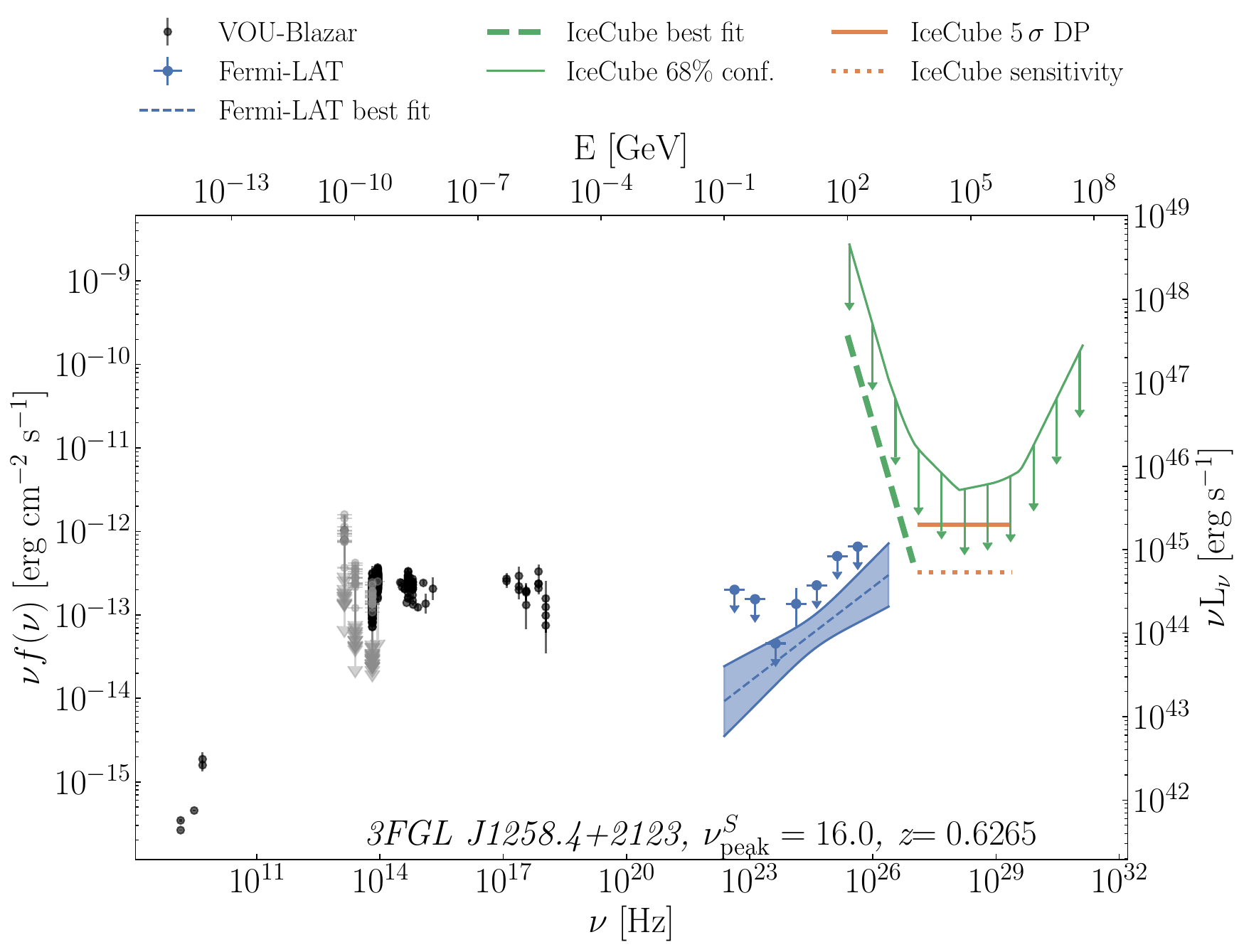}

\caption{- \textit{Continued}}
\end{figure*}

\setcounter{figure}{1}

\begin{figure*}
\includegraphics[width=0.49\textwidth]{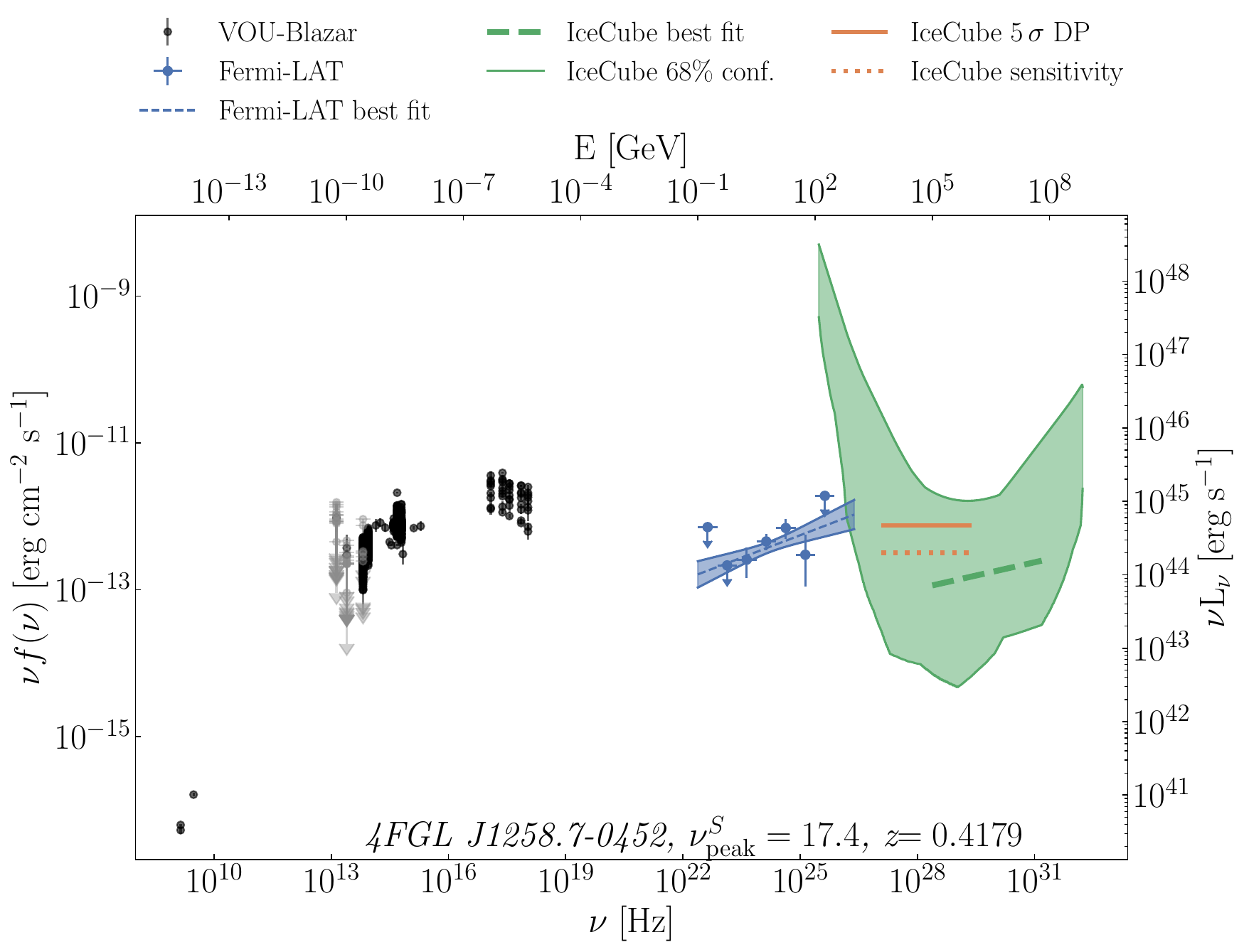}
\includegraphics[width=0.49\textwidth]{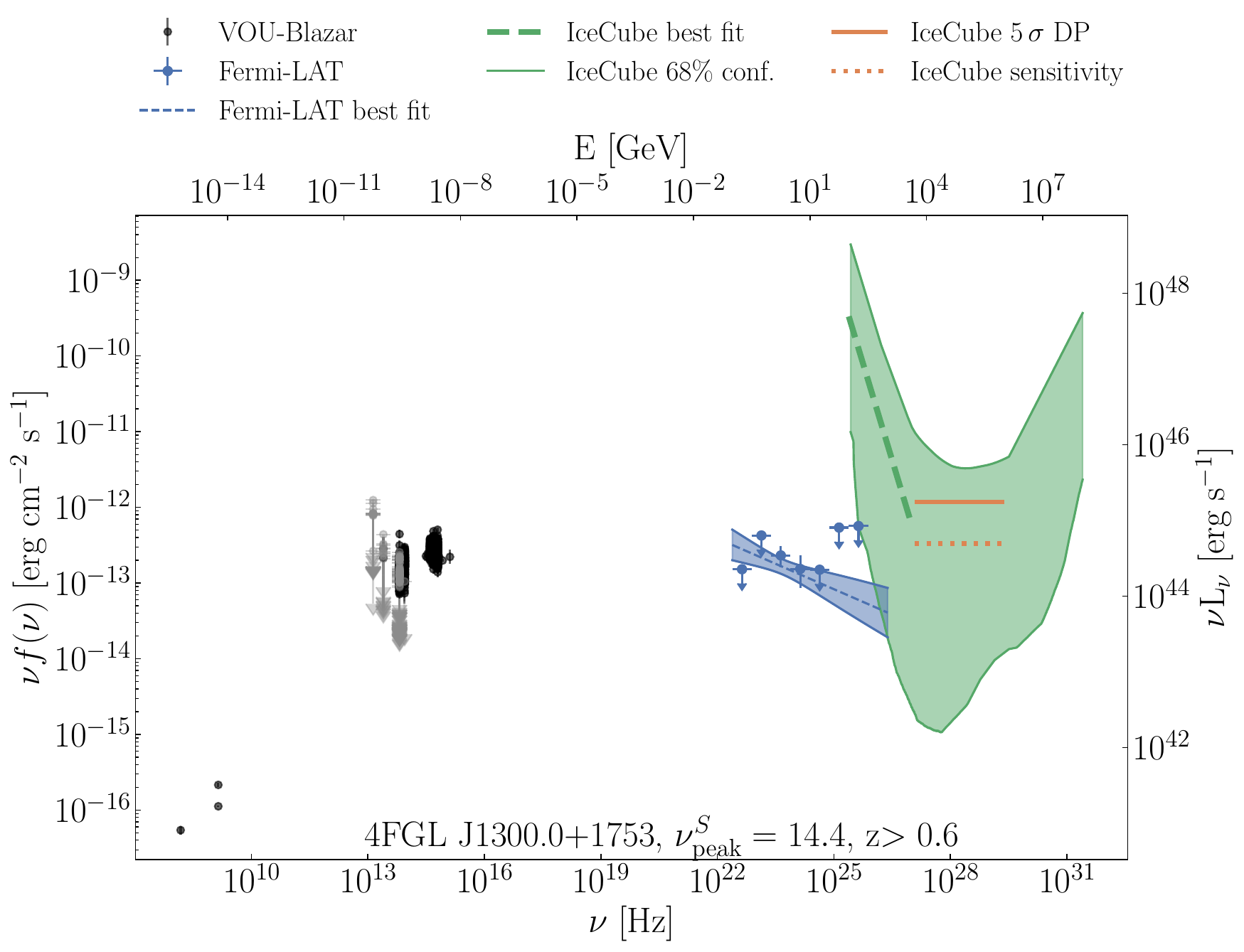}
\includegraphics[width=0.49\textwidth]{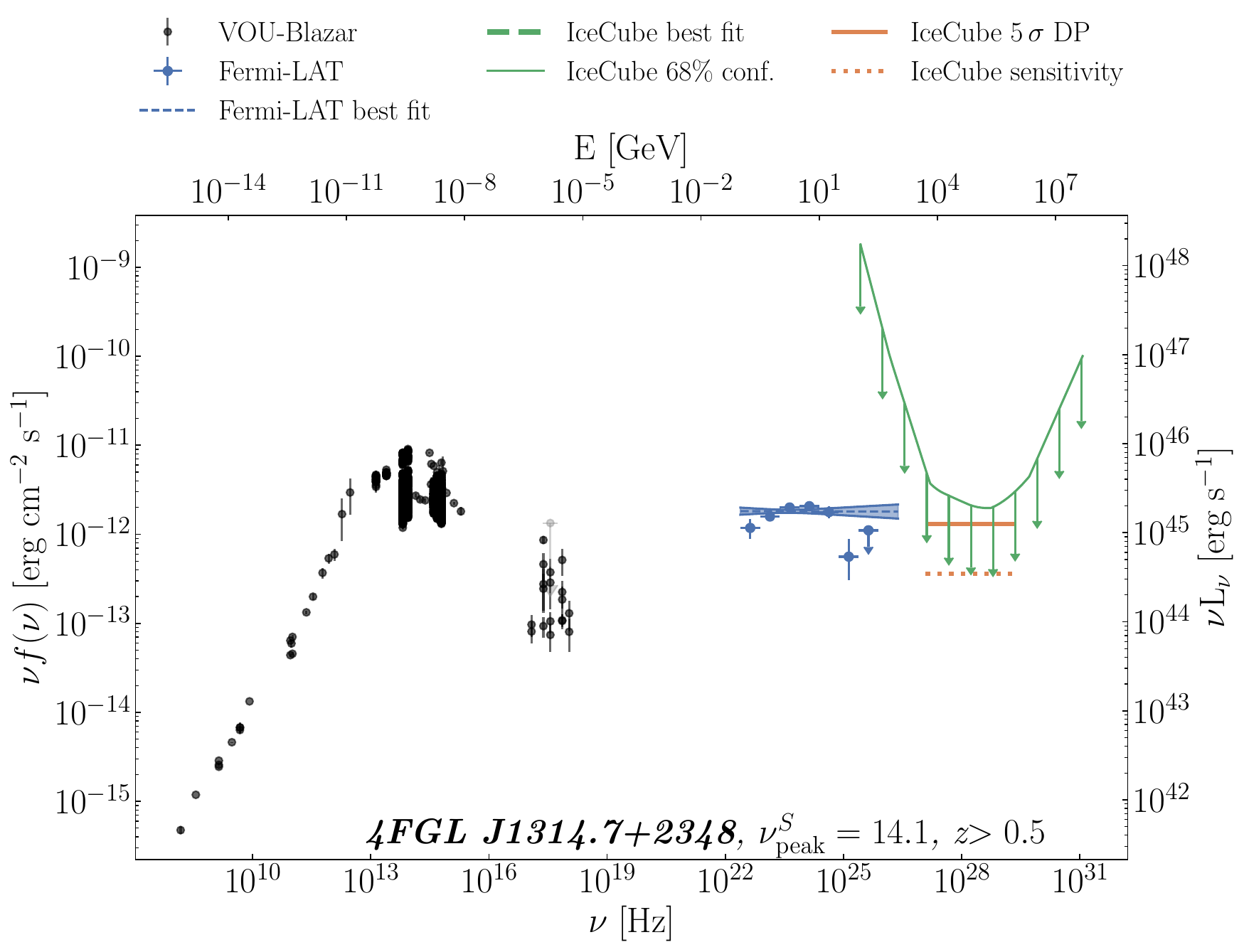}
\includegraphics[width=0.49\textwidth]{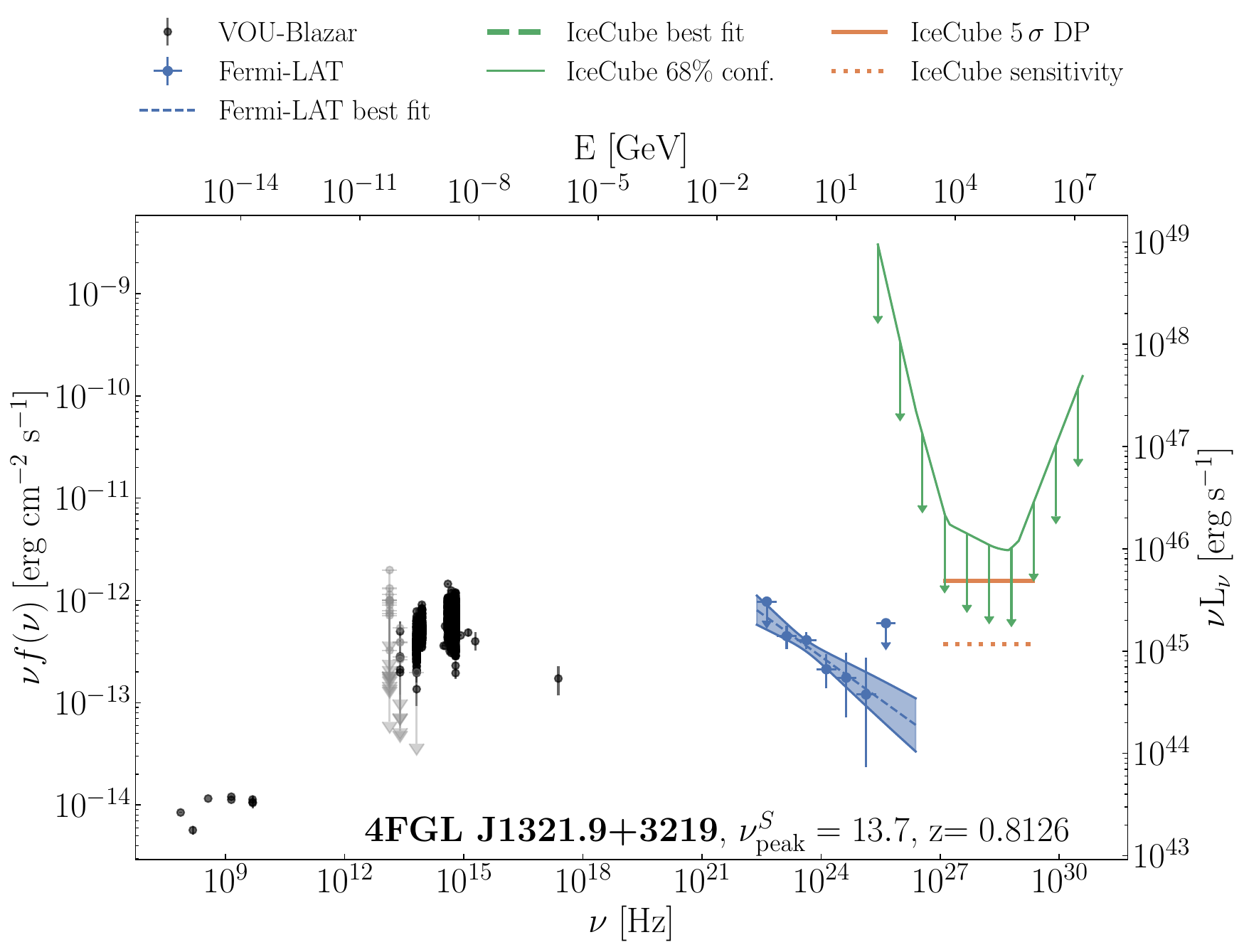}
\includegraphics[width=0.49\textwidth]{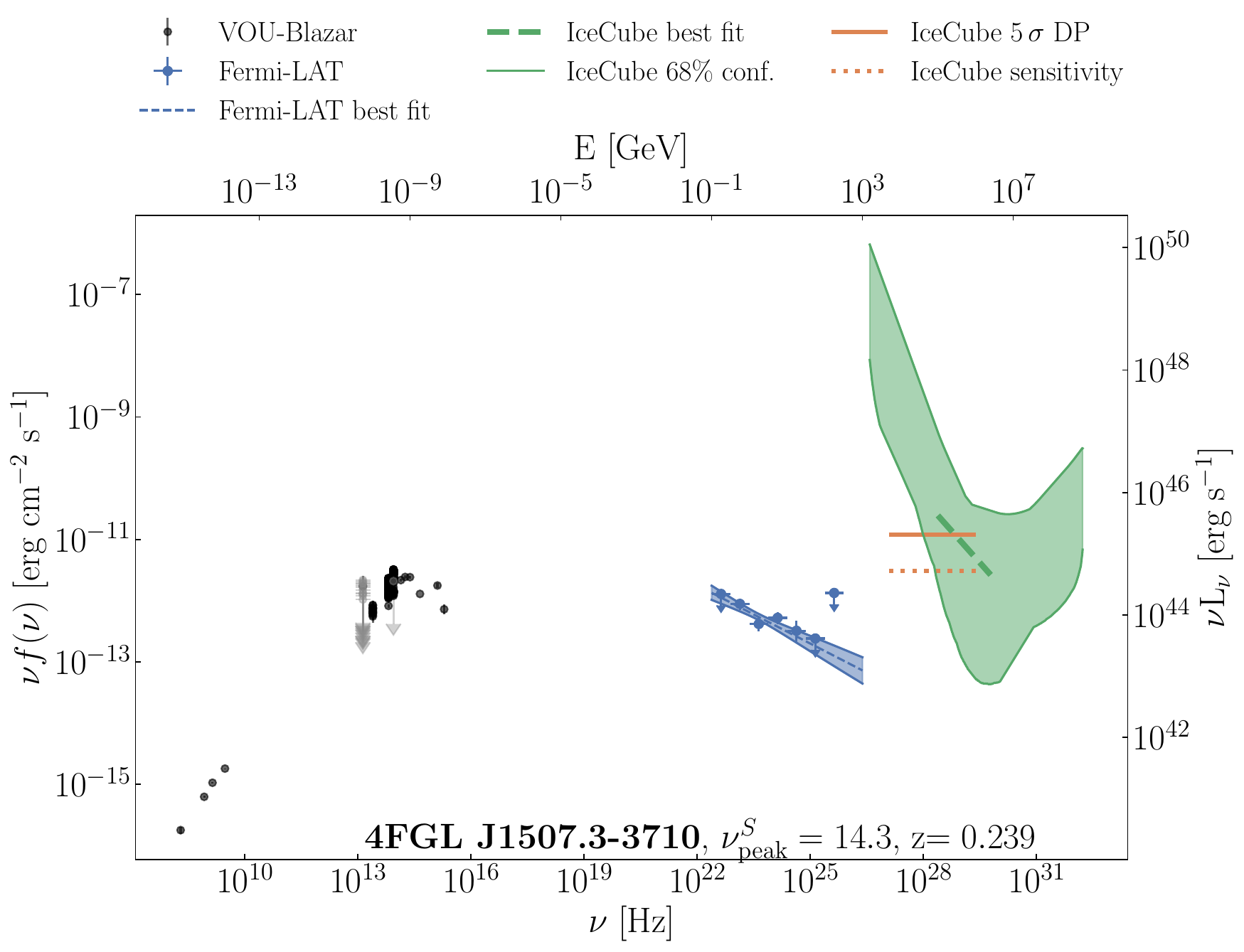}
\includegraphics[width=0.49\textwidth]{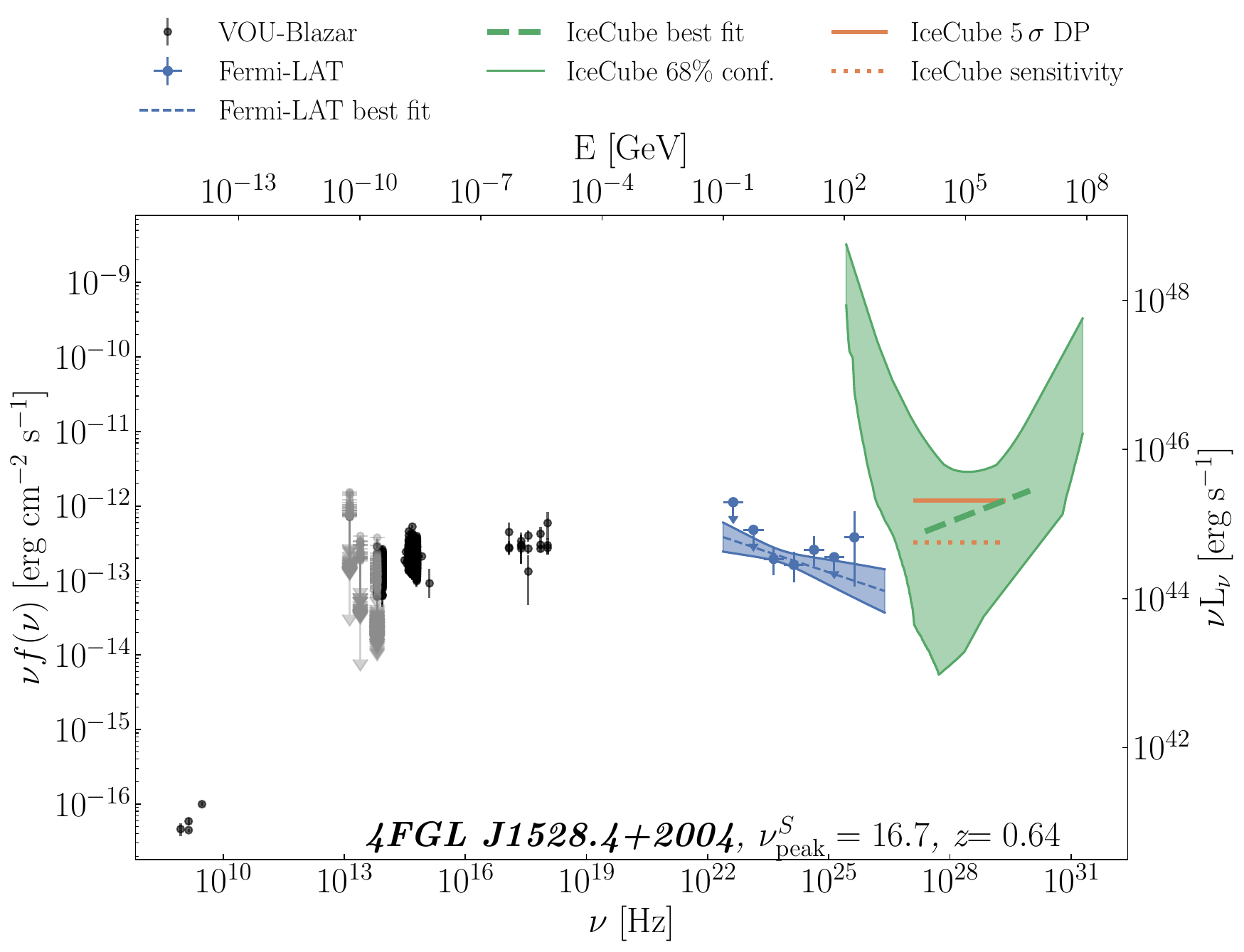}

\caption{- \textit{Continued}}
\end{figure*}

\setcounter{figure}{1}

\begin{figure*}
\includegraphics[width=0.49\textwidth]{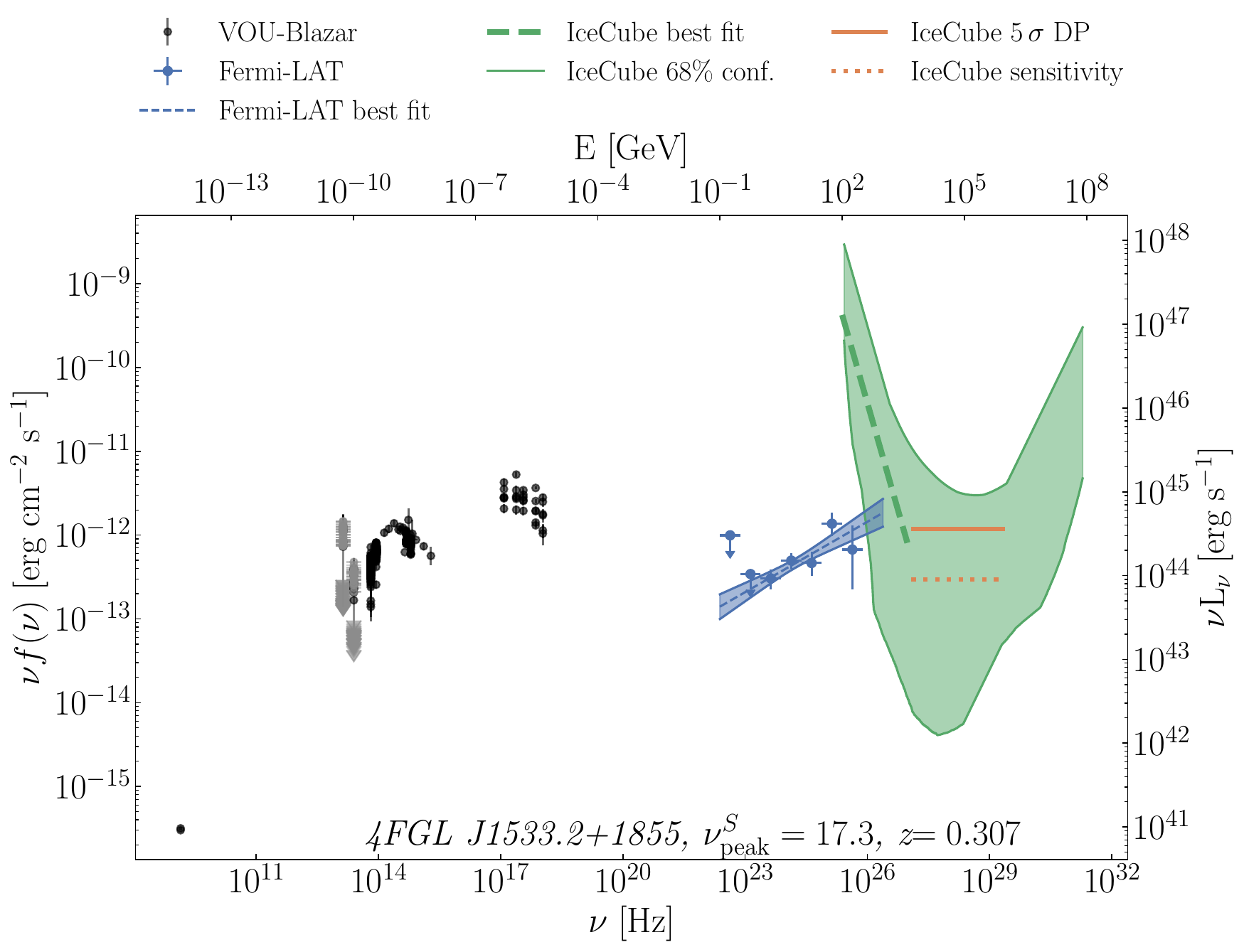}
\includegraphics[width=0.49\textwidth]{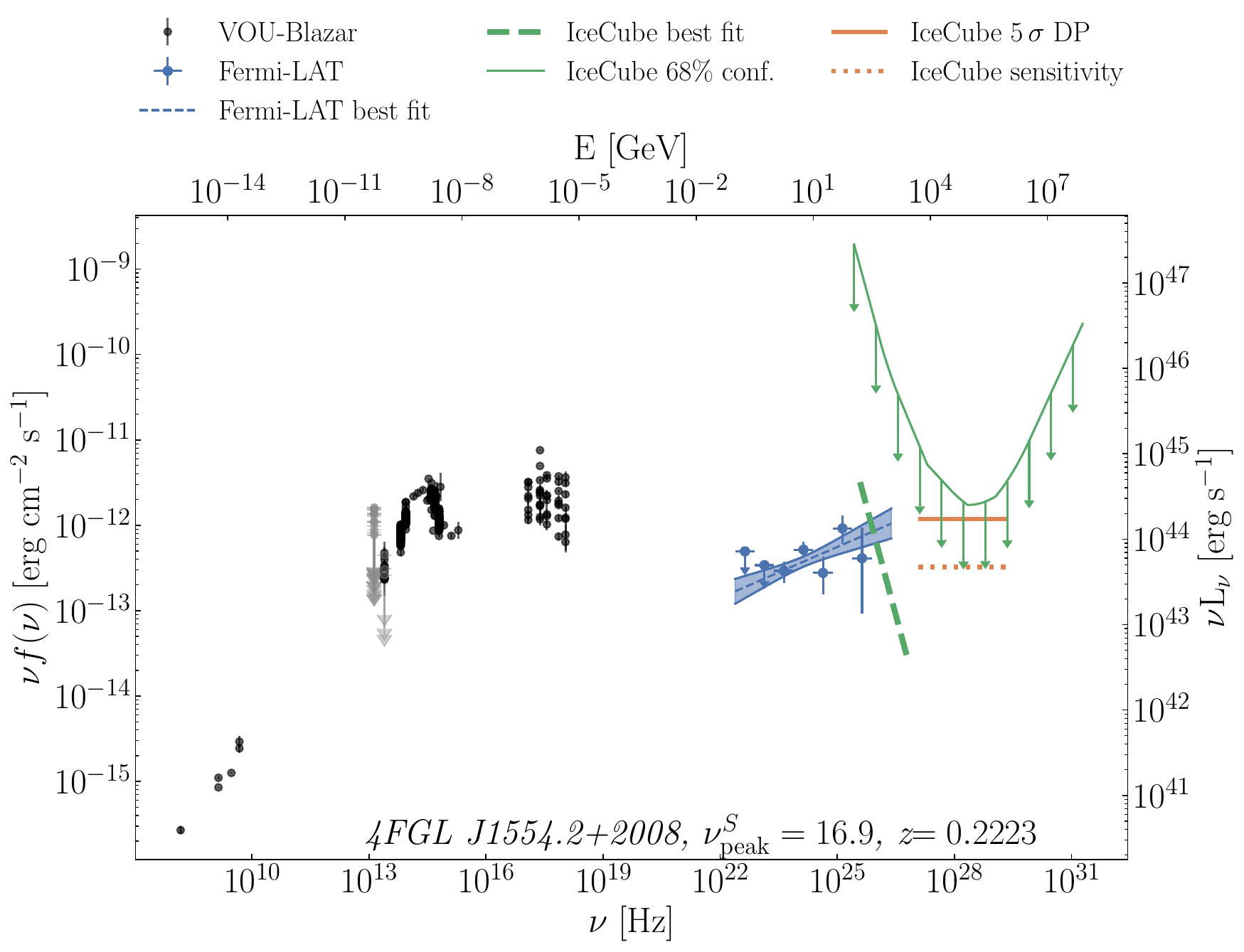}
\includegraphics[width=0.49\textwidth]{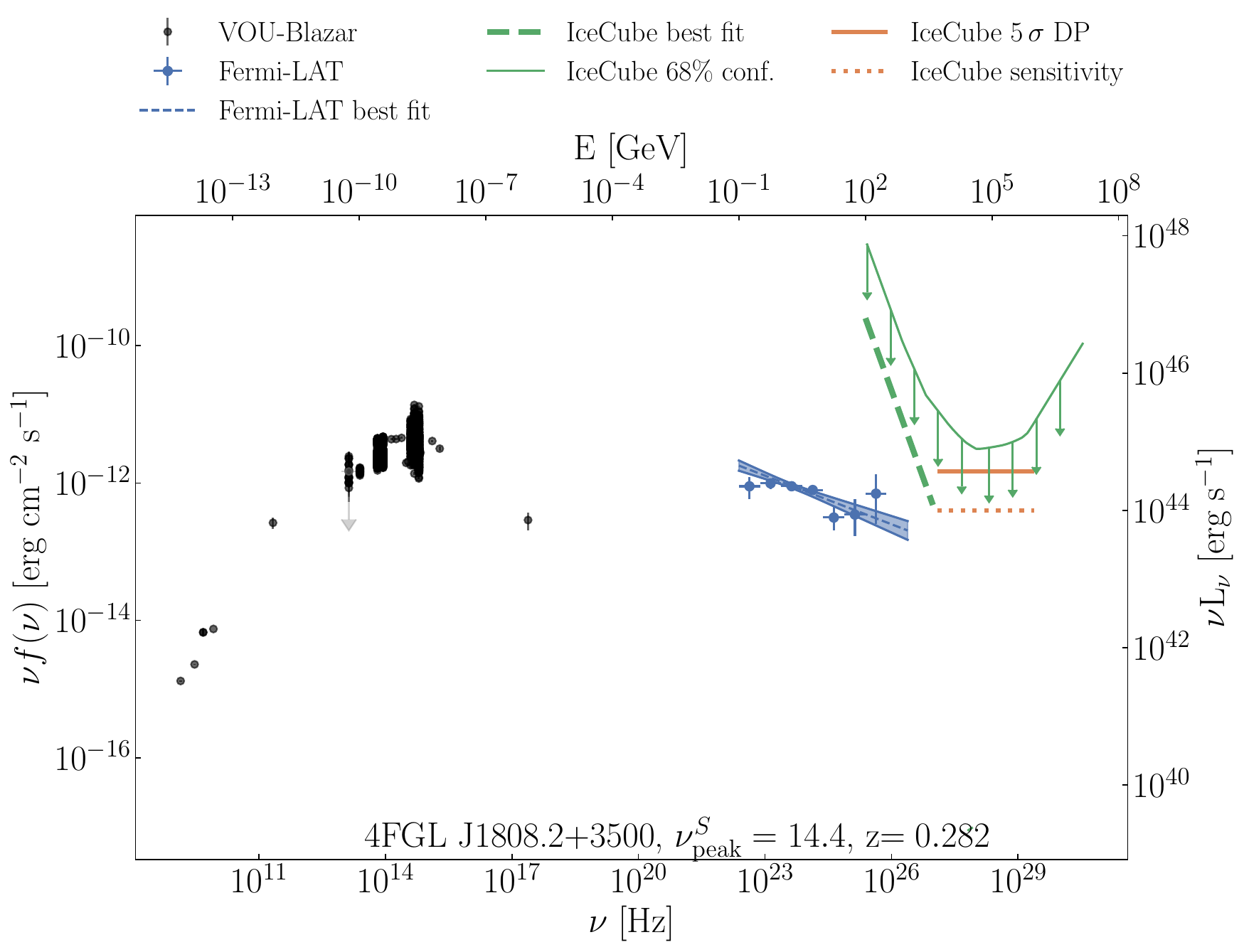}
\includegraphics[width=0.49\textwidth]{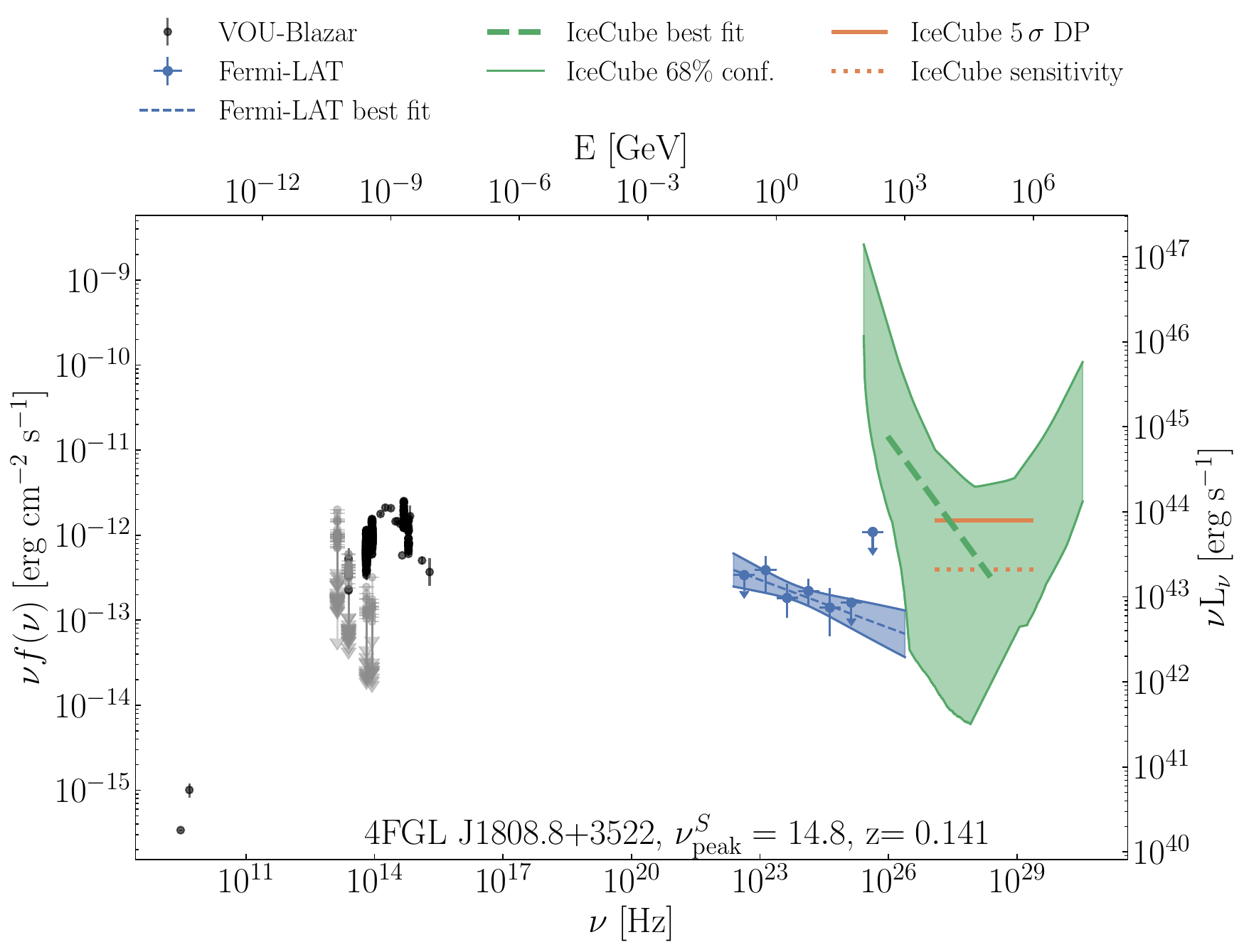}
\includegraphics[width=0.49\textwidth]{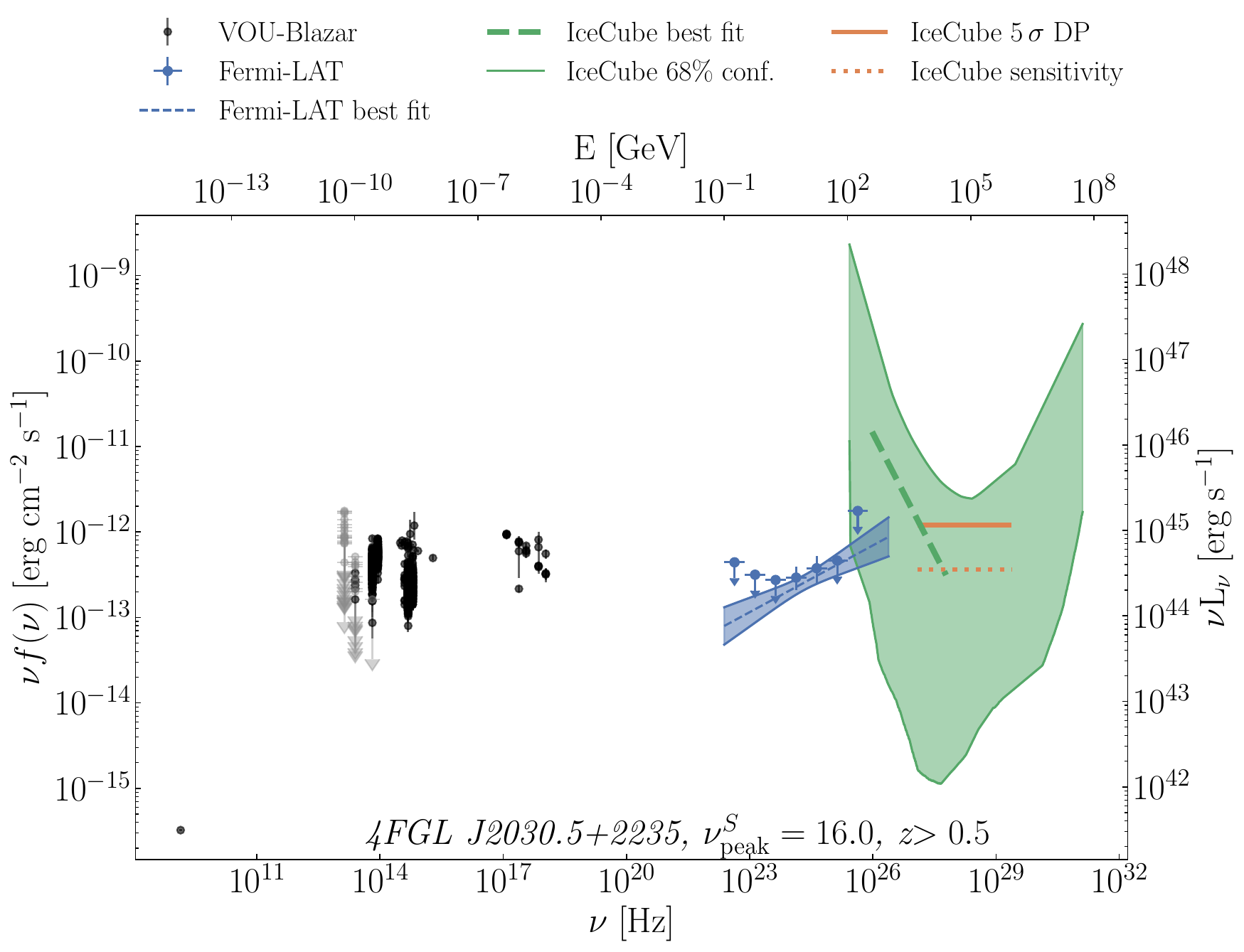}
\includegraphics[width=0.49\textwidth]{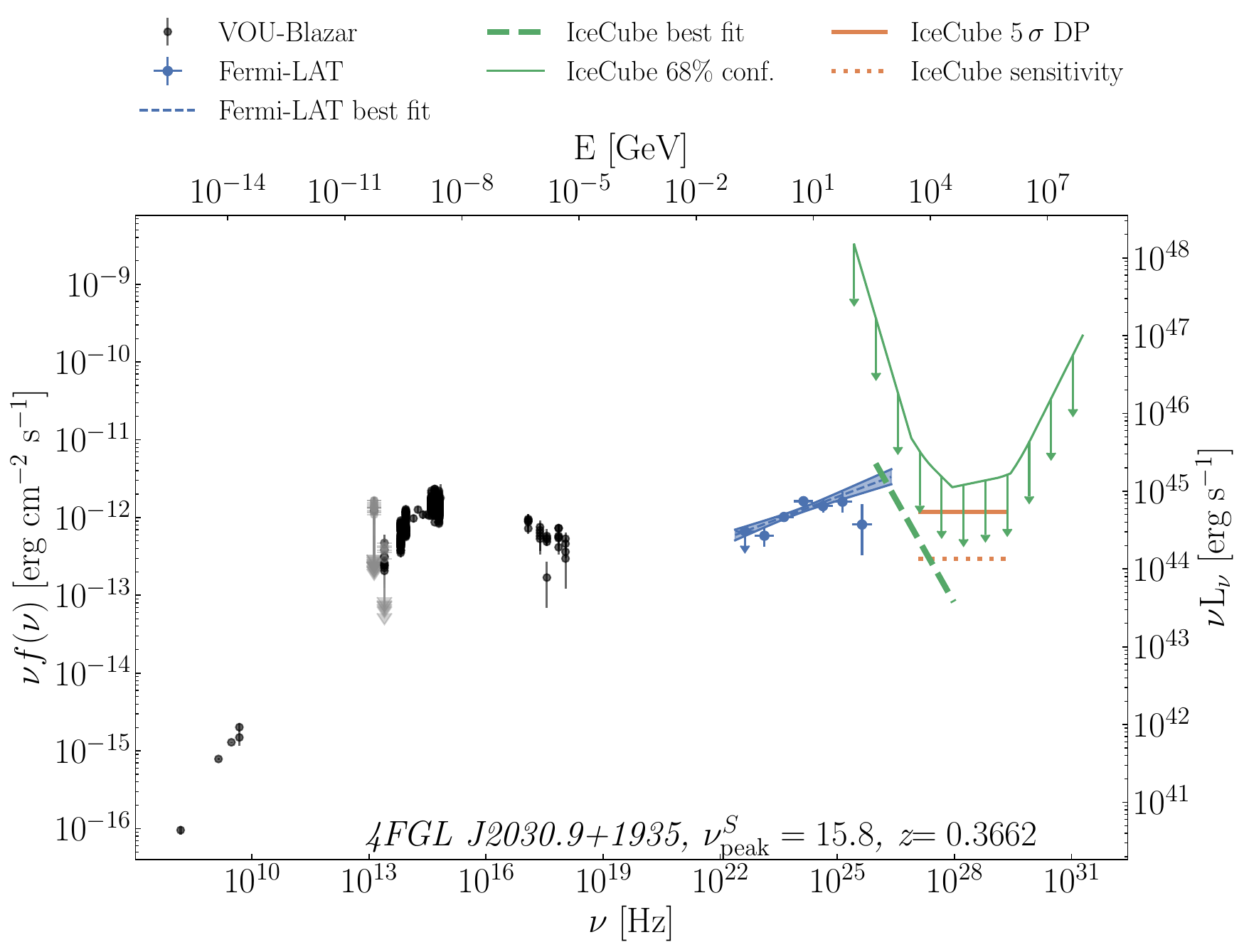}

\caption{- \textit{Continued}}
\end{figure*}

\setcounter{figure}{1}

\begin{figure*}
\includegraphics[width=0.49\textwidth]{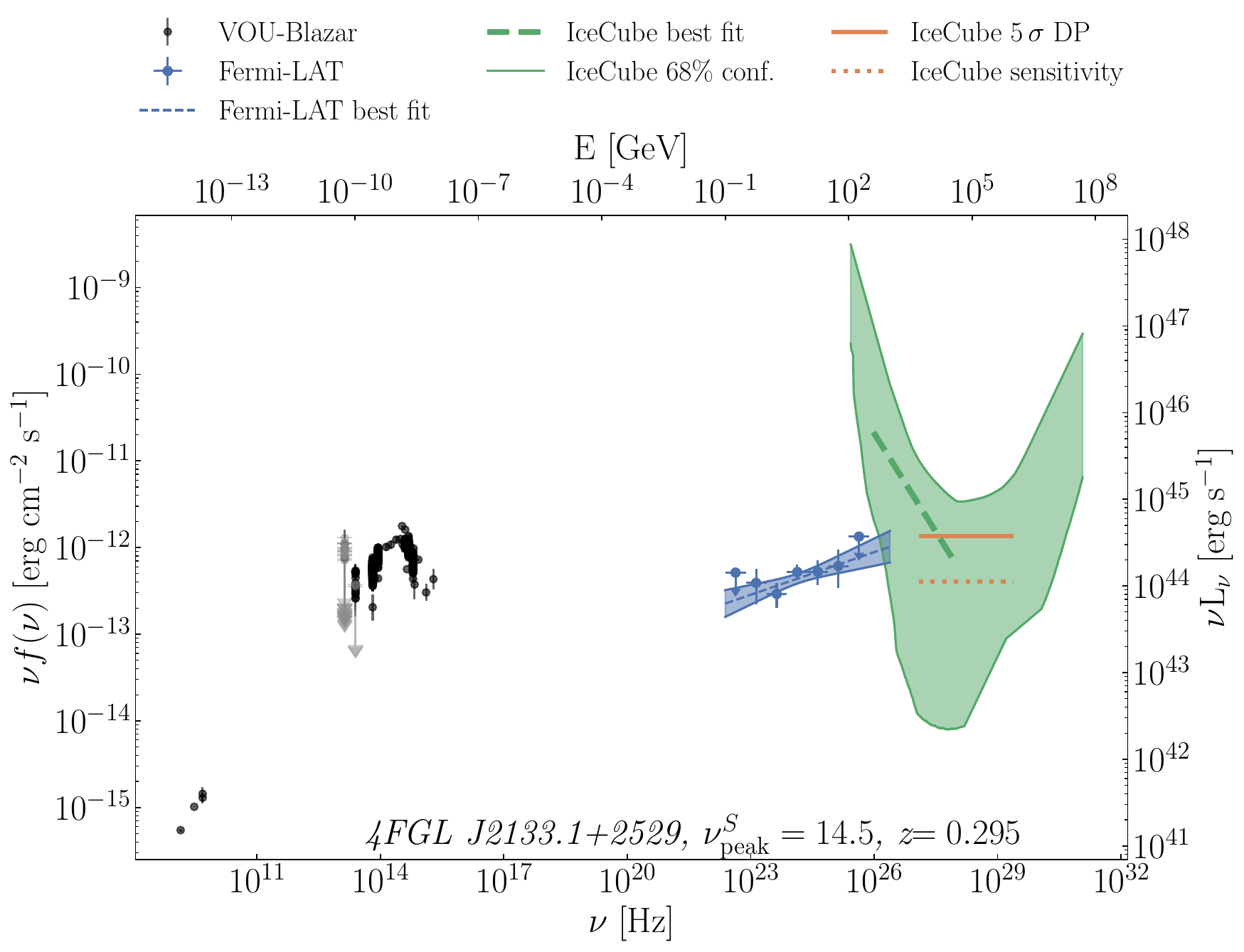}
\includegraphics[width=0.49\textwidth]{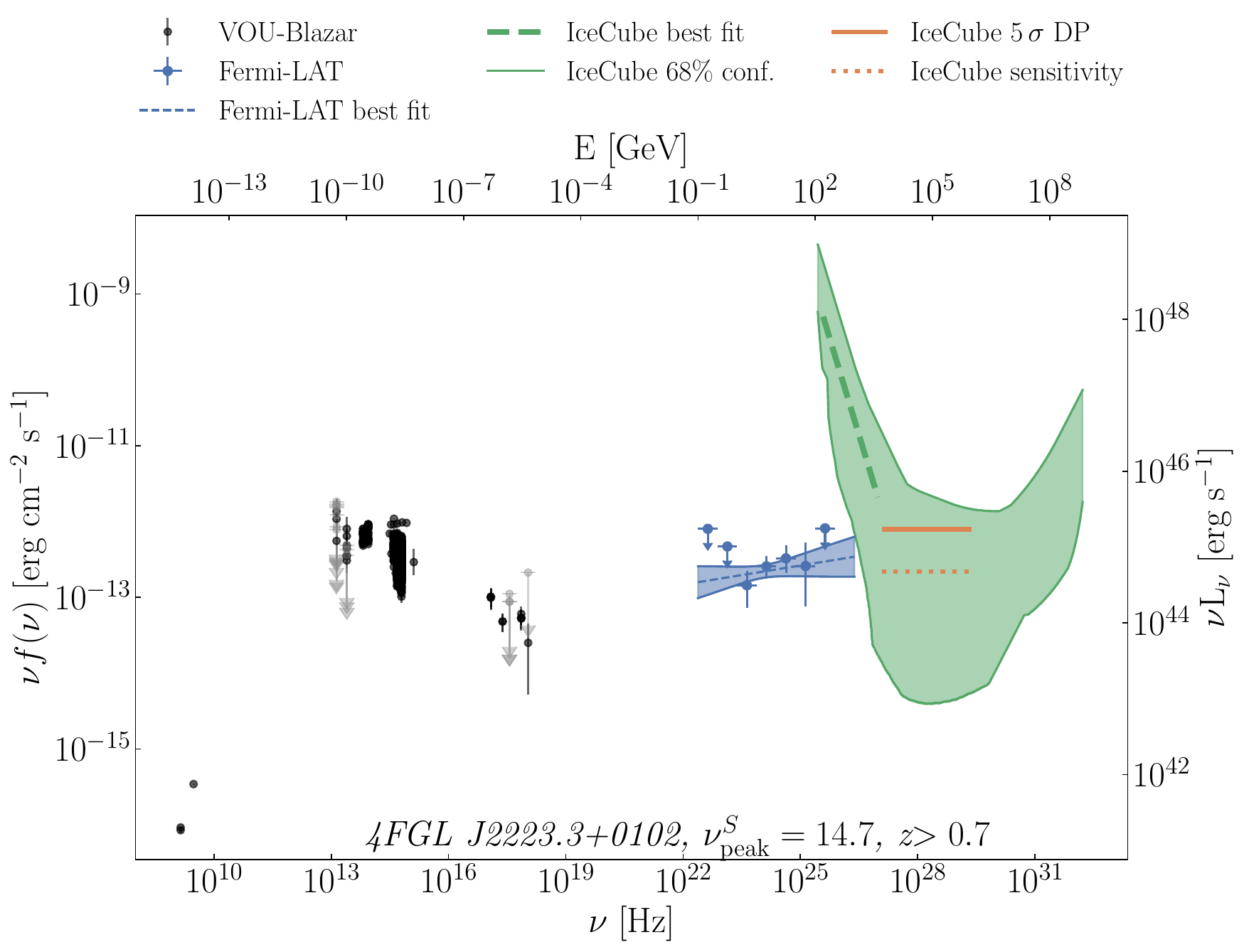}
\includegraphics[width=0.49\textwidth]{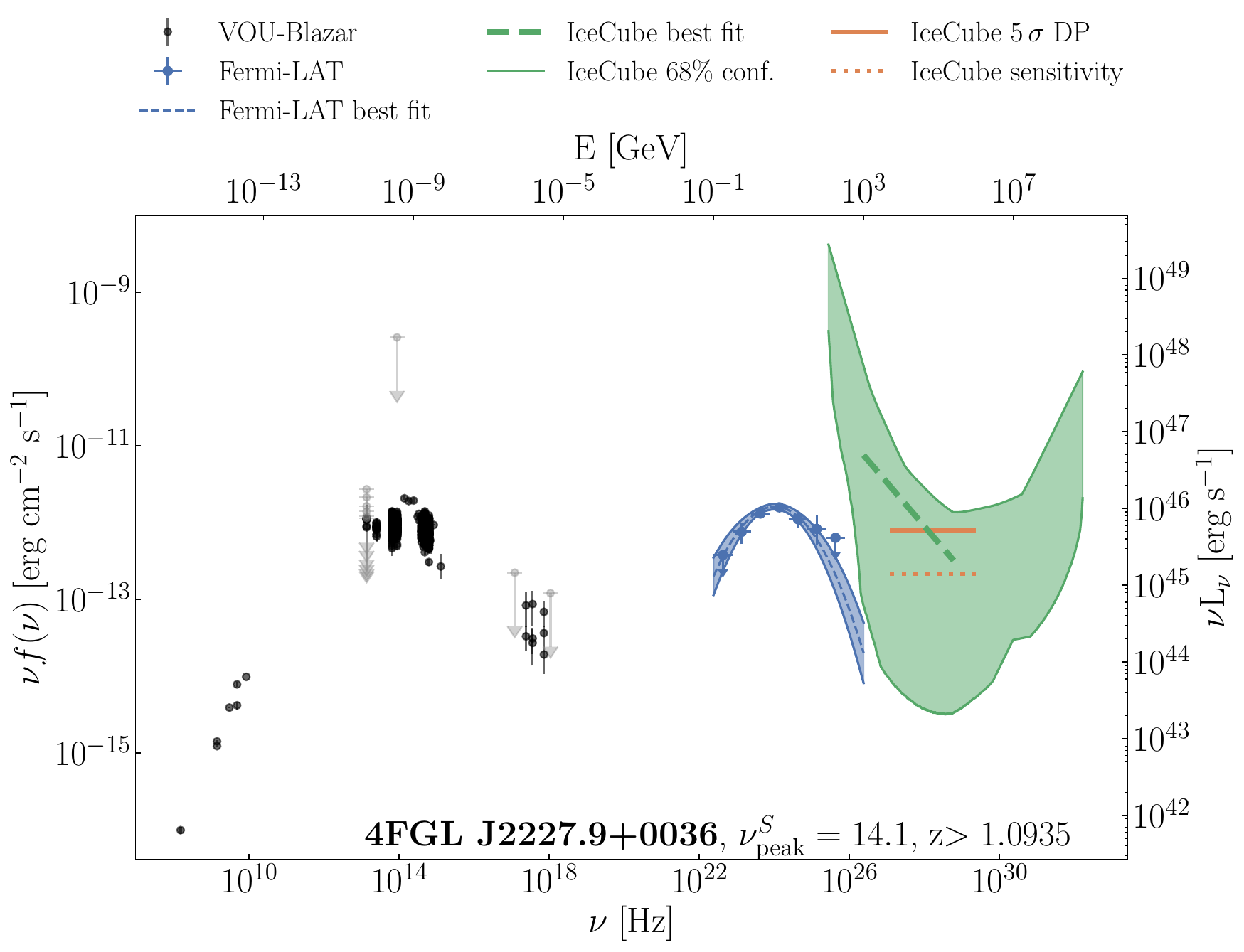}
\includegraphics[width=0.49\textwidth]{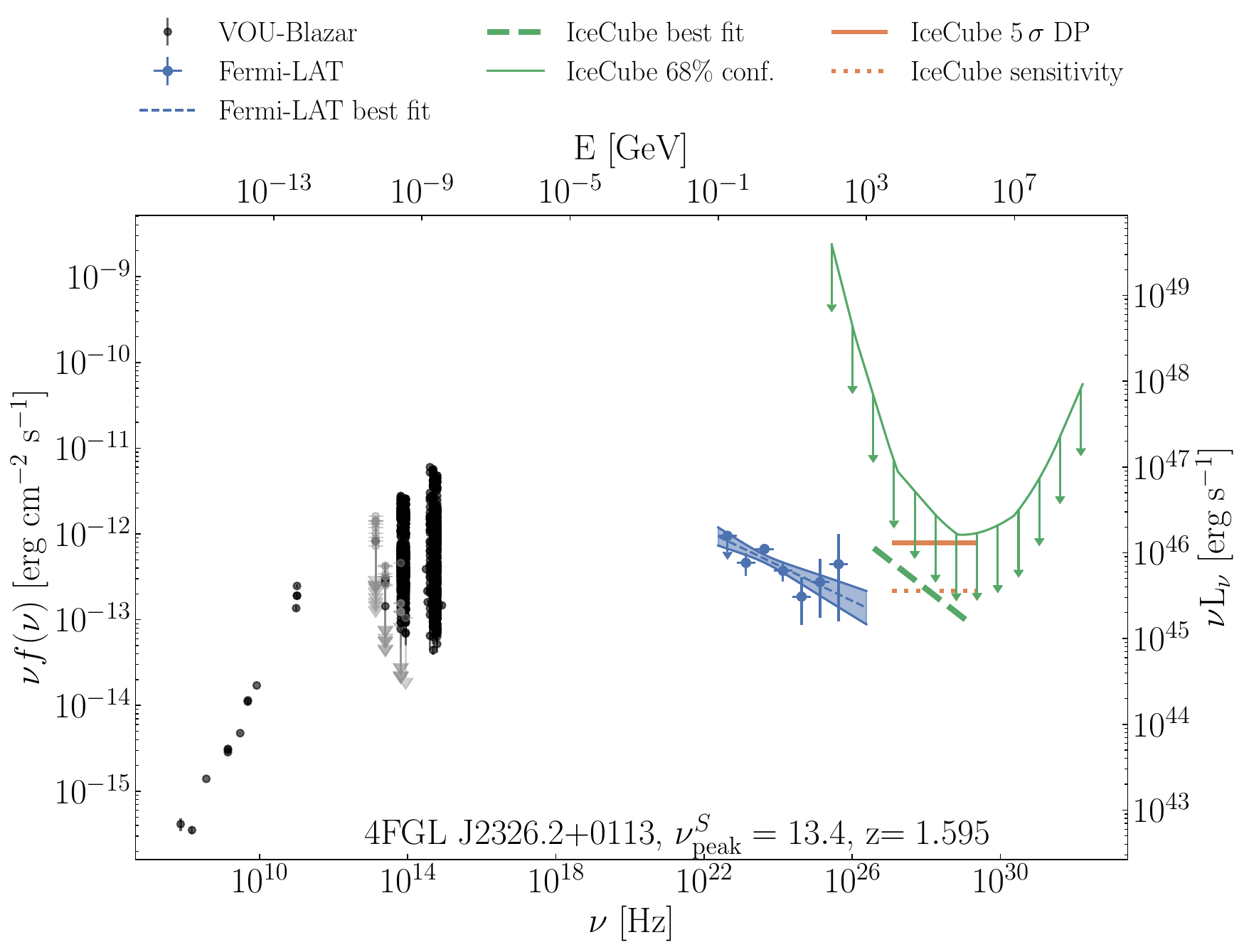}

\caption{- \textit{Continued}}
\end{figure*}

\newpage

 \begin{figure*}
 \includegraphics[width=0.49\textwidth]{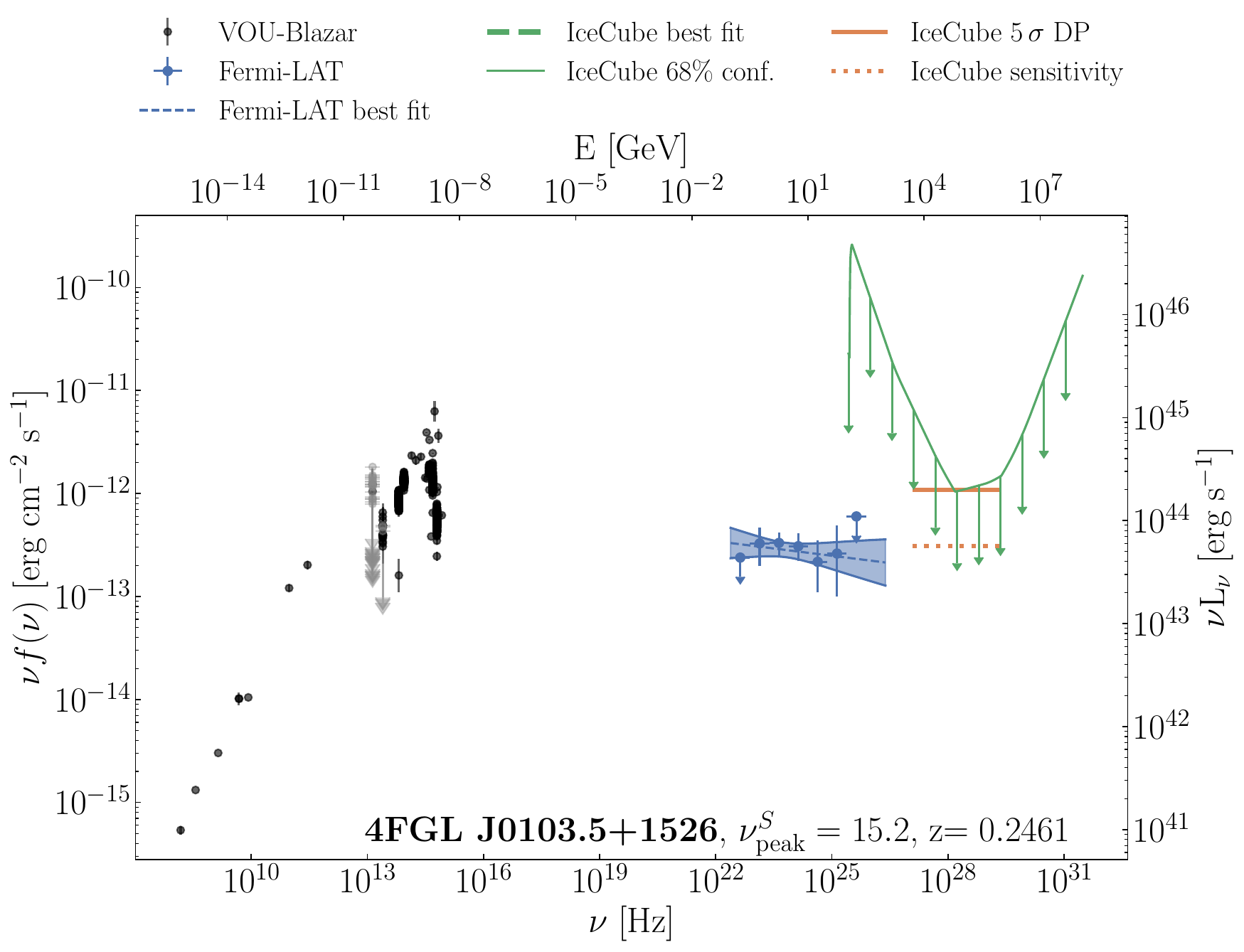}
\includegraphics[width=0.49\textwidth]{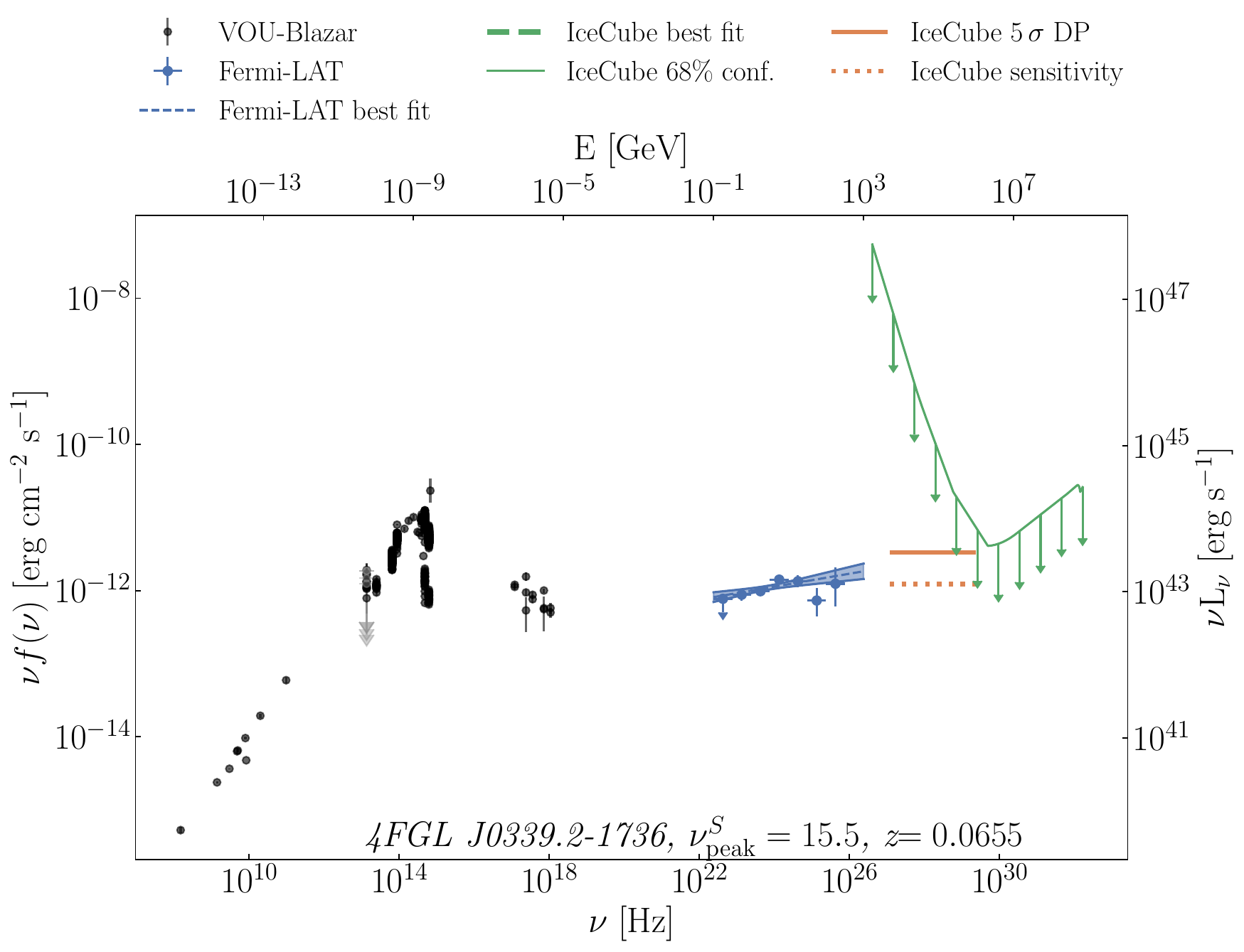}
 \includegraphics[width=0.49\textwidth]{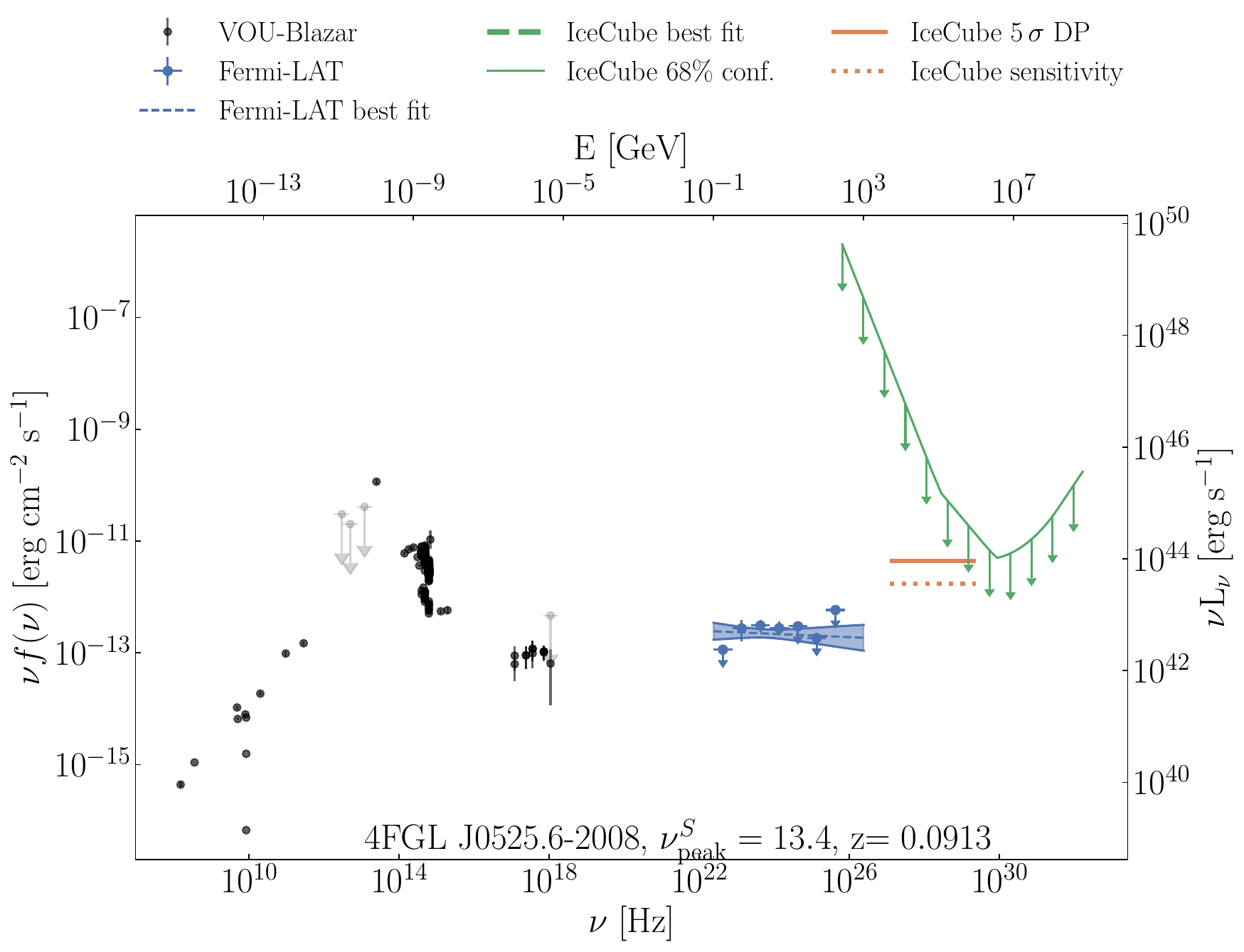}
\includegraphics[width=0.49\textwidth]{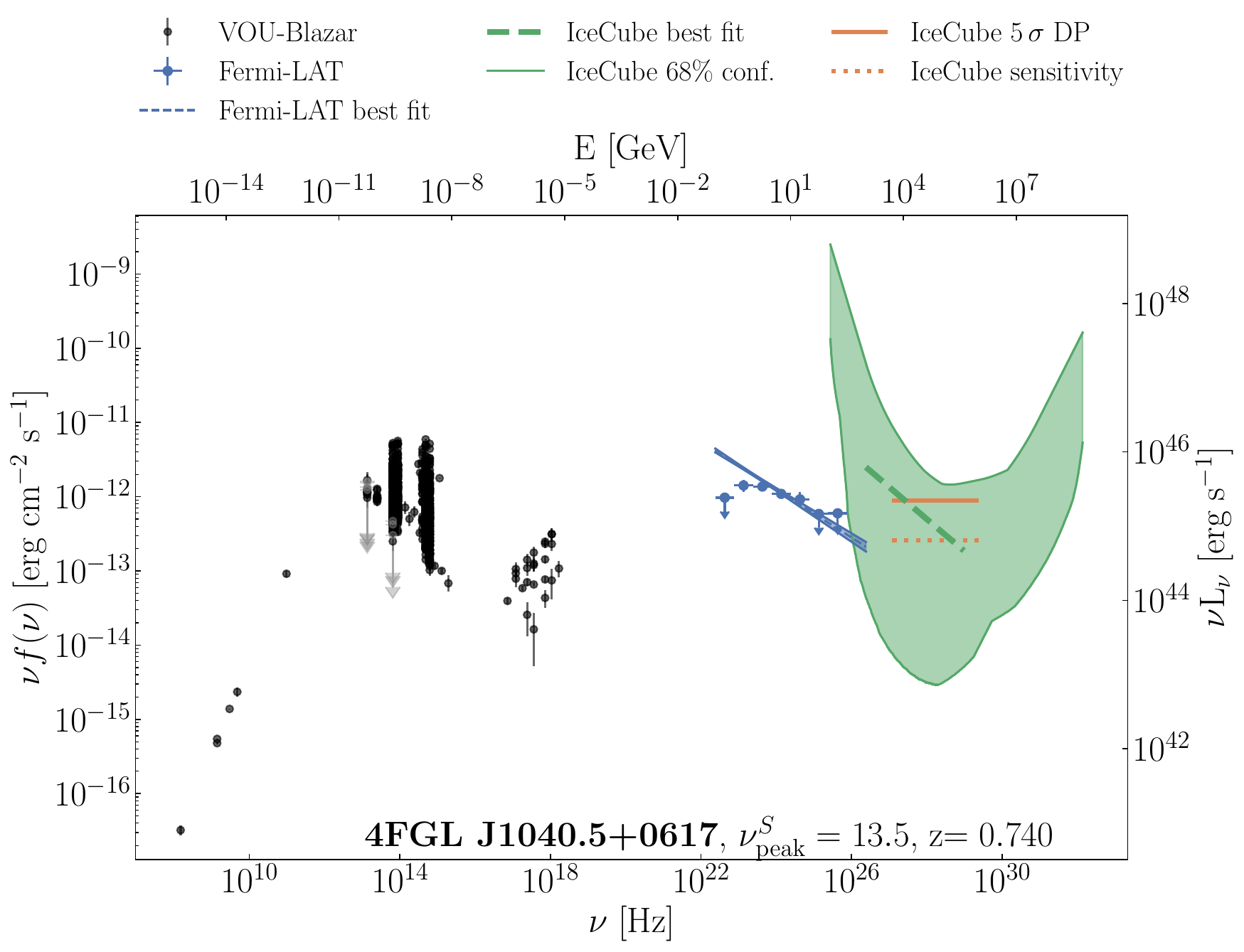}
\includegraphics[width=0.49\textwidth]{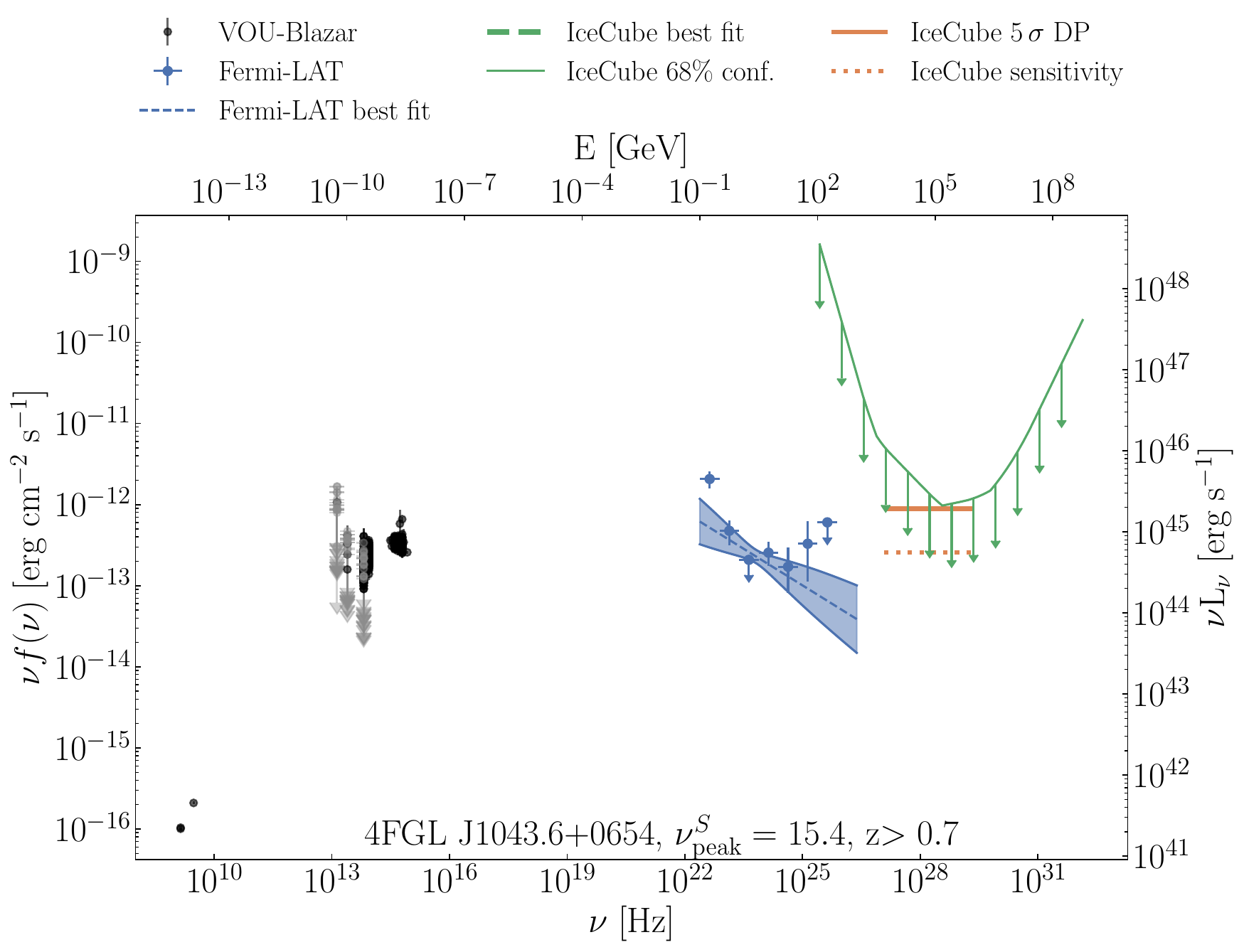}
 \includegraphics[width=0.49\textwidth]{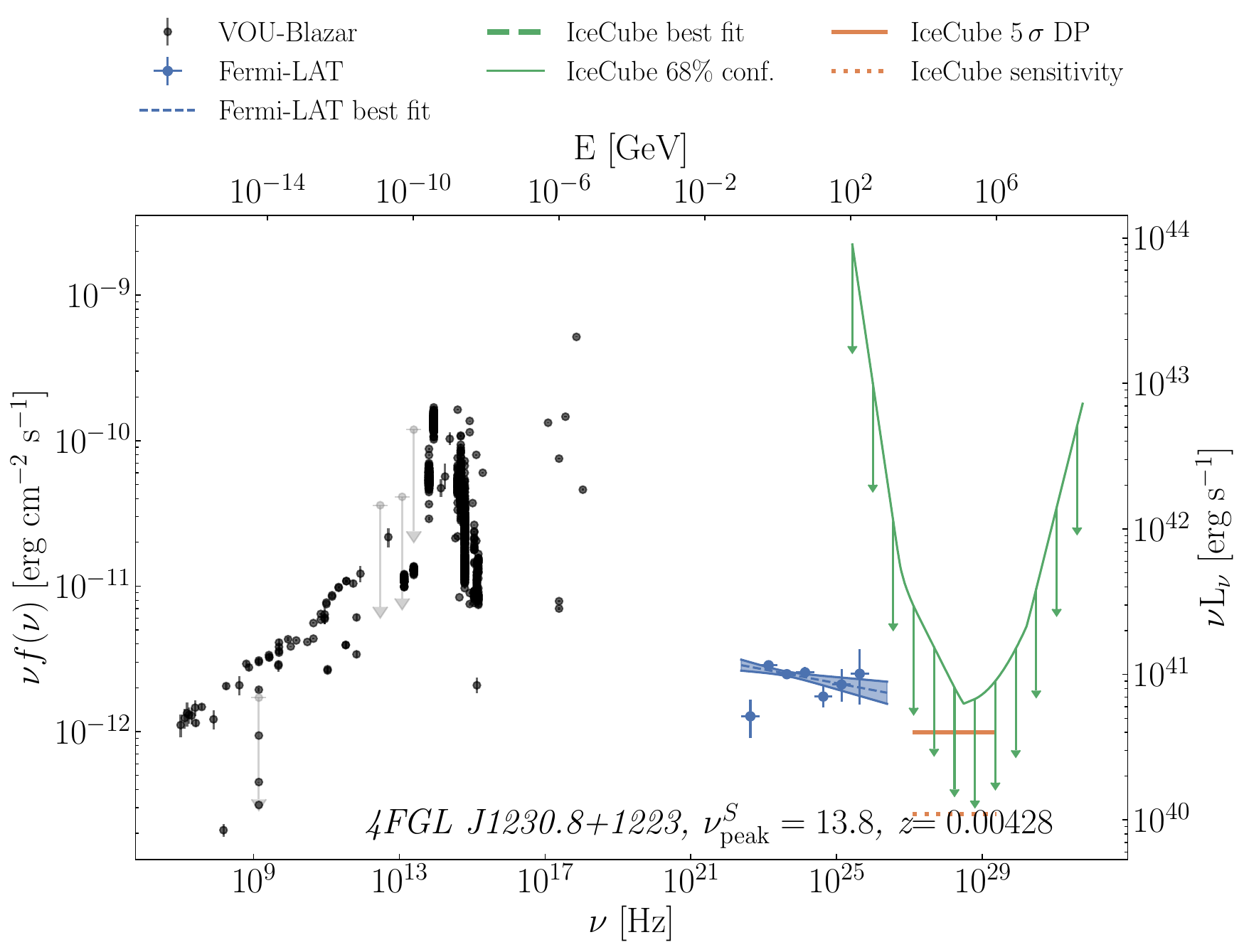}
 \caption{SEDs of all excluded objects. The black dots show the multi-wavelength data, and upper limits are displayed in grey. The blue dots display the best fit of the \fermi-LAT SED, and the blue bowtie indicates the uncertainties for the fitted gamma-ray flux. The solid (dashed) orange line shows the $5\,\sigma$ discovery potential (the sensitivity) for neutrino emission from \citet{Aartsen2020}. The discovery potential (DP) is the flux (with an $E^{-2}$ spectrum) necessary to detect the source with $5\,\sigma$ significance at 50 per cent confidence level. The sensitivity shows the 90 per cent confidence flux limit in case of no neutrino flux (also assuming an $E^{-2}$ spectrum). The green line (band) shows the 68 per cent confidence limits (band) on the best-fit neutrino flux based on public IceCube data \citep{IceCube_2021}. If the best-fit neutrino flux was $>0$, we show it as a green dashed line. All neutrinos fluxes are single flavour (muon neutrino and antineutrino) fluxes. Masquerading sources are marked in bold, and sources where the extension of the $\gamma$-ray flux meets the $5\,\sigma$ discovery potential or the best-fit neutrino flux are marked in italics.}
 \label{fig:excludedSEDs}
\end{figure*}

\setcounter{figure}{2}
\begin{figure*}
\includegraphics[width=0.49\textwidth]{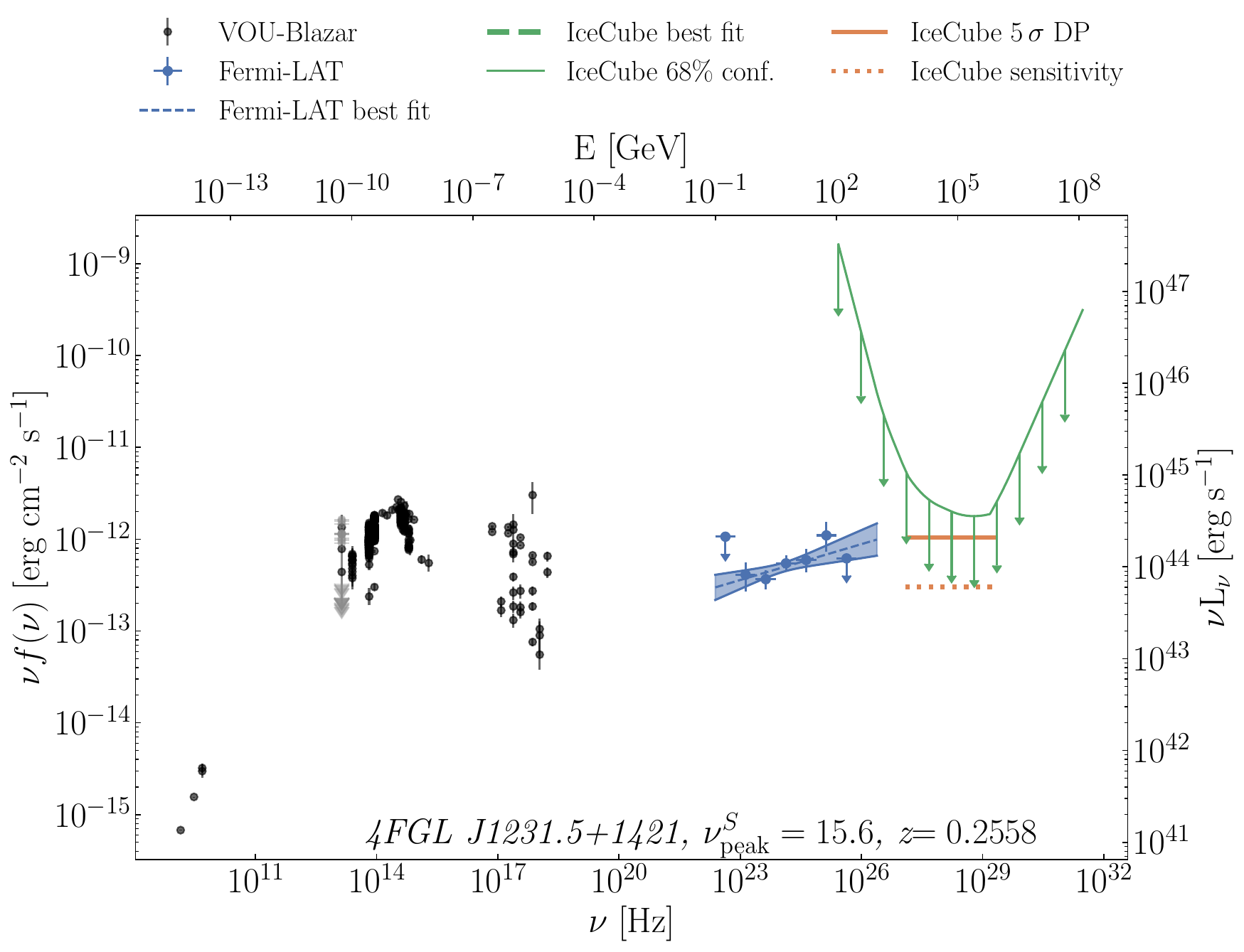}
\includegraphics[width=0.49\textwidth]{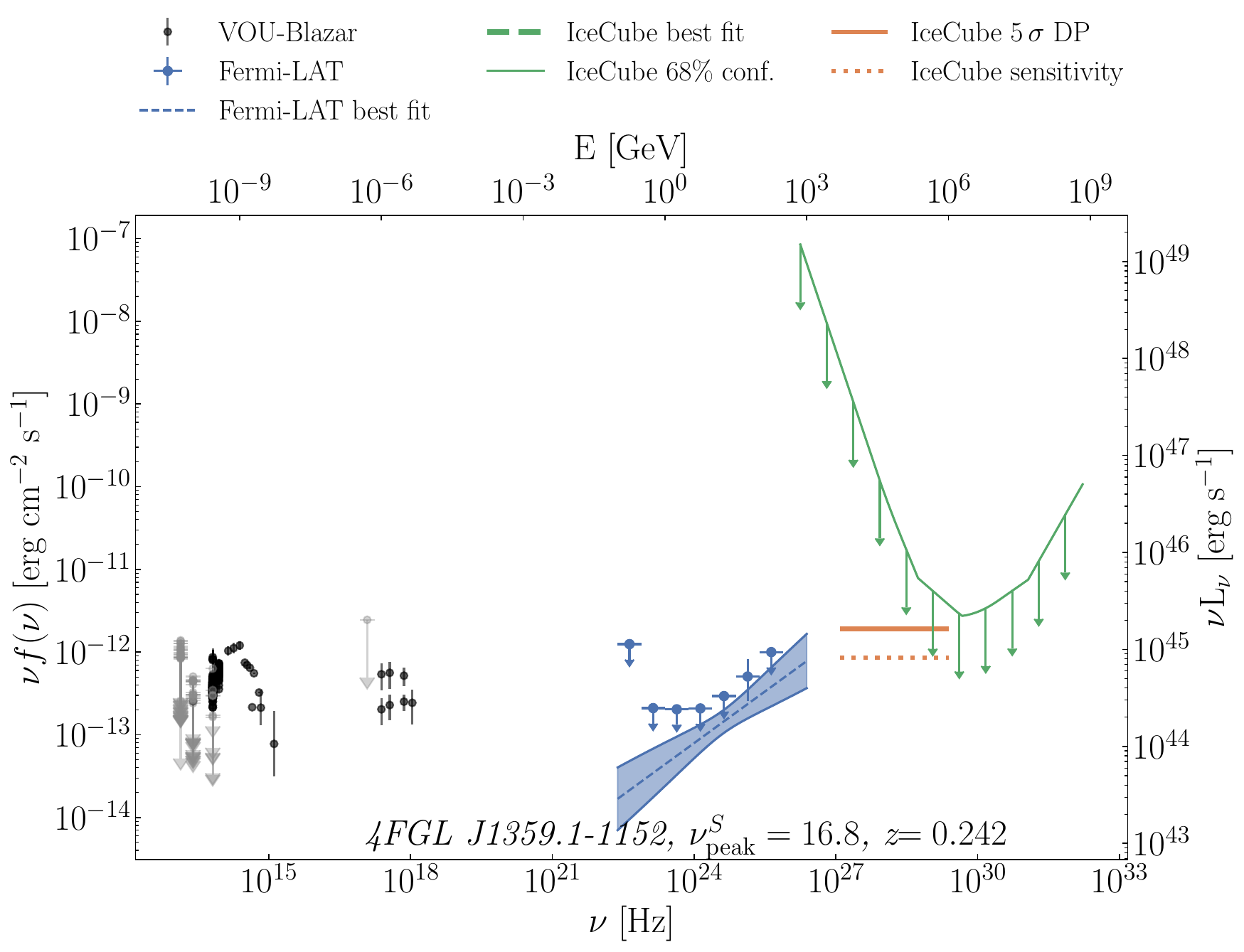}
\includegraphics[width=0.49\textwidth]{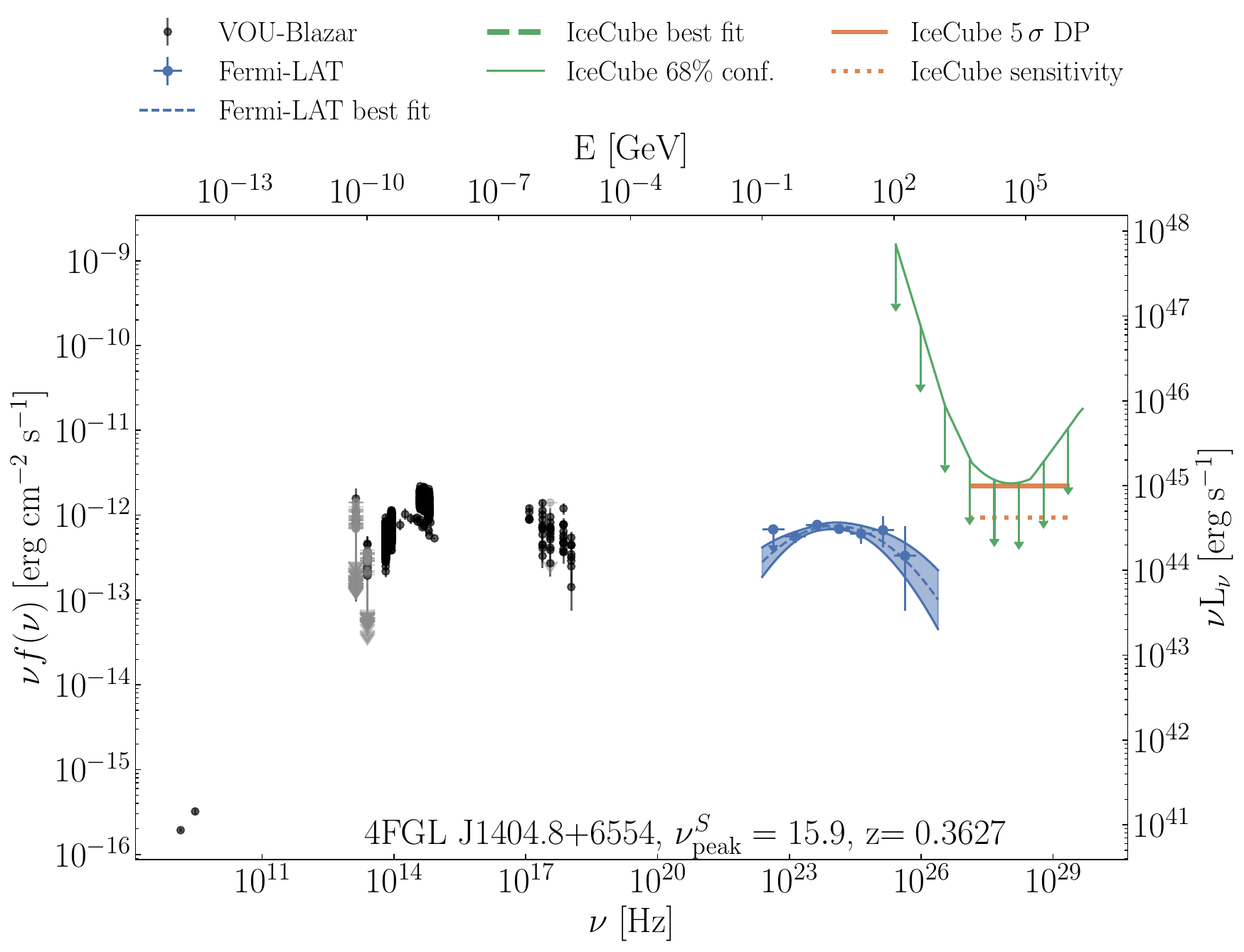}
\includegraphics[width=0.49\textwidth]{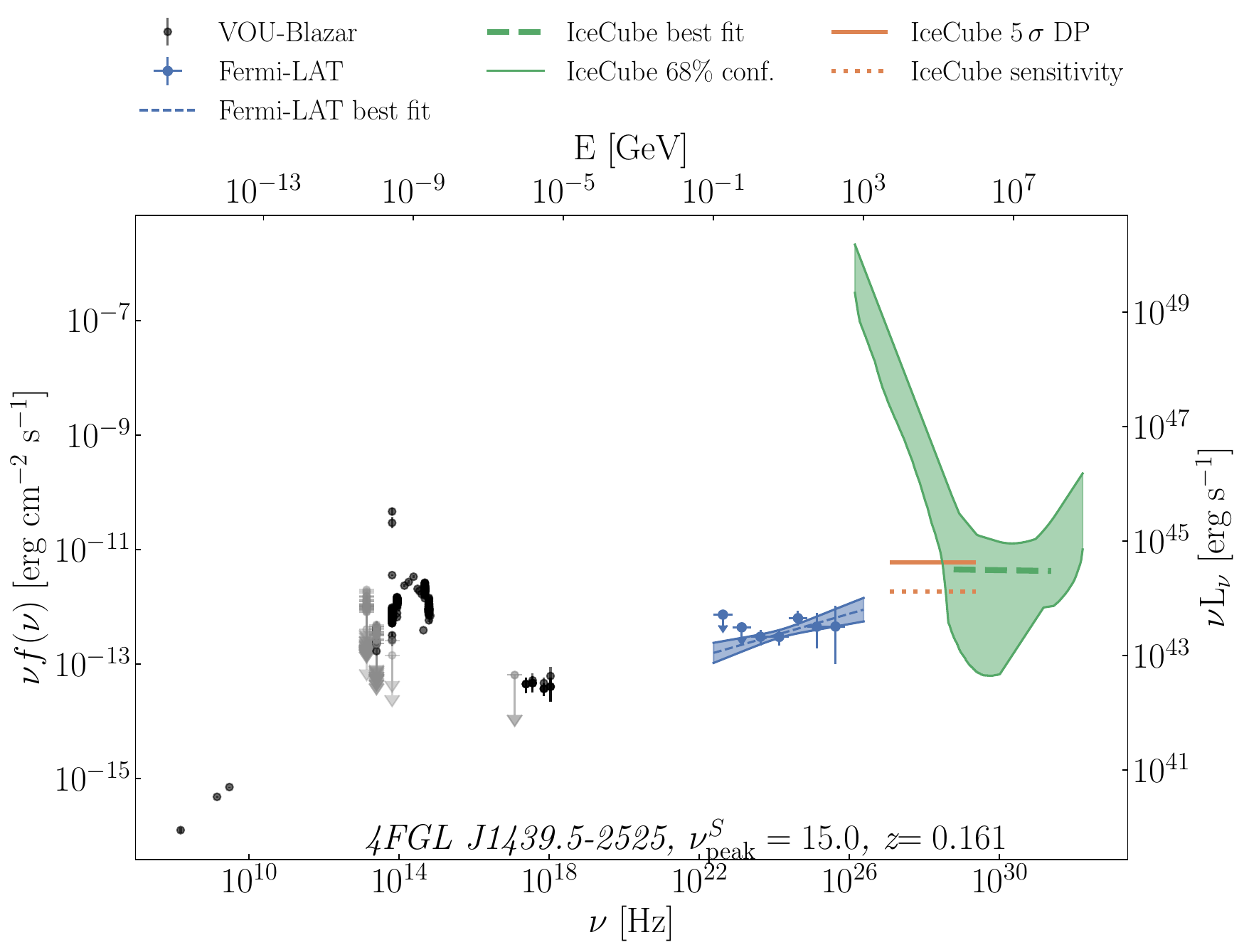}
\includegraphics[width=0.49\textwidth]{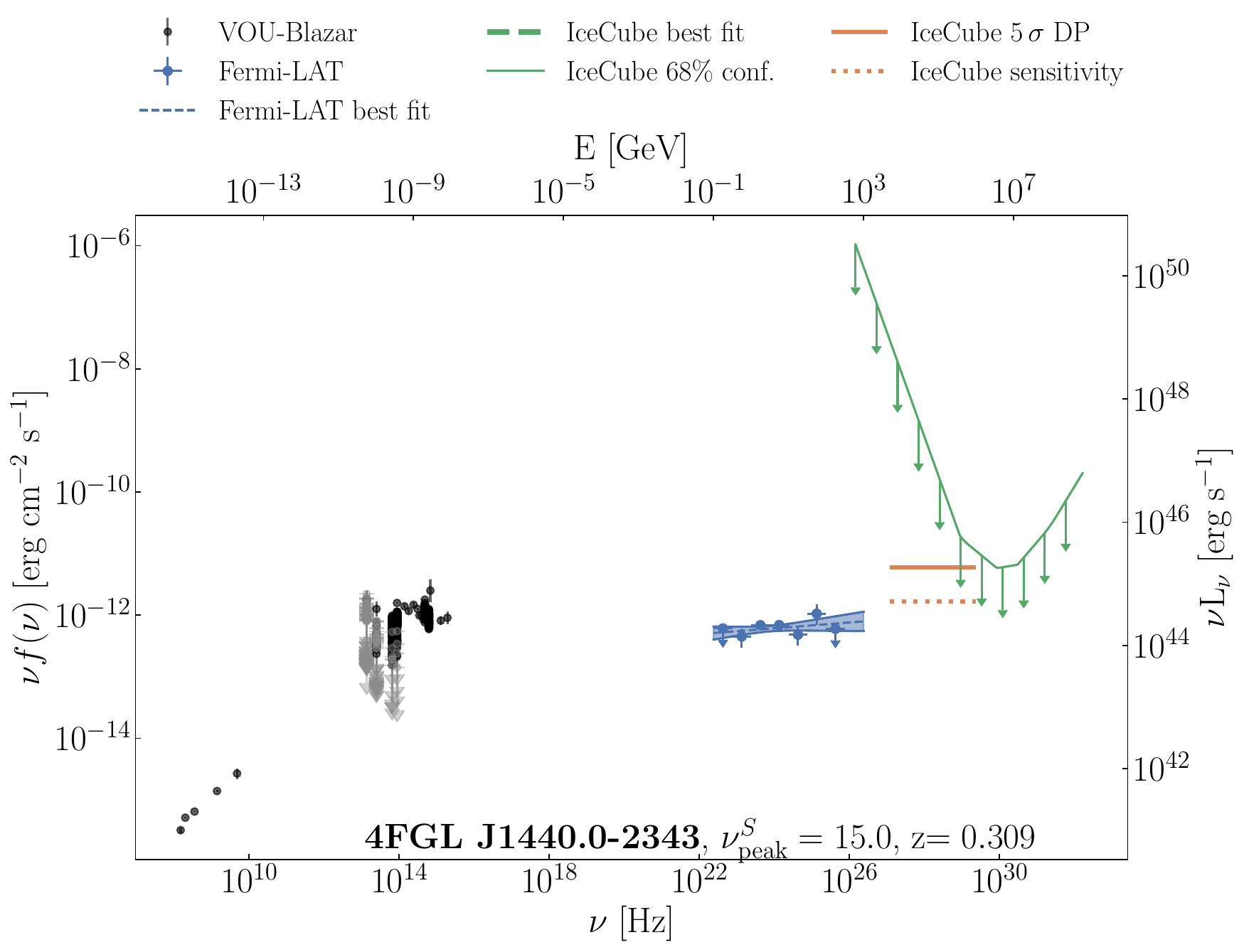}
\includegraphics[width=0.49\textwidth]{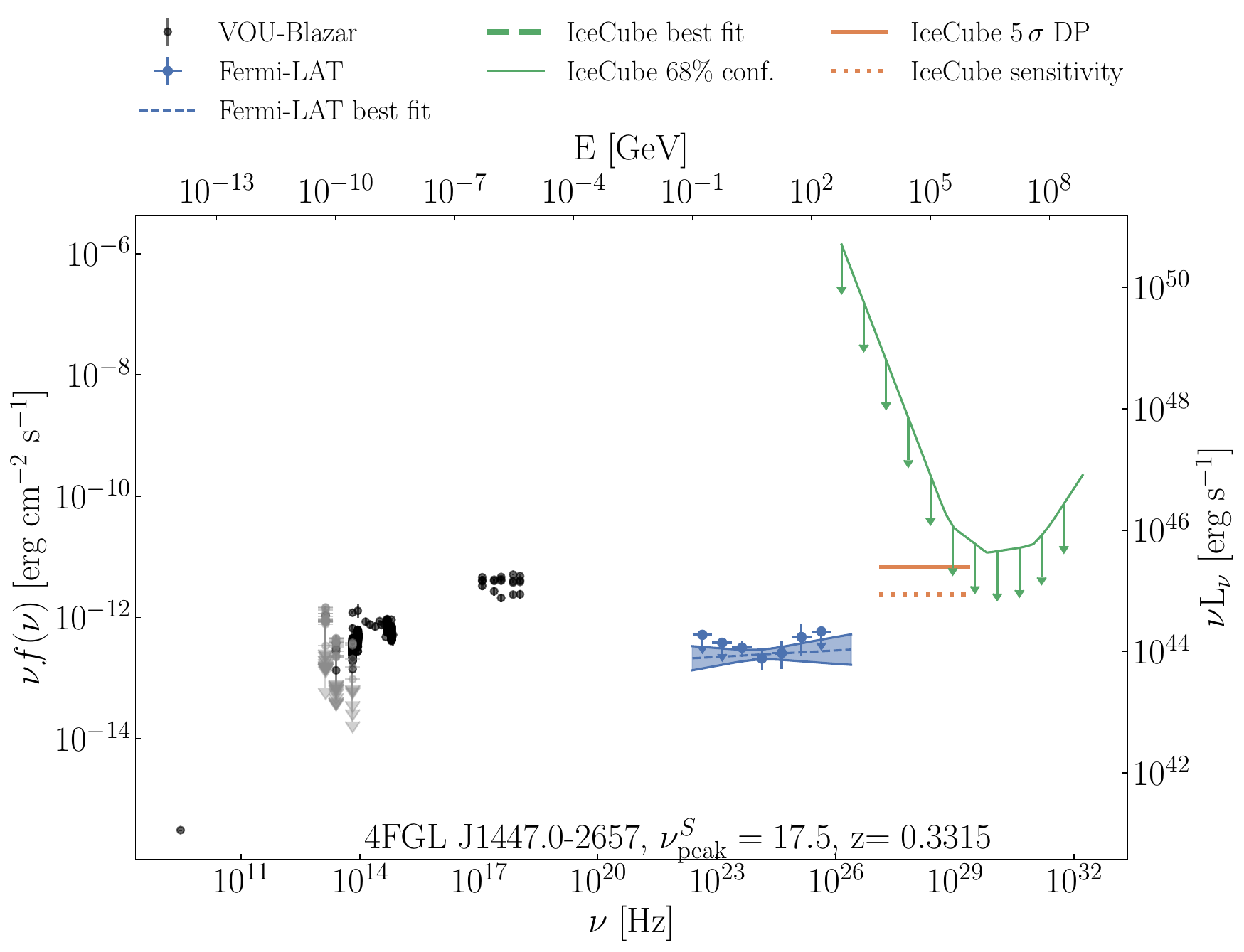}
\caption{- \textit{Excluded Sources -- Continued}}

\end{figure*}

\setcounter{figure}{2}
\begin{figure*}
\includegraphics[width=0.49\textwidth]{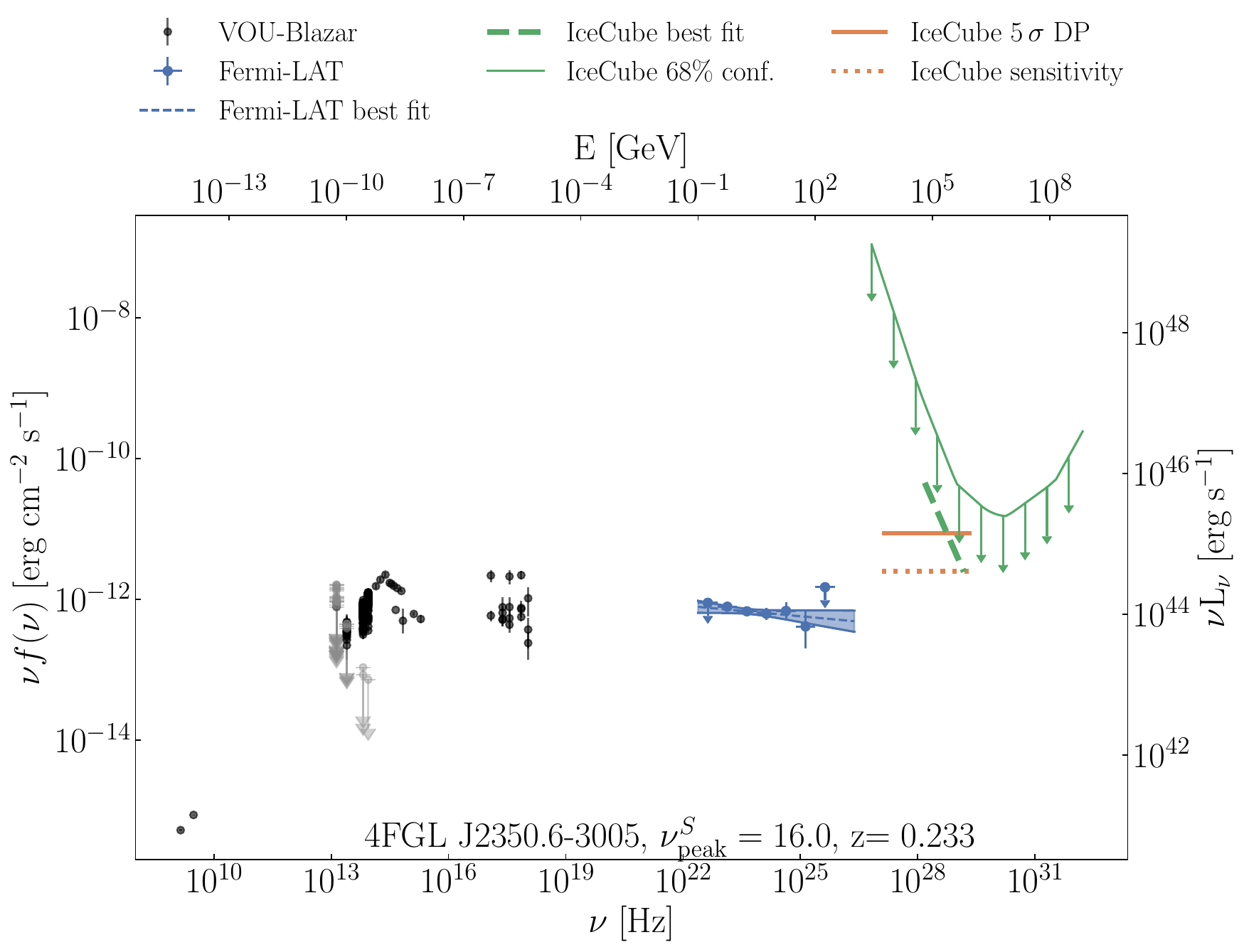}
\includegraphics[width=0.49\textwidth]{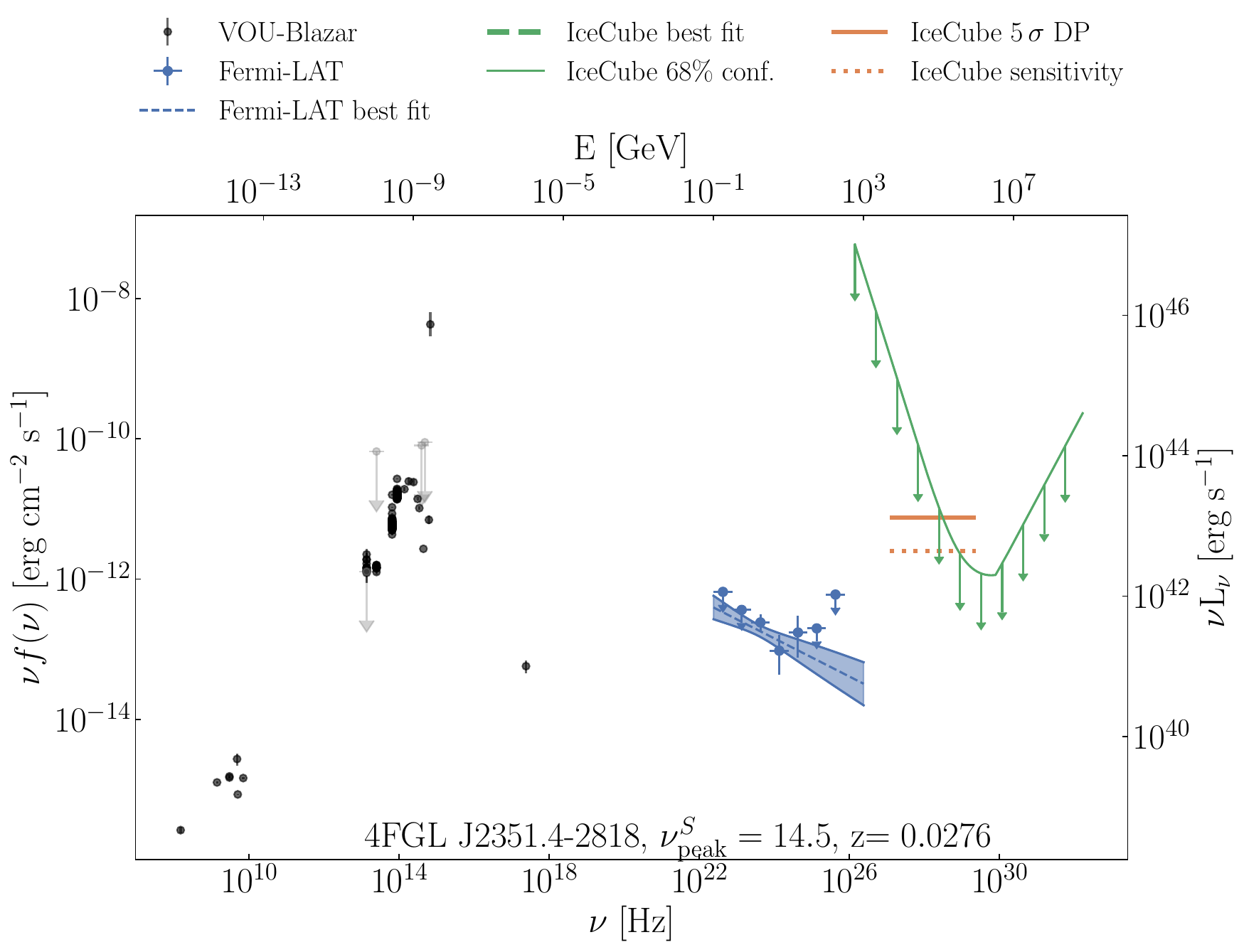}
\includegraphics[width=0.49\textwidth]{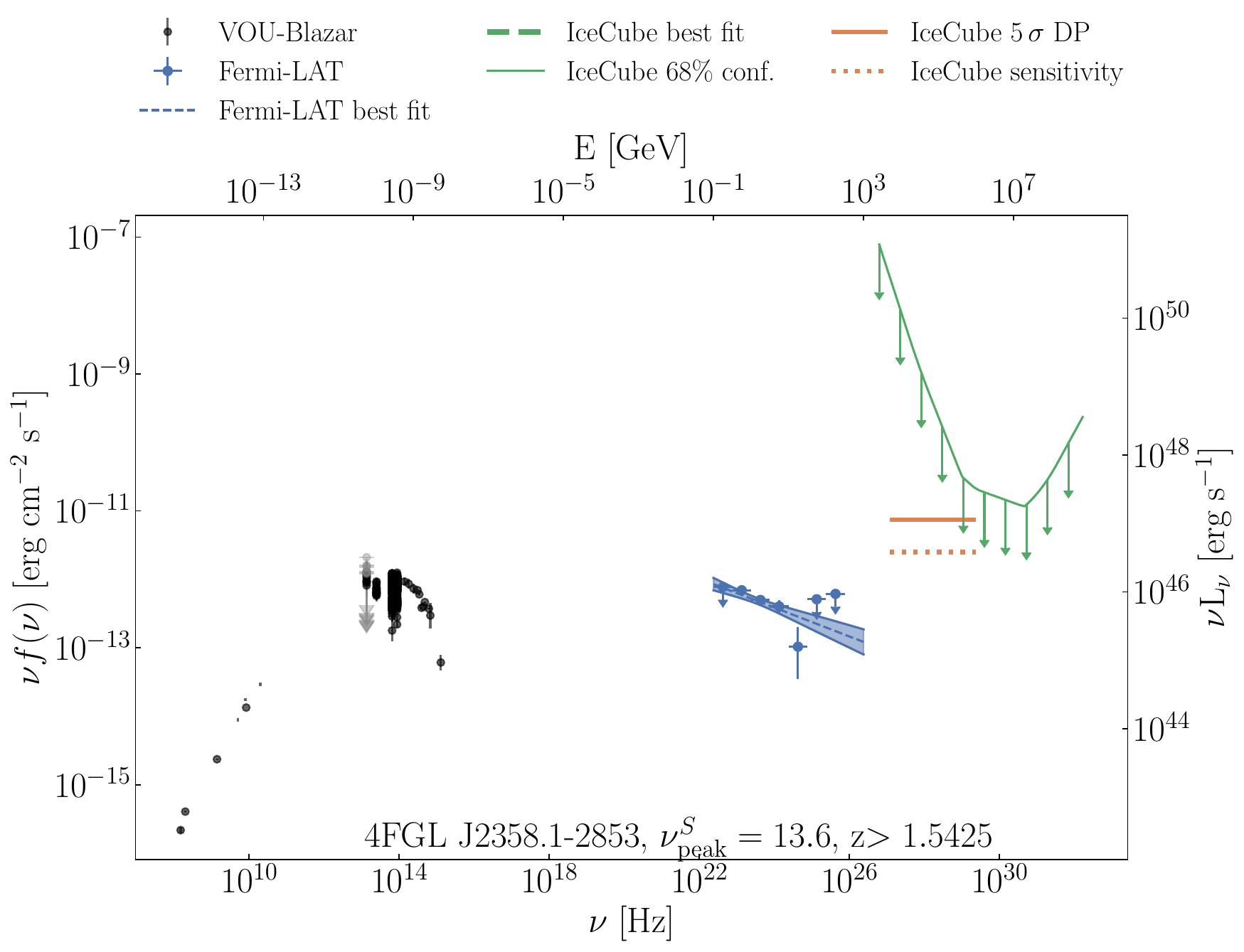}
\caption{- \textit{Excluded Sources -- Continued}}

\end{figure*}

\begin{scriptsize}
\begin{table*}
\setlength{\tabcolsep}{1.7pt}
\caption{Overview of the time span and sampling of the light curve data.}
\begin{center}
\begin{tabular}{llccccccccc}
\hline\hline
4FGL Name & Name & \multicolumn{3}{c}{IR} & \multicolumn{3}{c}{optical} & \multicolumn{3}{c}{X-ray} \\
 & & First & Last & N$_{\text{Obs}}$ & First & Last & N$_{\text{Obs}}$ & First & Last & N$_{\text{Obs}}$ \\
\hline
4FGL J0158.8+0101 & 5BZU J0158+0101 & 55210.33 & 59415.80 & 213 & 58315.44 & 59883.23 & 265 & - & - & -   \\ 
4FGL J0224.2+1616 & VOU J022411+161500 & 55221.32 & 59429.39 & 219 & 58301.48 & 59883.24 & 336 & - & - & -  \\ 
4FGL J0232.8+2018 & 3HSP J023248.5+20171 & 55223.83 & 59432.39 & 218 & 58301.48 & 59888.24 & 421 & 54683.60 & 59824.56 & 97  \\ 
4FGL J0239.5+1326 & 3HSP J023927.2+13273 & 55223.30 & 59431.74 & 228 & 58336.49 & 59874.39 & 119 & - & - &  \\ 
4FGL J0244.7+1316 & CRATESJ024445+132002 & 55224.36 & 59432.92 & 247 & 58305.47 & 59889.25 & 249 & - & - &  \\ 
4FGL J0344.4+3432 & 3HSP J034424.9+34301 & 55241.43 & 59450.43 & 237 & 58313.47 & 59889.24 & 326 & - & - & -   \\ 
4FGL J0509.4+0542 & TXS 0506+056 & 55257.57 & 59467.56 & 246 & 58204.14 & 59889.25 & 279 & 54882.43 & 59841.30 & 108   \\ 
3FGL J0627.9-1517 & 3HSP J062753.3-15195 & 55277.68 & 59490.70 & 287 & 58373.50 & 59882.42 & 111 & - & - & -  \\ 
4FGL J0649.5-3139 & 3HSP J064933.6-31392 & 55288.66 & 59499.59 & 352 & - & - & - & 56096.97 & 59026.75 & 12   \\ 
4FGL J0854.0+2753 & 3HSP J085410.1+27542 & 55310.11 & 59520.36 & 230 & 58206.19 & 59882.51 & 159 & - & - & -   \\ 
4FGL J0946.2+0104 & 3HSP J094620.2+01045 & 55329.42 & 59545.45 & 230 & 58202.25 & 59881.46 & 284 & 56110.50 & 57572.08 & 10   \\ 
4FGL J0955.1+3551 & 3HSP J095507.9+35510 & 55319.64 & 59529.38 & 272 & 58202.26 & 59889.49 & 342 & 56036.35 & 58900.05 & 35   \\ 
4FGL J1003.4+0205 & 3HSP J100326.6+02045 & 55332.86 & 59545.71 & 226 & 58202.26 & 59888.46 & 223 & 56671.93 & 59558.89 & 4   \\ 
4FGL J1055.7-1807 & VOU J105603-180929 & 55354.95 & 59357.92 & 232 & 58244.19 & 59722.20 & 57 & - & - & -   \\ 
4FGL J1117.0+2013 & 3HSP J111706.2+20140 & 55344.64 & 59553.55 & 233 & 58202.27 & 59743.26 & 273 & 54941.55 & 56491.25 & 4 \\ 
4FGL J1124.0+2045 & 3HSP J112405.3+20455 & 55345.96 & 59554.73 & 237 & 58202.29 & 59889.50 & 155 & - & - & -  \\ 
4FGL J1124.9+2143 & 3HSP J112503.6+21430 & 55345.83 & 59554.46 & 245 & 58235.22 & 59304.35 & 26 & - & - & -  \\ 
3FGL J1258.4+2123 & 3HSP J125821.5+21235 & 55366.21 & 59368.77 & 240 & 58235.23 & 58967.25 & 14 & - & - & -  \\ 
4FGL J1258.7-0452 & 3HSP J125848.0-04474 & 55205.27 & 59379.10 & 226 & 58203.31 & 59758.22 & 157 & 56327.21 & 57728.78 & 11  \\ 
4FGL J1300.0+1753 & 3HSP J130008.5+17553 & 55370.44 & 59370.61 & 260 & 58202.31 & 59770.19 & 334 & - & - & - \\ 
4FGL J1314.7+2348 & 5BZB J1314+2348 & 55371.11 & 59371.26 & 283 & 58202.31 & 59771.25 & 558 & - & - & -  \\ 
4FGL J1321.9+3219 & 5BZB J1322+3216 & 55366.08 & 59368.78 & 266 & 58202.35 & 59771.25 & 488 & - & - & -  \\ 
4FGL J1507.3-3710 & VOU J150720-370902 & 55240.88 & 59422.10 & 139 & - & - & - & - & - & -  \\ 
4FGL J1528.4+2004 & 3HSP J152835.7+20042 & 55228.03 & 59408.52 & 256 & 58203.40 & 59812.22 & 268 & 58677.59 & 58760.32 & 6   \\ 
4FGL J1533.2+1855 & 3HSP J153311.2+18542 & 55229.61 & 59414.27 & 321 & 58333.19 & 59390.25 & 21 & 55725.55 & 55942.06 & 5   \\ 
4FGL J1554.2+2008 & 3HSP J155424.1+20112 & 55232.86 & 59418.45 & 287 & 58203.42 & 59853.14 & 400 & 57004.02 & 58767.17 & 8  \\ 
4FGL J1808.2+3500 & CRATESJ180812+350104 & 55278.31 & 59461.32 & 430 & 58204.43 & 59887.16 & 963 & - & - & -  \\ 
4FGL J1808.8+3522 & 3HSP J180849.7+35204 & 55278.57 & 59461.72 & 442 & 58245.49 & 59402.33 & 38 & - & - & -  \\ 
4FGL J2030.5+2235 & 3HSP J203031.6+22343 & - & - & - & 58230.50 & 59883.16 & 229 & - & - & - \\ 
4FGL J2030.9+1935 & 3HSP J203057.1+19361 & 55321.31 & 59147.21 & 65 & 58211.51 & 59890.13 & 455 & - & - & -  \\ 
4FGL J2133.1+2529 & 3HSP J213314.3+25285 & 55342.35 & 59524.32 & 287 & 58277.48 & 59075.39 & 51 & - & - & - \\ 
4FGL J2223.3+0102 & 3HSP J222329.5+01022 & - & - & - & 58262.47 & 59883.23 & 499 & - & - & - \\ 
4FGL J2227.9+0036 & 5BZB J2227+0037 & 55345.27 & 59531.44 & 238 & 58263.48 & 59883.17 & 397 & - & - & -    \\ 
4FGL J2326.2+0113 & CRATESJ232625+011147 & 55361.01 & 59542.22 & 233 & 58288.42 & 59889.23 & 276 & - & - &   \\ 
\hline\hline
\end{tabular}
\end{center}
\footnotesize {\textit{Notes.} Overview of the time span and the sampling of the light curve data used to calculate the FVs in the IR, optical, and X-ray bands for the sources that meet the newly revised criteria (Section \ref{sec:update-tracks}). For each object, we list the dates of the first and last observation available and the number of observed data points N$_{\text{Obs}}$. All dates are given in MJD format.}
\label{Tab:LCDataOverview}
\end{table*}
\end{scriptsize}

\begin{scriptsize}
\begin{table*}
\caption{Overview of neutrino events and their coincidence with flares in different bands.}
\begin{center}
\begin{tabular}{llcccccccc}
\hline\hline
4FGL Name & Name &\multicolumn{2}{c}{IR} &\multicolumn{2}{c}{optical} &\multicolumn{2}{c}{X-ray}  &\multicolumn{2}{c}{$\gamma$-ray} \\
 & &flare &$P_{\text{BG}}$ &flare &$P_{\text{BG}}$ &flare &$P_{\text{BG}}$ &flare &$P_{\text{BG}}$ \\
\hline
4FGL J0158.8+0101 &  5BZU J0158+0101 & \xmark  & 0.103 & - & - & \xmark & 0.0 & - & -  \\ 
4FGL J0224.2+1616 &  VOU J022411+161500 & \xmark  & 0.1447 & - & - & - & - & - & -  \\ 
4FGL J0232.8+2018 &  3HSP J023248.5+20171 & \xmark  & 0.0004 & - & - &  \checkmark  & 0.3647 & - & -  \\ 
4FGL J0239.5+1326 &  3HSP J023927.2+13273 & \xmark  & 0.1001 & \xmark & 0.0761 & \xmark & 0.0 & \xmark & 0.0  \\ 
4FGL J0244.7+1316 &  CRATESJ024445+132002 & \xmark  & 0.086 & \xmark & 0.0188 & \xmark & 0.0 & \xmark & 0.0  \\ 
4FGL J0344.4+3432 &  3HSP J034424.9+34301 & \xmark  & 0.2684 & - & - & \xmark & 0.0 & - & -  \\ 
4FGL J0509.4+0542 &  TXS 0506+056 &  \checkmark  & 0.1813 &  \checkmark  & 0.209 & \xmark & 0.0967 &  \checkmark  & 0.0459  \\ 
3FGL J0627.9-1517 &  3HSP J062753.3-15195 & \xmark  & 0.0 & - & - & \xmark & 0.0 & - & -  \\ 
4FGL J0649.5-3139 &  3HSP J064933.6-31392 &  \checkmark  & 0.4092 & \xmark & 0.0141 & \xmark & 0.0517 & \xmark & 0.0  \\ 
4FGL J0854.0+2753 &  3HSP J085410.1+27542 & \xmark  & 0.0 & - & - & - & - & - & -  \\ 
4FGL J0946.2+0104 &  3HSP J094620.2+01045 & \xmark  & 0.0584 & \xmark & 0.1708 & \xmark & 0.0 & - & -  \\ 
4FGL J0955.1+3551 &  3HSP J095507.9+35510 & \xmark  & 0.1996 & \xmark & 0.1798 &  \checkmark  & 0.4555 & - & -  \\ 
4FGL J1003.4+0205 &  3HSP J100326.6+02045 & \xmark  & 0.1649 &  \checkmark  & 0.2492 & \xmark & 0.0 & - & -  \\ 
4FGL J1055.7-1807 &  VOU J105603-180929 & \xmark  & 0.0 & - & - & \xmark & 0.0 & - & -  \\ 
4FGL J1117.0+2013 &  3HSP J111706.2+20140 & \xmark  & 0.2804 & - & - & \xmark & 0.7629 & \xmark & 0.0  \\ 
4FGL J1124.0+2045 &  3HSP J112405.3+20455 & \xmark  & 0.0801 & - & - & - & - & - & -  \\ 
4FGL J1124.9+2143 &  3HSP J112503.6+21430 & \xmark  & 0.1198 & - & - & - & - & - & -  \\ 
3FGL J1258.4+2123 &  3HSP J125821.5+21235 & \xmark  & 0.0 & - & - & \xmark & 0.0 & - & -  \\ 
4FGL J1258.7-0452 &  3HSP J125848.0-04474 &  \checkmark  & 0.167 & - & - & \xmark & 0.1965 & - & -  \\ 
4FGL J1300.0+1753 &  3HSP J130008.5+17553 & \xmark  & 0.125 & - & - & - & - & - & -  \\ 
4FGL J1314.7+2348 &  5BZB J1314+2348 & \xmark  & 0.295 & \xmark & 0.05 & \xmark & 0.0 & \xmark & 0.0  \\ 
4FGL J1321.9+3219 &  5BZB J1322+3216 & \xmark  & 0.0836 & - & - & - & - & - & -  \\ 
4FGL J1507.3-3710 &  VOU J150720-370902 & \xmark  & 0.0 & - & - & - & - & - & -  \\ 
4FGL J1528.4+2004 &  3HSP J152835.7+20042 & \xmark  & 0.0602 & - & - & \xmark & 0.0371 & - & -  \\ 
4FGL J1533.2+1855 &  3HSP J153311.2+18542 & \xmark  & 0.0 & - & - & \xmark & 0.0 & - & -  \\ 
4FGL J1554.2+2008 &  3HSP J155424.1+20112 & \xmark  & 0.0959 & - & - & \xmark & 0.0012 & - & -  \\ 
4FGL J1808.2+3500 &  CRATESJ180812+350104 & \xmark  & 0.222 & - & - & - & - & \xmark & 0.0  \\ 
4FGL J1808.8+3522 &  3HSP J180849.7+35204 & \xmark  & 0.3084 & - & - & \xmark & 0.0 & - & -  \\ 
4FGL J2030.5+2235 &  3HSP J203031.6+22343 & - & - & - & - & \xmark & 0.4595 & - & -  \\ 
4FGL J2030.9+1935 &  3HSP J203057.1+19361 & \xmark  & 0.286 & - & - & \xmark & 0.0 & - & -  \\ 
4FGL J2133.1+2529 &  3HSP J213314.3+25285 & \xmark  & 0.0814 & - & - & - & - & - & -  \\ 
4FGL J2223.3+0102 &  3HSP J222329.5+01022 & - & - & - & - & \xmark & 0.0 & - & -  \\ 
4FGL J2227.9+0036 &  5BZB J2227+0037 & \xmark  & 0.2049 & - & - & \xmark & 0.0 & - & -  \\ 
4FGL J2326.2+0113 &  CRATESJ232625+011147 & \xmark  & 0.1996 & \xmark & 0.0417 & \xmark & 0.0 & \xmark & 0.0  \\ 
\hline\hline
\end{tabular}
\end{center}
\footnotesize {\textit{Notes.} Overview of all G20 sources meeting the revised criteria in Section \ref{sec:update-tracks} with information on their flaring state at the time of the neutrino arrival (\checkmark ~if yes, \xmark ~if no, - if no data was available) for the different bands. $P_{\text{BG}}$ gives the chance probability of finding the source in a flaring state (fraction of the time a source was flaring vs. the whole observation time), with $P_{\text{BG}} = 0$ if there was no flare in the light curve. TXS~0506+056 is the only object found in a flaring state in several wavelengths during the neutrino arrival time.}
\label{Tab:COINCresults}

\end{table*}
\end{scriptsize}

\label{lastpage}

\bsp	

\end{document}